\documentclass[a4paper,fleqn,usenatbib
											]{mnras}

\usepackage{
	newtxmath}

\usepackage[T1]{fontenc}
\usepackage{ae,aecompl}

\usepackage{graphicx}	
\usepackage{amsmath}	
\usepackage{amssymb}	
\usepackage{accents} 
\usepackage[mathscr]{euscript}
\usepackage[group-digits=false]{siunitx} 
\usepackage{bigints} 
\usepackage{mathtools} 
\usepackage{xfrac} 
\usepackage{centernot} 

\def\be#1\ee{\begin{equation}#1\end{equation}}
\def\bml#1\eml{\begin{multline}#1\end{multline}}

\newcommand{\opb}{}
\newcommand{\clb}{}

\newcommand{\eref}[1]{Eq.~(\ref{#1})}

\newcommand{\dotp}{\boldsymbol{\cdot}}
\newcommand{\infrac}[2]{{\sfrac{#1}{#2}}}
\renewcommand{\vec}[1]{\mathbfit{#1}}

\newcommand{\bell}[1]{\accentset{\cong}{#1}}

\newcommand{\ii}{\mathrm{i}}
\newcommand{\ex}{\mathrm{e}}
\newcommand{\bego}{\bell{g}_1}

\newcommand{\x}{\vec{x}}
\renewcommand{\u}{\vec{u}}
\renewcommand{\v}{\vec{v}}

\renewcommand{\k}{\vec{k}}
\newcommand{\kz}{\vec{k}_0}

\newcommand{\xo}{\x_1}
\newcommand{\vo}{\v_1}
\newcommand{\ko}{\k_1}
\newcommand{\kp}{\k_+}
\newcommand{\kot}{\k_{12}}
\newcommand{\kzo}{\k_{01}}
\newcommand{\xt}{\x_2}
\newcommand{\vt}{\v_2}

\newcommand{\kt}{\k_2}
\newcommand{\xth}{\x_3}
\newcommand{\vth}{\v_3}

\newcommand{\fof}{f_{1,\text{a}}}

\newcommand{\bfo}{\bar{f}_1}
\newcommand{\bfof}{\bar{f}_{1,\text{a}}}

\newcommand{\bbg}{\bar{\bar{g}}}
\newcommand{\bbgz}{\bar{\bar{g}}_0}
\newcommand{\bbgo}{\bar{\bar{g}}_1}
\newcommand{\ggoa}{g_{1,\text{a}}}

\newcommand{\bbgof}{\bar{\bar{g}}_{1,\text{a}}}
\newcommand{\gof}{\gamma_\text{a}}

\newcommand{\bfino}{{\bar{f}_{1,\text{init}}}}
\newcommand{\bt}[1]{\tilde{\bar{#1}}}

\newcommand{\btfo}{\bt{f}_1}
\newcommand{\btgo}{\bt{g}_1}
\renewcommand{\r}{\mathbf{r}}

\newcommand{\ato}{\tilde{\gamma}}
\newcommand{\Ato}[1]{\tilde{\Gamma}^{(#1)}}
\newcommand{\muo}[1]{\mu^{(#1)}}
\newcommand{\nuo}[1]{\nu^{(#1)}}

\newcommand{\bGz}{\bar{G}_0}

\newcommand{\zvec}{\vec{0}}
\newcommand{\pd}[2]{\frac{\partial #1}{\partial #2}}
\newcommand{\inpd}[2]{\infrac{\partial #1}{\partial #2}}
\renewcommand{\d}[2]{\frac{\mathrm{d} #1}{\mathrm{d} #2}}
\newcommand{\ind}[2]{\infrac{\mathrm{d} #1}{\mathrm{d} #2}}

\newcommand{\kJ}{k_\text{J}}

\newcommand{\vol}{V}

\newcommand{\acln}{\vec{a}}
\newcommand{\ddth}{{\delta^{(3)}}}
\newcommand{\propa}{\mathcal{G}}
\newcommand{\qb}{\bar{q}}

\DeclareMathOperator{\maxwell}{\mathcal{M}}
\DeclareMathOperator{\order}{O}
\DeclareMathOperator{\sinc}{sinc}

\renewcommand{\Im}{\operatorname{Im}}

\DeclareMathOperator{\res}{Res}
\newcommand{\residue}[1]{\res_{\,#1}}

\newcommand{\Scg}{S_\text{acg}}
\newcommand{\Scgr}{S_{\text{acg}}^\circ}
\newcommand{\Scgrt}{S_{\text{acg,}2}^\circ}
\newcommand{\Scgrntl}{S_{\text{acg}}^{\circ\text{ntl}}}
\newcommand{\Shcgr}{\hat{S}_{\text{acg}}^\circ}
\newcommand{\Shcgrntl}{\hat{S}_{\text{acg}}^{\circ\text{ntl}}}

\newcommand{\dd}{\text{d}}

\newcommand{\Tstrut}{\rule{0pt}{2.6ex}}         
\newcommand{\Bstrut}{\rule[-0.9ex]{0pt}{0pt}}   

\newcommand{\mc}[1]{\multicolumn{1}{c}{#1}}

\title[Entropy and gravitational collapse]{A core-halo pattern of entropy creation \\in gravitational collapse}

\author[Andrew J. Wren]{Andrew J. Wren\thanks{andrew.wren@ntlworld.com}}

\date{Accepted 2018 March 22. Received 2018 March 4; in original form 2017 July 16}

\pubyear{2018}

\begin{document}
\label{firstpage}
\pagerange{1--29}
\maketitle

\begin{abstract}
This paper presents a kinetic theory model of gravitational collapse due to a small perturbation. Solving the relevant equations yields a pattern of entropy destruction in a spherical core around the perturbation, and entropy creation in a surrounding halo. This indicates collisional ``de-relaxation'' in the core, and collisional relaxation in the halo.  Core-halo patterns are ubiquitous in the astrophysics of gravitational collapse, and are found here without any of the prior assumptions of such a pattern usually made in analytical models. Motivated by this analysis, the paper outlines a possible scheme for identifying structure formation in a set of observations or a simulation. This scheme involves a choice of coarse-graining scale appropriate to the structure under consideration, and might aid exploration of hierarchical structure formation, supplementing the usual density-based methods for highlighting astrophysical and cosmological structure at various scales.
\end{abstract}




\section{Introduction} \label{sec:introduction}

An early landmark in the study of kinetic theory entropy and gravitational collapse was the consideration by \citet{antonov1962vest}\footnote{There is an English translation by Antonov in \citet{1985IAUS..113.....G}.} of the thermodynamics of a model in which self-gravitating particles are confined within a (reflecting) sphere, in particular examining the ``Antonov instability'' associated with the absence of a global state of maximum kinetic theory entropy.  Following this, \citet{1968MNRAS.138..495L} looked at the link between thermodynamics, entropy and the formation of a core-halo pattern, showing numerically that, in their model of finite volume, the inner part of the system loses energy to the outer part, but that the kinetic energy~-- the temperature~-- of the inner part increases as its particles fall into the potential energy well.  This negative specific heat capacity can drive a continuing ``gravothermal catastrophe'' which  increases the flow of energy from the higher-temperature core to the lower-temperature halo.  This can be illustrated \citep[see also, for example,][p572]{BT} as an expression of the virial theorem, albeit making an artificial division of the system into core and halo as an assumption for the argument, rather than its conclusion.  This paper describes a kinetic theory model and analysis which avoids that artificial division, and sees a core-halo pattern emerge naturally in terms of a quantity we will construct, the \emph{asymptotic course-grained entropy creation rate}, which indicates the effects of two-body collisions, and the rate of the system's collisional relaxation.  We then consider the potential physical implications of this result in terms of astrophysical and cosmological structure, in particular for supplementing the identification of structure based on patterns in density.\\

There are now a variety of approaches to considering the kinetic theory of gravitational collapse in cosmology and astrophysics. Some examples of useful sources include: \citet{2017rkt..book.....V} for a cosmological context; \citet{BT}, the standard text on galactic dynamics; \citet{2013degn.book.....M} on galactic nuclei; and \citet{2003gmbp.book.....H} on star clusters.

Collisionless dynamics neglects the specific interactions between specific particles, focusing only on the evolution of their distribution under the ``Vlasov equation'' in the ``mean-field'' created by the average effect of all particles.  In our current context, it is worth noting \citet{1967MNRAS.136..101L}'s characterisation of ``violent relaxation'' through a coarse-graining of the collisionless Vlasov equation. Violent relaxation sees a coarse-grained Boltzmann entropy increasing not through entropy creation but via so-called ``phase-mixing''. The fine-grained Boltzmann entropy remains unchanged, as for any evolution of the Vlasov equation. The evolution of entropy-like ``H-functions'' during violent relaxation was explored in \citet{1986MNRAS.219..285T}.  Violent relaxation was more recently considered in, for example, \citealp{1996ApJ...471..385C} and \citealp{2005MNRAS.360..892D}.

Collisional dynamics is usually approached by adding a ``collisional'' term to the Vlasov equation to give, for example, the Fokker-Planck equation \citep[\citealp{1980ApJ...242..765C}, and as focused on in][]{BT}, the Klimontovich equation \citep[see, for example,][]{2009PhR...480...57C}, the Smoluchowski equation (see \citealp{2002PhRvE..66c6105C} on the emergence of a core-halo pattern in that context), Lenard-Balescu-type equations \citep[see, for example,][]{2012PhyA..391.3680C}, a generalized Landau equation \citep[as in, for example,][]{2013A&A...556A..93C} or a hierarchy of ``BBGKY'' equations linking $n$-particle distribution functions for $n=1,2,3...$ (see, for example,  \citealp{1968ApJ...152.1043G}, and other references mentioned below).  The BBGKY hierarchy is often truncated to provide a closed set of equations.  A form of truncated BBGKY hierarchy is used in this paper.   It is also possible to consider collisional dynamics from the point of view of applied mathematics, as reviewed in \citet{villani2002review}.

Statistical mechanics also provides approaches to modelling distributions of self-gravitating particles~-- see, for example, \citet{1990PhR...188..285P} or \citet{campa2014physics}. \citet{2006IJMPB..20.3113C} reviews the statistical mechanics of self-gravitating systems, illustrating their core-halo structure.  One statistical mechanics approach is to use the one-dimensional Hamiltonian Mean Field (HMF) as a toy model to explore systems with long-range interactions. Exploration of the HMF,  and its analogies with astrophysics, in \citet{2009PhRvE..80b1138S} illustrates the  thermodynamics that can be associated with self-gravitating systems and, as set out in \citet{2014PhR...535....1L}, the appearance of a core-halo form in the HMF, starting from a particularly simple (so-called ``water bag'') type of initial condition.  

As described in, for example, \citet{BT} or \citet{2010gfe..book.....M}, there is also a widely-used approach of modelling astrophysical self-gravitating systems through fairly \emph{ad hoc}, but useful, density distributions. These frequently have a core-halo form. \citet{1980MNRAS.190..497K} considers gravitational instabilities in connection with a selection of such models.\\

As mentioned, the approach used in this paper is based on the well-known BBGKY hierarchy.  The BBGKY hierarchy is named from the initials of its  pioneers: \citet{1946bogbovart}, \citet{born1946general}, \citet{kirkwood1946statistical}, and \citet{yvon1935theorie}.  It is reviewed in, for example, \citet{balescu1997statistical}, \citet{1987stme.book.....H}, and in an astrophysical context in \citet{1968ApJ...152.1043G}.  A notable astrophysical application was made in \citet{1993ApJ...410..543W}, which highlighted the importance of large-scale fluctuations of growing amplitude in the relaxation of a self-gravitating system.  A broadly similar approach to ours, differing considerably in context and detail, is found in \citet{2010MNRAS.407..355H}, which examines the collisional evolution and entropy of otherwise stable self-gravitating systems.

In this paper, we explore how much, and where, entropy is created and destroyed during gravitational collapse. The underlying notion of entropy is the Boltzmann entropy of the one-particle distribution function in kinetic theory phase \opb{}space.  To\clb{} facilitate the derivation of Boltzmann entropy and the mapping of core-halo patterns, this paper uses physical space and physical velocity co-ordinates, and the Fourier transforms of the space co-ordinates\opb{}. An alternative approach of so-called \emph{angle-action} (or \emph{action-angle}) variables is often used in the context of less homogeneous systems~-- see for example, \citet{2003gmbp.book.....H}, \citet{BT}, \citet{2010MNRAS.407..355H}, and \citet{2012PhyA..391.3680C,2013A&A...556A..93C}.\clb{}

Section~\ref{sec:the-bbgky-hierarchy} describes the kinetic theory model used in this paper.  The approach is to use the first two equations of the BBGKY hierarchy, under an assumption that the number of particles is sufficiently large that account need not be taken of the effect of collisions on one-particle distribution functions, while the collisional term is still useful for calculating the rate of creation of entropy.  Our model system consists of a homogeneous Maxwellian distribution with a small central, nearly point-like, perturbation. Care is taken to define the underlying distribution so it is held equilibrium by external forces~-- an approach equivalent to the well-known Jeans swindle.  

Section~\ref{sec:a-coarse-grained-entropy} constructs the rate of entropy creation, and then extracts the ``asymptotic coarse-grained'' part of that rate.  It is ``asymptotic'' in the sense that it uses the term in the entropy creation rate with the strongest exponential time dependence, and so will asymptotically over time become the dominant part of entropy creation (provided the perturbation is small enough that perturbation theory is still valid when that dominance begins). ``Coarse-grained'' means that only small wave-number parts are retained~-- these represent the fastest-growing asymptotic parts, and also make calculations relatively tractable.

Section~\ref{sec:solving-the-evolution-equations} then addresses the relevant evolution equations: recalling the well-known \citep{landau1946vibrations} solution for the first order perturbation of the distribution function, and then dealing with the zeroth and first order perturbations of the correlation function.  Section~\ref{sec:the-entropy-increase-for-the-whole-system} then calculates the creation rate for the asymptotic coarse-grained entropy in our system, both over all space, and for its distribution in space, before noting some simple variants of the main model and further avenues for exploration. Section~\ref{sec:discussion-of-physical-implications} discusses physical implications, proposing a use for our approach in identifying structure formation, before a brief conclusion in Section~\ref{sec:conclusions}.

Many detailed considerations and calculations are set out in the appendices to the paper. Appendix~\ref{sec:approximating-the-initial-delta-function-perturbation-by-a-gaussian} explains a technicality in the definition of our initial point-like perturbation.  Appendix~\ref{sec:behaviour-at-small-wave-numbers} looks at the plasma dispersion function, which plays a key role in the formula for the distribution function associated with the perturbation. It gives an asymptotic formula for the ``Landau zeros" of the plasma dispersion function, motivating a proof that, as  often assumed, the residue sum rule for the inverse Laplace transform applies for the one-particle distribution function.  Appendix~\ref{sec:asymptotic-coarse-grained-number-density-and-entropy-density-of-the-first-order-perturbation-function} shows that the coarse-grained asymptotic number density and entropy density both have the same shape, strongly peaked near the initial central perturbation's location: so the entropy density's pattern is not a useful supplement to the number density's. Appendix~\ref{sec:calculations-for-the-zeroth-order-correlation-function} calculates the correlation function associated with the underlying homogeneous distribution function, whilst Appendix~\ref{sec:calculations-for-the-first-order-correlation-function} derives the equation for the correlation function's perturbation. Appendices~\ref{sec:exploring-the-correlation-equation-using-a-propagator} and~\ref{sec:the-landau-approach-to-deriving-the-first-order-correlation-function} provide two different approaches for finding an integral involving that correlation perturbation, which is needed to obtain the asymptotic coarse-grained entropy rate. Appendix~\ref{sec:exploring-the-correlation-equation-using-a-propagator} employs a propagator method, whilst Appendix~\ref{sec:the-landau-approach-to-deriving-the-first-order-correlation-function} stays closer to \citet{landau1946vibrations}'s technique used to derive the perturbed distribution function.  Appendix~\ref{sec:entropy-calculations} sets out some detailed work needed to complete the entropy creation rate calculations.  Further details of many calculations, and associated numerical integrations, are set out in a \emph{Mathematica} notebook at \citet{mycalcs}.

\section{The BBGKY hierarchy and the modelled system} \label{sec:the-bbgky-hierarchy}
	
\subsection{Distributions and the BBGKY equations}
\label{sec:the-underlying-equations}

This subsection establishes notation and recalls well-known equations and terminology. Following \citet{1968ApJ...152.1043G}, we let $f(j)$ be the \emph{one-particle distribution function} (DF), the probability density that a given particle is at the phase space point $(\x_j,\v_j).$ This implies that $\int\!f(j)\,d(j)=1,$ where $d(j)$ is short-hand for $\dd^3\x_j\dd^3\v_j,$ and, unless otherwise indicated, in this paper integrals will always be over the whole range of the relevant variable(s) of integration. Similarly, let $f(1,2)$ be be the \emph{two-particle distribution function}, the probability that two given distinct particles are respectively at the phase space points $(\xo,\vo)$ and $(\xt,\vt).$
	
Let $N$ be the total number of particles, which we will assume is very large.  We define the (two-particle) \emph{correlation function}, $g(1,2)\equiv (N-1)\left[f(1,2)-f(1)f(2)\right]$ and note that
\be 
f(1,2)=f(1)f(2)+\frac{1}{N}g(1,2)+\order\!\left(\frac{1}{N^2}\right)
\ee
so that in practice we will assume
\be 
f(1,2)
\approx
f(1)f(2)+\frac{1}{N}g(1,2)
.
\ee
Let $\acln(1,2)$ be $(N-1)$ times the acceleration of particle $1$ due to particle $2,$ in other words the acceleration if all the other particles were at position $2$ in phase space:
\be \label{eq:acln-defn}
\acln(1,2)
=
-Gm(N-1)\frac{(\xo-\xt)}{|\xo-\xt|^3}
\approx
-GmN\frac{(\xo-\xt)}{|\xo-\xt|^3}
,
\ee
which $G$ is Newton's gravitational constant and $m$ is the small mass of each particle (we assume that each particle has the same mass).

For self-gravitating particles, from \citet{1968ApJ...152.1043G} we then have the pair of equations, truncated to leading order in $\infrac{1}{N},$
\bml \label{eq:Gilbert-f-eq}
\pd{f(1)}{t}
+
\vo\dotp\pd{f(1)}{\xo}
+
\int\!\acln(1,2)f(2)\,\dd(2)\dotp\pd{f(1)}{\vo}
\\
=
-
\frac{1}{N}\pd{}{\vo}\int\!\acln(1,2)g(1,2)\,\dd(2)
\,,
\eml
and, also truncating to disregard three-particle correlations,
\bml \label{eq:Gilbert-g-eq}
\pd{g(1,2)}{t} 
+
\Bigg\lbrace
\vo\dotp\pd{g(1,2)}{\xo}
+\int\!\acln(1,3)f(3)\,\dd(3)\dotp\pd{g(1,2)}{\vo}
\\
+\pd{f(1)}{\vo}\dotp  \int \acln(1,3)\,g(3,2)\,\dd(3)
\\+\left(\acln(1,2)-\int\!\acln(1,3)f(3)\,\dd(3)\right)
\dotp\pd{f(1)}{\vo}f(2)
+
\ (1)\leftrightarrow(2)
\Bigg\rbrace
=0
,
\eml
where, for brevity, we used the abbreviation $+\ (1)\leftrightarrow(2)$ to indicate that we need to add terms which repeat all the terms in the braces, but with the variables $(1)=(\xo,\vo)$ and $(2)=(\xt,\vt)$ swapped over. Eqs.~\eqref{eq:Gilbert-f-eq} and~\eqref{eq:Gilbert-g-eq} represent the first two equations of the BBGKY hierarchy. The form of the acceleration, from \eref{eq:acln-defn}, implies that Eqs.~\eqref{eq:Gilbert-f-eq} and~\eqref{eq:Gilbert-g-eq} suffer from short-range ``ultraviolet'' divergences, but these are not relevant when we focus on only large-scale, or \emph{coarse-grained}, effects.

\opb{}If the \emph{collisional term} on the right-hand side of \eref{eq:Gilbert-f-eq} is  omitted,\clb{} it becomes the \emph{Vlasov equation},
\be \label{eq:Vlasov}
\pd{f(1)}{t}
+
\vo\dotp\pd{f(1)}{\xo}
+
\int\!\acln(1,2)f(2)\,\dd(2)\dotp\pd{f(1)}{\vo}
=
0
.
\ee
The \emph{collisional assumption} for our model is that $N$ is large enough, as for many physical systems, so the effect of the correlation function $g(1,2)$ on the one-particle DF $f(1)$ is minimal, even integrated over the whole of the time period we shall consider. In contrast, the one-particle DF will still drive the evolution of the correlation function.  We will only take account of the correlation function's effect on the one-particle DF when formulating our definition of entropy, or more precisely our definition of entropy creation.  In that context, we will see in Section~\ref{sec:a-coarse-grained-entropy} that the collisional term will play a key role because, as is well known, Boltzmann entropy\opb{}, as for any functional of the DF\footnote{\opb{}Referred to as a \emph{Casimir functional.}\clb{}},\clb{} is invariant under the Vlasov equation. \\

For notational convenience we will often write expressions such as $x_j=|\x_j|, v_j=|\v_j|$ and write the \emph{Maxwellian} velocity distribution as
\be
\maxwell(\v_j)=\frac{\exp\left[-\infrac{v_j^2}{2\sigma^2}\right]}{(2\pi\sigma^2)^\infrac{3}{2}}
.
\ee
We can also identify some key  parameters for our system,  supplementing  the \emph{Maxwellian velocity parameter} $\sigma.$  Suppose there is a volume  $\vol$ associated with our system (this volume will be specified in the next subsection).  We write $n\equiv\infrac{N}{\vol}$ for the corresponding average \emph{number density}.  Recalling that we have assumed all particles have the same mass, we write that mass as $m.$  Our system has a characteristic length scale $\kJ^{-1},$ where $\kJ$ is the \emph{Jeans wave-number},
\be\label{eq:Jeans-wave-number}
\kJ^2
\equiv
\frac{4\pi G m  n}{\sigma^2}
=
\frac{4\pi G m N}{\sigma^2\,\vol}
.
\ee
Our system also has a characteristic time scale, its \emph{dynamical time}, given by $(\kJ\sigma)^{-1}.$

\subsection{The perturbed model}
\label{sec:the-perturbed-model}

Our model consists of an underlying DF and a perturbation.  This subsection sets out our model, and recalls the BBGKY equations corresponding to the underlying and perturbation DFs. In light of our normalisation convention that DFs are probability densities, in particular integrating to unity over all phase space, we set
\be \label{eq:epsilon}
f(1)=(1-\epsilon)f_0(1)+\epsilon f_1(1)
,
\ee
where $\epsilon\ll 1$ is the \emph{perturbation parameter}. Similarly we will also write the correlation function as
\be
g(1,2)=(1-\epsilon)g_0(1,2)+\epsilon g_1(1,2).
\ee

We now define our underlying one-particle DF $f_0$.  As in, for example, \citet{BT}, we assume that the velocity dependence of our underlying DF is Maxwellian and that it is also homogeneous across a large spherical $\vol=\infrac{4\pi R^3}{3}$ beyond which it vanishes,
\be \label{eq:f0-defn}
f_0(\xo,\vo)
=
\begin{dcases}
	\frac{\exp\left[{-\infrac{\vo^2}{2\sigma^2}}\right]}{\vol\,(2\pi\sigma^2)^\infrac{3}{2}}
	& \text{if } |\xo| < R \\
	0
	& \text{if } |\xo| \ge R 
\end{dcases}
\ .
\ee
We will regard $R$ and $\vol$ as so large that, for many practical purposes, we are taking the limit $R\to\infty.$  

The large, but finite, volume $\vol$ is needed to give us a non-zero probability density for the location of a particle at a given point. If we  interpreted this as meaning that there is \emph{no} mass beyond the volume $\vol,$ then the underlying distribution would itself not be in equilibrium, but would be undergoing gravitational collapse.  Instead, we assume that it is held in equilibrium by external accelerations.  These accelerations are set to be such that the acceleration integral terms in the BBGKY equations, Eqs.~\eqref{eq:Gilbert-f-eq}-\eqref{eq:Vlasov}  are not limited to the volume $\vol$ but extend over all space. We are, in effect, assuming that there is mass beyond the volume $\vol,$ but that its response to the perturbation will be ignored. As noted in \citet[p.~403]{BT}, this kind of approach is a version of the well-known \emph{Jeans swindle} \citep{1902RSPTA.199....1J}. In particular, this allows us to assume that $f_0$ is time-invariant under \opb{}an \clb{}evolution governed by the Vlasov equation, \eref{eq:Vlasov}, because it enables us to disregard that equation's acceleration integral.

We assume, and later confirm,  that there is a ``translation-invariant'' $g_0$ which depends on position only through the distance $r\equiv|\r|\equiv|\xt-\xo|.$  Such a choice implies that 
\be \label{eq:collisionally-stable}
\int\!\acln(1,2)g_0(1,2)\,\dd(2)
=
GmN\int\!\frac{\r}{r^3}g_0(r,\vo,\vt)\,\dd^3\r\,\dd^3\vt
=
0
,
\ee
where the last equality follows from the anti-symmetry of the integrand in $\r.$  This means that such a $g_0$ makes our $f_0$ not only time-invariant under the Vlasov equation, \eref{eq:Vlasov}, but also \emph{collisionally time-invariant} under evolution via the full one-particle BBGKY equation, \eref{eq:Gilbert-f-eq}.  We also assume that $g_0$ is time invariant, in keeping with our aim that unperturbed functions represent an equilibrium state. Section~\ref{sec:zeroth-and-first-order-correlation-functions} checks that there is indeed a translation- and time-invariant choice of $g_0,$ consistent with \eref{eq:Gilbert-g-eq}.

We now define the first order perturbation $f_1$ of the one-particle distribution function.  We assume the perturbation consists of particles of the same mass $m$ as the underlying distribution.  The perturbation is defined by the initial value $f_{1,\text{init}}$ at $t=0$ of the perturbation $f_1.$  We shall use the idealised expression
\be \label{eq:f1-init}
f_{1,\text{init}}(1)=\ddth(\xo)\maxwell(\vo)
\,,
\ee
representing a sharp perturbation concentrated entirely at the origin $\xo=\zvec,$ with a Maxwellian velocity distribution which has the same parameter $\sigma$ as the underlying DF.\footnote{In Subsection~\ref{sec:further-avenues-for-exploration-including-of-alternative-systems} and Appendix~\ref{sec:modifications-to-the-main-model}, we consider the case of choosing a different Maxwellian parameter for the perturbation.}  The formulation of $f_{1,\text{init}}$ via a spatial Dirac delta function is an approximation.  The delta function takes an infinite value at the origin and so, in principle, is not compatible with perturbation theory.  This technicality is dealt with in Appendix~\ref{sec:approximating-the-initial-delta-function-perturbation-by-a-gaussian}. We will also take the initial perturbation to be \emph{uncorrelated}, that is $g_1=0$ at time $t=0.$

We now, as is standard, write the BBGKY equations in terms of the underlying and perturbation functions, $f_0,f_1,g_0$ and $g_1.$ Using the Jeans swindle's implications for acceleration integrals of $f_0$ and $g_0,$ noted in and before \eref{eq:collisionally-stable}, we have: the \emph{first order\footnote{In this context ``order'' refers to perturbation order in $\epsilon.$} \opb{}Vlasov equation\clb{}}\opb{},
\be \label{eq:f1-cbe}
\pd{f_1(1)}{t}
+
\vo\dotp\pd{f_1(1)}{\xo}
+
\int\!\acln(1,2)f_1(2)\,\dd(2)\dotp\pd{f_0(1)}{\vo}
=
0
\,;
\ee
the \clb{}\emph{zeroth order  correlation equation}
\bml \label{eq:g0-eq}
\pd{g_0(1,2)}{t} 
+
\Bigg\lbrace
\vo\dotp\pd{g_0(1,2)}{\xo}
+\pd{f_0(1)}{\vo}\dotp  \int \acln(1,3)\,g_0(3,2)\,\dd(3)
\\
+\acln(1,2)\dotp\pd{f_0(1)}{\vo}f_0(2)
+
\ (1)\leftrightarrow(2)
\Bigg\rbrace
=
0
\,;
\eml
and the \emph{first order correlation equation}
\bml \label{eq:g1-eq}
\pd{g_1(1,2)}{t} 
+
\Bigg\lbrace
\vo\dotp\pd{g_1(1,2)}{\xo}
+\int\!\acln(1,3)f_1(3)\,\dd(3)\dotp\pd{g_0(1,2)}{\vo}
\\
+\pd{f_0(1)}{\vo}\dotp  \int \acln(1,3)\,g_1(3,2)\,\dd(3)
+\pd{f_1(1)}{\vo}\dotp  \int \acln(1,3)\,g_0(3,2)\,\dd(3)
\\ 
+\acln(1,2)\dotp\pd{f_0(1)}{\vo}f_1(2)
+\acln(1,2)\dotp\pd{f_1(1)}{\vo}f_0(2)
\\
-\int\!\acln(1,3)f_1(3)\,\dd(3)\dotp\pd{f_0(1)}{\vo}f_0(2)
+
\ (1)\leftrightarrow(2)
\Bigg\rbrace
=
0
.
\eml
Note that in the last two equations although $g_0$ is invariant under translations of both variables, it is not homogeneous in a single variable, implying that we cannot drop terms involving integrals of $g_0$ of forms like $\int\!\! \acln(2,3)\,g_0(1,3)\,\dd(3).$ \\

In summary, our assumptions are as follows.  That the total number of particles $N$ is large enough for the collisional assumption, described after \eref{eq:Vlasov}, to hold.  That the perturbation parameter $\epsilon,$ see \eref{eq:epsilon}, is small enough for first order perturbation theory to work.  That the volume $\vol,$ of \eref{eq:f0-defn} is large enough for the effects of the exterior beyond $\vol$ to be ignored, at least in a large region around the centre of the initial perturbation.  All these assumptions can be presumed to have a finite, but conceivably very long, lifetime from the introduction of the initial perturbation.

It is well known \citep[see, for example,][and \eref{eq:f1-fast-early} below]{BT} that the solution to \opb{}the first order Vlasov equation, \eref{eq:f1-cbe},\clb{}  is dominated by components of $f_1$ with small wave-numbers, which grow exponentially with time.  We will call these the \emph{asymptotically-dominant}, or simply \emph{asymptotic}, parts of the solution.

\section{Asymptotic coarse-grained entropy creation} \label{sec:a-coarse-grained-entropy}

\subsection{A Boltzmann entropy rate formula}
\label{sec:a-boltzmann-entropy-rate-formula}

We now derive a formula for the rate of creation of the standard  Boltzmann entropy in our model.  This will motivate a more tractable definition of asymptotic coarse-grained entropy creation in the following subsection.

For a one-particle distribution function, and $N$ particles in total, the standard definition of entropy is given by the \emph{Boltzmann entropy}
\be \label{eq:St-1}
S
\equiv
- 
N\int\!\!f(1) \ln\left[ f(1)\right]\,\dd(1)
.
\ee
It is well known that the overall entropy remains constant if $f$ is governed by the Vlasov equation, \eref{eq:Vlasov}, and the creation of total entropy over time comes entirely from \eref{eq:Gilbert-f-eq}'s collisional term, with 
\bml \label{eq:collisional-entropy}
\d{S}{t}
=
-
N\int\!\!\left(\pd{f(1)}{t}\right)_\text{coll} \ln\left[f(1)\right]\,\dd(1)
\\
=
\int\!\! \ln\left[f(1)\right]
\pd{}{\vo}\dotp\int\!\acln(1,2)g(1,2)\,\dd(1,2)
.
\eml
For a given physical point $\xo,$ similar arguments show that the rate of flow of entropy into $\xo$ is given by
\be \label{eq:entropy--physical-space-flow}
\left(\pd{S_{\xo}}{t}\right)_\text{flow} 
=
-
N\pd{}{\xo}\dotp
\int\!\vo 
f(1) \ln\left[f(1)\right]
\,\dd^3\vo
,
\ee
which for our perturbation, $f=(1-\epsilon)f_0+\epsilon f_1,$ will be of order $\epsilon,$ and that the entropy creation rate is given, from \eref{eq:Gilbert-f-eq}, by 
\bml \label{eq:entropy-creation}
\left(\pd{S_{\xo}}{t}\right)_\text{creation} 
=
\int\! \ln\left[ f(1)\right]\pd{}{\vo}\dotp\int\!\acln(1,2)g(1,2)\,\dd(2)\,\dd^3\vo
\\
=
-
N\int\! \ln\left[ f(1)\right]\left(\pd{f(1)}{t}\right)_\text{coll} \,\dd^3\vo
.
\eml
Superficially, the sign of the entropy creation rate depends on whether $\ln[f(1)]$ is positive or negative, that is whether $f(1)$ is greater or less than $1.$ However, this is misleading: the velocity derivative in the collisional term implies that we can replace $\ln[f(1)]$ on the right-hand side of \eref{eq:entropy-creation} with $\ln[f(1)/C]$ for any positive constant $C$ which is independent of velocity.  This indicates that the entropy creation rate measures the tendency of the collisional term to push the DFs $f(\xo,\vo)$ at fixed $\xo$ towards a constant value independent of $\vo$ (the logarithm implying this measurement is relative to $f(\xo,\vo)$'s size). In summary, the entropy creation rate probes the tendency of collisions to flatten the velocity distribution~-- in other words, the rate of collisional relaxation at $\xo,$ with a positive entropy rate indicating increasing collisional relaxation.

For now, we consider only the total creation of entropy over all space $\vol.$ From \eref{eq:collisional-entropy}, we find
\bml  \label{eq:S-t}
\d{S}{t}
=
\int\!\! \ln\left[f_0(1)\right]
\pd{}{\vo}\dotp\acln(1,2)g(1,2)\,\dd(1,2)
\\
+
\int\!\! \ln\left[1+\frac{\epsilon f_1(1)}{f_0(1)}+\order\left(\epsilon^2\right)\right]
\pd{}{\vo}\dotp\acln(1,2)\,\epsilon g_1(1,2)\,\dd(1,2)
\\
=
-
\frac{1}{2\sigma^2}
\int\!\! v_1^2
\pd{}{\vo}\dotp\acln(1,2)g(1,2)\,\dd(1,2)
\\
+
\epsilon^2 \int\! \frac{f_1(1)}{f_0(1)}
\pd{}{\vo}\dotp\acln(1,2) g_1(1,2)\,\dd(1,2)
+\order(\epsilon^3)
,
\eml
where the $g_0$ term in the second line was eliminated using \eref{eq:collisionally-stable}, making \eref{eq:S-t} exact up to, and including, order $\epsilon^2.$

We can see that the first summand in the last expression of \eref{eq:S-t} is zero, as follows. By conservation of energy, the total energy change \opb{}for \eref{eq:Gilbert-f-eq} must\clb{} be zero, as shown in \citet{1950JChPh..18..817I} \citep[and see also][]{1999IJMPC..10.1367M}.  Making sure we count the gravitational potential from each particle interaction only once, we then have, after division by $N,$  
\bml\label{eq:b-e-change}
0
=
\d{}{t}\left\lbrace 
\int\!\left[-\frac{Gm^2\left(N-1\right)}{2}\int\!\frac{f(3)}{|\xo-\xth|}\,\dd(3)+\frac{1}{2}m v_1^2\right]f(1)\,\dd(1)\right\rbrace
\\
=
\frac{Gm^2\left(N-1\right)}{2}\int\! \left[\frac{\vth\dotp\pd{f(3)}{\xth}f(1)}{|\xo-\xth|}
+
\frac{f(3)\vo\dotp\pd{f(1)}{\xo}}{|\xo-\xth|} \right]   \,\dd(1,3)
\\
-
\frac{1}{2}m\int\! v_1^2 f(2)\acln(1,2)\dotp\pd{f(1)}{\vo}\,\dd(1,2)
\\
-
\frac{m}{2N}\int\! v_1^2\pd{}{\vo}\dotp\acln(1,2)g(1,2)\,\dd(1,2)
\\
=
-
\frac{m}{2N}\int\! v_1^2\pd{}{\vo}\dotp\acln(1,2)g(1,2)\,\dd(1,2)
,
\eml
where we used the Vlasov equation, \eref{eq:Gilbert-f-eq}, to substitute for $\inpd{f(1)}{t}$ and $\inpd{f(3)}{t},$ omitting the terms from that equation which vanish as they give total derivatives, and then used substitution and partial integration to reach the final result. We can interpret \eref{eq:b-e-change} as reflecting the well-known result \citep[see, for example,][pp557-558]{BT} that two-particle collisional interactions do not result in the particles becoming bound.  This implies that all purely collisional interactions begin and end with the particles relatively far apart with effectively zero mutual potential energy and hence the collision itself does not alter the two particles' total kinetic energy.

Applying \eref{eq:b-e-change} to \opb{}\eref{eq:S-t}, we find,
\bml  \label{eq:S-t-2}
\d{S}{t}
=
\epsilon^2 \int\! \frac{f_1(1)}{f_0(1)}
\pd{}{\vo}\dotp\int\!\acln(1,2) g_1(1,2)\,\dd(1,2)
+\order(\epsilon^3)
\\
\equiv
-\epsilon^2\, N \int\! \left[\frac{f_1(1)}{f_0(1)}\right]
\left(\pd{f_1(1)}{t}\right)_\text{coll}\!\!\dd(1)
+\order(\epsilon^3)
,
\eml
where we have defined the first order collisional term $\left(\inpd{f_1(1)}{t}\right)_\text{coll},$ which arises from perturbation expansion of \eref{eq:Gilbert-f-eq}'s right-hand side.\clb{}
Note that there is no order $\epsilon$ term in this equation for the rate of entropy creation, so under-densities (negative $\epsilon$) have the same entropy creation as equal and opposite over-densities (positive $\epsilon$).

As for \eref{eq:entropy-creation}, we can interpret \eref{eq:S-t-2} in terms of collisional relaxation. The integrand in \eref{eq:S-t-2}'s final expression consists of a weighting (the factor in square brackets) multiplied by the \opb{}first order collisional term.\clb{}  This means that \eref{eq:S-t-2}'s entropy creation rate measures the (weighted average) tendency of collisions to suppress the perturbation (for positive rates) or enhance it (for negative rates).  For example, if, at a phase space point $(1),$ we have $f_1(1)$ positive, then a positive entropy creation rate at that point implies the collisional term is negative, tending the eliminate the perturbation.  The weighting $ \left[\infrac{f_1(1)}{f_0(1)}\right]$ used in the integral's averaging of such point rates over all phase space recognises changes affecting the perturbation $f_1$ \emph{relative} to the size of the underlying distribution $f_0$ at the velocity concerned.  It is worth recalling that in our model, the $\infrac{1}{N}$ factor in \opb{}\eref{eq:Gilbert-f-eq}'s\clb{} collisional term implies that its tendency to enhance or suppress the perturbation is very slight~-- our model assumed very large $N$ and hence a very long relaxation time.

\subsection{Defining the asymptotic coarse-grained entropy}
\label{sec:defining-the-coarse-grained-entropy creation}

As indicated at the end of Section~\ref{sec:the-bbgky-hierarchy}, it is well known the perturbing DF $f_1$ is dominated by exponentially-growing asymptotic components, associated with small wave-numbers.  It seems plausible, and we shall confirm  in Subsections~\ref{sec:zeroth-and-first-order-correlation-functions} and~\ref{sec:the-first-order-correlation-function}  below, that $g_1$ also has this behaviour.  Motivated by \eref{eq:S-t-2}, we introduce a \emph{definition} of the rate of creation of asymptotic coarse-grained entropy along the following lines
\be  \label{eq:Scg-follow}
\d{\,\Scg}{t}
\sim
\epsilon^2 \int_{\mathcal{K}}\! \frac{\fof(1)}{f_0(1)}
\pd{}{\vo}\dotp\acln(1,2)\, \ggoa(1,2)\,\dd(1,2)
,
\ee
where $\mathcal{K}$ indicates that we do the integral over some region involving only small wave-numbers and the subscript ``a'' indicates that we only take the asymptotically-dominant term of each of $f_1$ and $g_1.$ We will indeed assume that $\fof$ and $\ggoa$ are also defined so that their Fourier transforms are non-zero only for small wave-numbers compatible with the region  $\mathcal{K}.$ 

We accordingly now introduce Fourier transforms to isolate the small wave-number components.  We also define our Laplace transform convention which we will need subsequently.  In writing transforms, a single bar on a function indicates a Fourier transform with respect to one space variable, for example
\be
\bfo(1)\equiv\bfo(\ko,\vo)\equiv\int\!f_1(\xo,\vo)\,\ex^{-\ii\ko\dotp\xo}\,\dd^3\xo
,
\ee 
and a double bar, such as in $\bbgo(1,2),$ indicates a Fourier transform with respect to two space variables using the same convention.  A tilde then further indicates a Laplace transform, as in 
\be \label{eq:Laplace-conv}
\btfo(1)\equiv\btfo(\ko,\vo,\omega)\equiv\int_0^\infty\!\bfo(\ko,\vo,t)\,\ex^{\ii\omega t}\,\dd t
,
\ee
or, with the same Laplace convention, $\bego(1,2).$ 
These Fourier and Laplace conventions are as in \citet{BT}, and it is worth noting that the Fourier and Laplace exponentials have differing signs. For concision, we will generally suppress both the time variable $t$ and its Laplace conjugate $\omega$ in our functions.

To define the region of interest, $\mathcal{K},$ we convert the expression of \eref{eq:Scg-follow} into an integral over the Fourier space of wave-numbers. To do this, we use a standard approach to express the acceleration via the Fourier transform of Poisson's equation for a point mass.  This gives the Fourier transform with respect to $\xo,$
\bml \label{eq:double-argument-a-integral-ft}
\int\!\dd^3\xo\,\left[\int\!\dd(2)\,\acln(1,2)\ggoa(1,2)\right]\,\ex^{-\ii\ko\dotp\xo}
\\
=
-\ii\int\!\frac{\dd^3\kt}{(2\pi)^3}\,\frac{\kt\,\bbgof(\ko-\kt,\vo,\kt,\vt)}{k_2^2}
,
\eml
where $k_2=|\kt|$ is the wave-number associated with $\kt$ (and similarly below for other $k_j$).
We can now see that \eref{eq:Scg-follow} gives
\bml \label{eq:parted-ft}
\epsilon^2 \int_\mathcal{K}\! \frac{\fof(1)}{f_0(1)}
\pd{}{\vo}\dotp\acln(1,2) \ggoa(1,2)\,\dd(1,2)
\\
=
-
4\pi G mN \ii \epsilon^2
\int_\mathcal{K}\!\dd^3\vo\,\dd^3\vt\,\frac{\dd^3\ko}{(2\pi)^3}\,\frac{\dd^3\kt}{(2\pi)^3}
\\
\qquad
\pd{\left({\bfof(\ko,\vo)}/{f_0(\vo)}\right)}{\vo}
\dotp\frac{\kt\,\bbgof(-\ko-\kt,\vo,\kt,\vt)}{k_2^2}
,
\eml
using the convolution theorem and integrating by parts.

We now specify more precisely the notion of ``small'' wave-numbers. Recall from \eref{eq:Jeans-wave-number} that we have a fundamental dynamical length scale in our model, corresponding to the Jeans wave-number, $\kJ.$  For $\beta\ll 1,$ let $\mathcal{K}({\beta\kJ})$ be the region of Fourier-transformed two-particle phase space which includes points with arbitrary velocities $\vo$ and $\vt$ and includes only small wave-numbers $\ko,$ $\kt$ and $\kp\equiv\ko+\kt,$ with $0<k_1,\,k_2,\,k_+<\beta\kJ\ll \kJ.$  We coarse-grain by setting our functions $\fof$ and $\ggoa$ to be zero for any argument with wave-number greater than or equal to $\beta\kJ.$   Motivated by \eref{eq:parted-ft}'s last expression, we finally make our \emph{definition} of the \emph{asymptotic coarse-grained entropy creation rate} as
\bml  \label{eq:Scg-defn}
\d{\,\Scg}{t}
\equiv
-
4\pi G mN \ii \epsilon^2
\int_{\mathcal{K}({\beta\kJ})}\!\dd^3\vo\,\frac{\dd^3\ko}{(2\pi)^3}\,\frac{\dd^3\kt}{(2\pi)^3}
\\
\qquad\qquad\qquad
\pd{\left({\bfof(\ko,\vo)}/{f_0(\vo)}\right)}{\vo}
\dotp
\frac{\kt\,\gof(-\kp,\vo,\kt)}{k_2^2}
,
\eml
where, as before, the subscript ``a'' indicates that for each of $\bfo$ and $\bbgo$ we keep only the part with the asymptotically-dominant growth, which will come from the poles of the Laplace transforms $\btfo$ and $\bego$ with the most positive imaginary parts; we also wrote $\kp\equiv\ko+\kt;$ and, for brevity, we defined
\be \label{eq:gamma}
\gof(\ko,\vo,\kt)
\equiv
\int\!\bbgof(\ko,\vo,\kt,\vt)\,\dd^3\vt
.
\ee

In order to be consistent with our assumption that the system is within a finite volume $\vol$ of radius $R,$ we should require $\beta$ to be chosen so that $R^{-1}\ll\kJ\beta\ll\kJ.$  In Appendix~\ref{sec:entropy-calculations}, we note after \eref{eq:scg-calc-1} that the contribution of very small $k_1,k_2,k_+\sim R^{-1}$ does not materially affect the entropy calculation for $\beta$ satisfying $R^{-1}\ll\kJ\beta,$ so we do not need to account for the infrared cut-off at such small wave-numbers, and can take $0$ as the lower limit for the wave-number integrals.

Note that because $\Scg$ in \eref{eq:Scg-defn} is a measure of entropy creation, although it introduces a form of coarse-graining by including only small wave-numbers, it excludes phase-mixing which occurs in some other definitions of coarse-grained entropy \citep[see, for example, the original reference in][]{1967MNRAS.136..101L}.  As the name suggests, phase-mixing arises from particles with different velocities mixing closely together~-- an effect associated with entropy flow rather than entropy creation.

\section{Solving the evolution equations}
\label{sec:solving-the-evolution-equations}

\subsection{The Vlasov perturbation equation}\label{sec:the-boltzmann-perturbation-equation}

The standard approach for deriving the first order perturbation $f_1$ from the Vlasov perturbation equation, \opb{}\eref{eq:f1-cbe},\clb{} originated in \citet{landau1946vibrations}.  Its application to self-gravitating particles is reviewed in, for example, \citet{BT}. This approach takes both Fourier and Laplace transforms of the distributions and the associated equations.

With the convention set out in Subsection~\ref{sec:defining-the-coarse-grained-entropy creation} above, the Fourier-transformed initial perturbation is
\be
\bfino(\ko,\vo)
=
\maxwell(\vo)
\,
.
\ee
Fourier- and Laplace-transforming the Vlasov perturbation equation, \opb{}\eref{eq:f1-cbe},\clb{} then gives
\be \label{eq:f1-model-4}
\btfo(\ko,\vo)
=
\frac{-\ii \maxwell(\vo)}{\ko\dotp\vo-\omega}+\frac{ \kJ^2\,\ko\dotp\vo\maxwell(\vo) }{k_1^2\left(\ko\dotp\vo-\omega\right)}\,\int\!\btfo(\ko,\u)\,\dd^3\u
\,,
\ee
where Fourier transforming the acceleration term was handled via its relationship with the Poisson equation for a point mass, along the lines described above \eref{eq:double-argument-a-integral-ft}. Following the standard method of \citet{landau1946vibrations}, integrating \eref{eq:f1-model-4} with respect to $\vo$ gives the first order perturbation for $\ko\ne 0$ as
\be \label{eq:f1-model-8}
\btfo(\ko,\vo)
=
-
\frac{\ii\maxwell(\vo)}{\Big(\ko\dotp\vo-\omega\Big)}
-
\frac{\ii \kJ^2\,\ko\dotp\vo\maxwell(\vo)\,Y(k_1,\omega)}{\Big(\ko\dotp\vo-\omega\Big)\left(k_1^2-\kJ^2 P(k_1,\omega)\right)}
.
\ee
We used definitions, for $\omega$ with $\Im\omega>0,$ 
\be \label{eq:def-Y}
Y(k_1,\omega)
\equiv
\int\!\frac{\maxwell(\u)}{\ko\dotp\u-\omega}\,\dd^3\u
=
\frac{1}{\sqrt{2}\sigma k_1}Z\left(k_1,\omega\right)
,
\ee
and 
\be \label{eq:def-P}
P(k_1,\omega)
\equiv
\int\!\frac{ \ko\dotp\u\,\maxwell(\u)}{\ko\dotp\u-\omega}\,\dd^3\u 
=
1
+
\frac{\omega}{\sqrt{2}\sigma k_1}Z\left(k_1,\omega\right)
,
\ee
where the so-called \emph{plasma dispersion function} $Z$ is defined by
\be \label{eq:def-Z}
Z(k_1,\omega)
\equiv
\sqrt{2}\sigma k_1\!\int\!\frac{\maxwell(\u)}{\ko\dotp\u-\omega}\,\dd^3\u
.
\ee
These three functions $Y,P$ and $Z$ can each be extended to $\Im\omega\le 0$ by analytic continuation. Appendix~\ref{sec:behaviour-at-small-wave-numbers} recalls and explores relevant properties of these functions. 

\begin{figure}
	\centering
	\includegraphics[width=0.9\columnwidth]{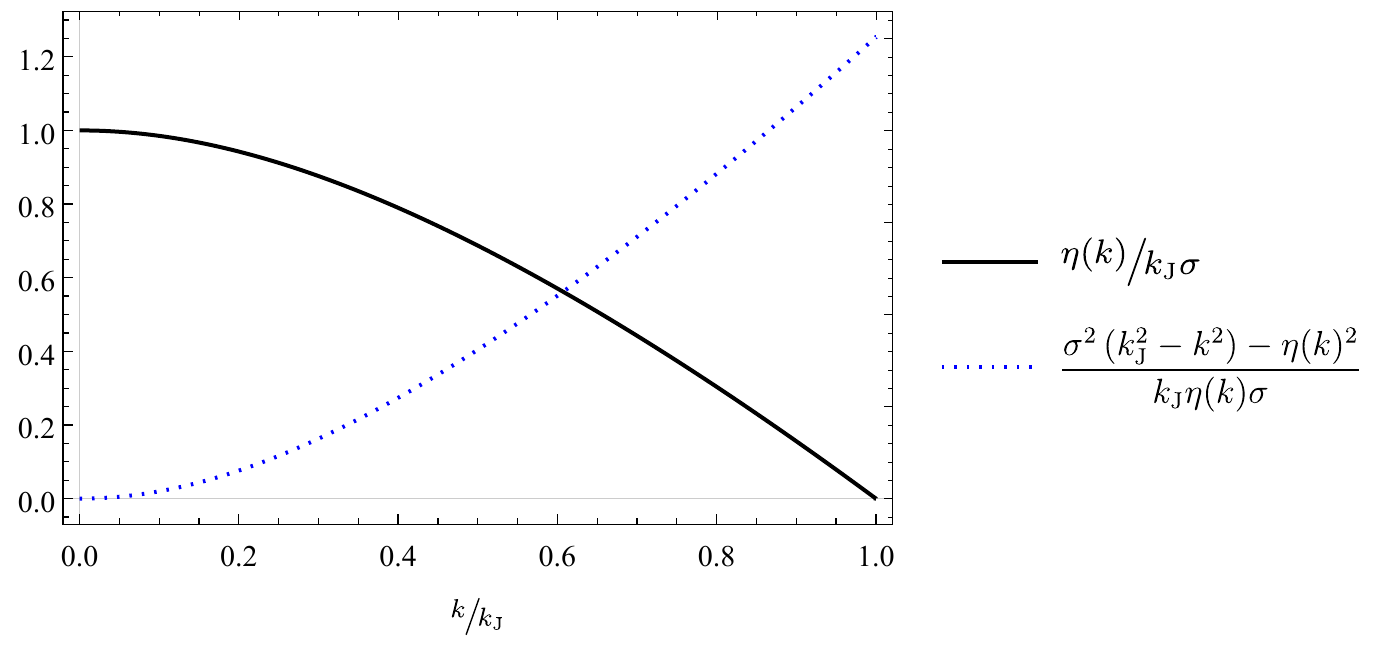}
	\caption{The black solid line shows purely imaginary solutions $\omega=\ii\,\eta(k)$ of the dispersion relation $k^2=\kJ^2 P(k,\omega).$  The blue dotted line is for use in Appendix~\ref{sec:the-residue-approach-for-the-inverse-laplace-transform-of-the-one-particle-distribution-function}. Values of $\eta$ are calculated using a method discussed in the paragraph containing \eref{eq:Z-conjugation}.}
	\label{fig:figdispersionrelation}
\end{figure}

As discussed in \citet{BT}, the key property is that, for a given $0<k_1<\kJ,$ there is exactly one value of $\omega$ with  $\Im\omega>0,$ satisfying the \emph{dispersion relation} $k_1^2=\kJ^2 P(k_1,\omega).$   For each such $0<k_1<\kJ$ and $\omega,$ we define the positive real number $\eta(k_1)$ by $\omega\equiv\ii\eta(k_1),$ which is shown by the solid black line in Figure~\ref{fig:figdispersionrelation}.  For other values of $k_1,$ there is no such $\omega$ with $\Im\omega>0.$

We now look at $\bfo,$ which is the inverse Laplace transform of $\btfo.$ To do this inverse transform, we use the well-known residue formula recalled in  \eref{eq:poles-app-2-simple}.   Appendix~\ref{sec:the-residue-approach-for-the-inverse-laplace-transform-of-the-one-particle-distribution-function} demonstrates that this formula does indeed work for $\btfo$ (this is often assumed without proof).  

The residue formula is the sum of terms each with exponential time-dependency with rate $\omega,$ for each solution $\omega$ of the dispersion relation. This implies that, for a given wave-number $k_1,$ the asymptotically-dominant part of $\bfo(\ko,\vo,t)$~-- the term with the fastest growth in the limit of large times~-- is given by the value of $\omega$ with the largest positive imaginary value.  We saw just above that, for $0<k_1< \kJ,$ this value is purely imaginary, and we then have
\be \label{eq:f1-fast-early}
\bfof(\ko,\vo,t)
\equiv
\frac{ \sigma^2\,\ko\dotp\vo\maxwell(\vo)\,\left(\kJ^2-k_1^2\right)\ \ex^{\eta(k_1)\,t}}{\Big(\ko\dotp\vo-\ii\,\eta(k_1)\Big)\left(\sigma^2\left(\kJ^2-k_1^2\right)-\eta(k_1)^2\right)}
\ee
for this asymptotically fastest-growing part of $\bfo(\ko,\vo,t).$ In Appendix~\ref{sec:asymptotic-coarse-grained-number-density-and-entropy-density-of-the-first-order-perturbation-function}, we see that the asymptotic coarse-grained number density (by volume) has a strong positive central peak, with a much weaker, oscillating tail, as shown in Figure~\ref{fig:figdensity}~(Right).  The entropy density pattern's provides no information additional to the pattern of the number density, because they are proportional.  This helps motivate us to find out whether the entropy creation rate gives a different density pattern: we shall find this is indeed the case, and that the new pattern has a clear core-halo configuration.

\subsection{The zeroth order correlation equation}
\label{sec:zeroth-and-first-order-correlation-functions}

We now solve the zeroth order correlation equation, \eref{eq:g0-eq}, to calculate the Fourier transform, $\bbgz,$ of the correlation function associated with the time-invariant equilibrium distribution function $f_0.$ In Subsection~\ref{sec:the-perturbed-model}, we chose $g_0$ to be both time and translation invariant, implying that we can write
\be \label{eq:G-0}
g_0(\xo,\vo,\xt,\vt)
=
G_0(\r=\xt-\xo,\vo,\vt)
\ee
and in turn this implies
\be \label{eq:g0-separation-1}
\bbgz(\ko,\vo,\kt,\vt)
=
\bGz(\k_-,\vo,\vt)\,(2\pi)^3\ddth(\k_+)
,
\ee
where $\bGz$ is the Fourier transform with respect to $\r,$ of $G_0,$  $\k_-=\infrac{(\kt-\ko)}{2},$ and we continue to write $\k_+=\ko+\kt.$ 

We substitute \eref{eq:G-0} into the zeroth order correlation equation, \eref{eq:g0-eq}, and use time and translation invariance to get
\bml \label{eq:g0-eq-r}
\left(\vt-\vo\right)\dotp\pd{G_0(\r,\vo,\vt)}{\r}
\\
+\pd{f_0(1)}{\vo}\dotp  \int \acln(1,3)\,G_0(\xt-\x_3,\v_3,\vt)\,\dd(3)
\\
+\pd{f_0(2)}{\vt}\dotp  \int \acln(2,3)\,G_0(\x_3-\xo,\vo,\v_3)\,\dd(3)
\\
+\acln(1,2)\dotp\pd{f_0(1)}{\vo}f_0(2)
+\acln(2,1)f_0(1)\dotp\pd{f_0(2)}{\vt}
=
0
.
\eml
In Appendix~\ref{sec:calculations-for-the-zeroth-order-correlation-function}, we Fourier transform \eref{eq:g0-eq-r} and show that $\bGz(\k_-,\vo,\vt)=\qb(\k_-)\maxwell(\vo)\maxwell(\vt),$ where, neglecting a term which is irrelevant for $k_1<\kJ,$ from \eref{eq:q-ansatz-result-pre} we have
\be \label{eq:q-ansatz-result-main}
\qb(\k_-)
=
-\frac{\kJ^2}{\vol\left(\kJ^2-k_-^2\right)}
+
\order\left(\vol^{-2}\right)
.
\ee
As mentioned in Appendix~\ref{sec:calculations-for-the-zeroth-order-correlation-function}, this expression for $\qb$ was derived in  \citet{1983Ap&SS..89..143K}. 

\subsection{The first order correlation function}
\label{sec:the-first-order-correlation-function}

We now discuss the first order correlation equation, \eref{eq:g1-eq}, for $g_1,$ or rather for its Fourier and Laplace transform, $\bego.$ Appendix~\ref{sec:calculations-for-the-first-order-correlation-function} works through the Fourier and Laplace transforms of the first order correlation equation, finding that
\bml \label{eq:g1-eq-flt-main}
\left(\omega-\ko\dotp\vo-\kt\dotp\vt\right)\bego(1,2)
\\
=
\Bigg\lbrace
-
\frac{\kJ^2\, \ko\dotp\vo}{k_1^2}\maxwell(\vo) \int\! \bego(\ko,\u,2)\,\dd^3\u
\\\qquad
-
\frac{\kJ^2\,\ko\dotp\vo}{k_1^2}\maxwell(\vo)\btfo(\kp,\vt)
\\\qquad
-
\frac{\kJ^2\,\vol \kp\dotp\vo}{k_+^2}\int\!\btfo(\kp,\u)\,\dd^3\u\,\qb(k_2)\maxwell(\vo)\maxwell(\vt)
\\\qquad
+
\frac{\kJ^2\,\ko\dotp\vo}{\vol\,k_1^2}\int\!\btfo(\ko,\u)\,\dd^3\u\,\maxwell(\vo)\maxwell(\vt)(2\pi)^3\ddth(\kt)
\\
-
\frac{\kJ^2\sigma^2\, \kt}{k_2^2}\dotp\pd{\btfo(\kp,\vo)}{\vo}\Big(\vol\,\qb(k_2)+1\Big)\maxwell(\vt)
+
\ (1)\leftrightarrow(2)
\Bigg\rbrace
.
\eml
Given the zeroth and first order distribution functions, \eref{eq:g1-eq-flt-main} is an integral equation for $\btgo$ in $\vo$ and $\vt,$ with $\omega,\ko$ and $\kt$ as parameters.

General integral equations can only be solved numerically, a procedure which would be cumbersome given that our equation, \eref{eq:g1-eq-flt-main}, has multiple parameters.  However, it is well known that integral equations such as this can be solved by use of a propagator approach, which, in essence, derives a Green's function for the equation.  This kind of approach is used in, for example, \citet{ichimaru1973basic} in the context of plasmas and \citet{2010MNRAS.407..355H} employing angle-action variables in the context of self-gravitating particles.  The propagator approach to solving \eref{eq:g1-eq-flt-main} is set out in Appendix~\ref{sec:exploring-the-correlation-equation-using-a-propagator}.  In principle, this could give us results (which might involve new special functions of origin similar to $P$ and $Y$) for any $k_1, k_2$ and $\omega.$ However, as indicated by \eref{eq:Scg-defn}, we focus on the case where $0<k_1, k_2 \ll \kJ.$  This gives us tractable analytical solutions.

Another method of addressing \eref{eq:g1-eq-flt-main} is based on the approach of \cite{landau1946vibrations} which was used to derive \eref{eq:f1-model-8} for $\btfo.$ Along those lines, we can rearrange \eref{eq:g1-eq-flt-main} and integrate it with respect to velocities. This approach is followed through in Appendix~\ref{sec:the-landau-approach-to-deriving-the-first-order-correlation-function} for $0<k_1, k_2 \ll \kJ.$  It is more complicated than the derivation of $\btfo,$ requiring calculation of a number of integrals through solving a set of seven simultaneous equations.

Using either the propagator method of  Appendix~\ref{sec:exploring-the-correlation-equation-using-a-propagator}, or the Landau method of Appendix~\ref{sec:the-landau-approach-to-deriving-the-first-order-correlation-function}, we get a power series in $k_j$ for $\gof$ as set out in \eref{eq:bigint-result}. This provides the key function needed to calculate the rate of asymptotic coarse-grained entropy creation.

\section{The rate of asymptotic coarse-grained entropy creation} \label{sec:the-entropy-increase-for-the-whole-system}

\subsection{The total entropy creation}
\label{sec:the-total-entropy-creation}

We are now in a position to calculate the rate of asymptotic coarse-grained entropy creation over all space, using the results of Section~\ref{sec:solving-the-evolution-equations} in the definition set out in \eref{eq:Scg-defn}. We start by looking at the factor in \eref{eq:Scg-defn} that comes from the velocity derivative related to the one-particle distribution function.   We find,
\bml\label{eq:factor-we-need} 
\pd{\left({\bfof(\ko,\vo)}/{f_0(\vo)}\right)}{\vo}
\\
=
-
\frac{\ii\,\sigma^2\,\vol\,\eta(k_1)\,\left(\kJ^2-k_1^2\right)\,\ko\, \ex^{\eta(k_1)\,t}}{\left(\sigma^2\left(\kJ^2-k_1^2\right)-\eta(k_1)^2\right)\Big(\ko\dotp\vo-\ii\,\eta(k_1)\Big)^2}
.
\eml
Appendix~\ref{sec:asymptotic-series-for-small-wave-numbers} reviews a relevant asymptotic series for small wave-numbers, and this provides a good approximation for the derivative, which is set out in \eref{eq:vel-div-deriv-app}.

Calculations in Appendix~\ref{sec:all-space} evaluate the asymptotic coarse-grained entropy creation rate of \eref{eq:Scg-defn}, using the formula for $\gof$ from \eref{eq:bigint-result}, and \eref{eq:vel-div-deriv-app}'s series approximation.  To leading order in $\epsilon$ and $k_j,$ we find that we have
\bml \label{eq:dScg-dt}
\d{\,\Scg}{t}
=
-\frac{\kJ^7\sigma\beta^6\,\vol^2\epsilon^2}{8\pi^4}
\times
\num{0.0116} \,\ex^{3\kJ\sigma t}
\\
=
-\frac{2\kJ\sigma\, \,N_1^2}{9\pi^2 n^2\,B^2}
\times
\num{0.0116} \,\ex^{3\kJ\sigma t}
.
\eml
We wrote $N_1\equiv\epsilon N$ for the number of particles associated with the perturbation, and recalled that $n=\infrac{N}{\vol}$ is the average number density of the system.  While $\epsilon$ is a measure of the size of the perturbation relative to the total number of particles, $N_1$ is a more absolute measure of that size.  We also wrote $B\equiv\infrac{4\pi}{3\left(\kJ\beta\right)^3}$ for the volume of a sphere associated with the coarse-graining \opb{}scale $\kJ\beta.$

The\clb{} total (net) asymptotic coarse-grained entropy creation from the initial time $0$ to some later time $t>0$ is then clearly
\be \label{eq:entropy-change}
\Delta\Scg
\approx
-\frac{2\,N_1^2}{27\pi^2 n^2\,B^2}
\times
\num{0.0116} \,\ex^{3\kJ\sigma t}
.
\ee
Note that our asymptotic coarse-grained entropy \emph{decreases} with time.  This is consistent with the second law of thermodynamics because our system is not isolated~-- as set out following \eref{eq:f0-defn}, it is subject to \opb{}Jeans swindle\clb{} forces from outside the system (beyond the radius $R$).  \opb{}These  forces ensure that the acceleration integrals of the perturbation evolution equations, Eqs.~\eqref{eq:f1-cbe}-\eqref{eq:g1-eq}, are not limited to the volume $\vol$ but may be taken to extend over all space.  If, as a thought experiment, we imagine these forces as being created by some actual physical machine, its generation of these forces must produce entropy which at least offsets the negative entropy creation within the system. The negative entropy creation can also be viewed as indicating that, without Jeans swindle forces, it would be entropically-favourable for the system to contract under gravity.\clb{}

\subsection{The distribution of entropy creation in space} \label{sec:the-varying-entropy-change-by-radius} 

We now look at the distribution of asymptotic coarse-grained entropy creation in space.  Clearly our system is spherically symmetric, so we look primarily at the \emph{shell density} of the entropy creation rate, measuring the entropy creation rate on a thin sphere of a given radius, $r,$ centred on the initial perturbation. Referring back to Section~\ref{sec:a-coarse-grained-entropy}, we see that the position of the entropy creation is indicated by the variable $\xo,$ and that to extract \opb{}that entropy creation's spatial distribution\clb{} we need to introduce a factor of $\delta(x_1-r)$ into the integral of \eref{eq:S-t-2}. On Fourier-transforming, this corresponds to inserting a convolution with $4\pi r^2\sinc(k_1 r)=4\pi r \sin(k_1 r)/k_1 $ into the integral of \eref{eq:Scg-defn}.  Writing $\kot=\ko+\kt$ and $\kzo=\kz+\ko,$ we therefore have,
\bml \label{eq:Scgrr}
\pd{^2\Scgr(r)}{t\,\partial r}
=
-4\pi G mN \,\ii \epsilon^2 \left(4\pi r^2\right)
\int_{\mathcal{K}'(\beta\kJ)}\!\frac{\dd^3\kz}{(2\pi)^3}\,\frac{\dd^3\ko}{(2\pi)^3}\,\frac{\dd^3\kt}{(2\pi)^3} 
\\
\int\dd^3\vo\,\sinc(k_0 r)\,\pd{\left({\bfof(\kzo,\vo)}/{f_0(\vo)}\right)}{\vo}
\dotp
\frac{\kt\,\gof(-\kot,\vo,\kt)}{k_2^2}
,
\eml
where \opb{}$\Scgr(r)$ is the asymptotic coarse-grained entropy within a sphere of radius $r,$ and\clb{} we defined $\mathcal{K}'(\beta\kJ)$ to be the region $0<k_0,k_1,k_2<\kJ\beta.$ Note that $k_0$ only appears in the integrand's first and second factors because the associated convolution links together these two factors, whilst a convolution via $k_1$ links together the second and third. 

\begin{figure*} 
	\centering
	\includegraphics[width=1.0\textwidth]{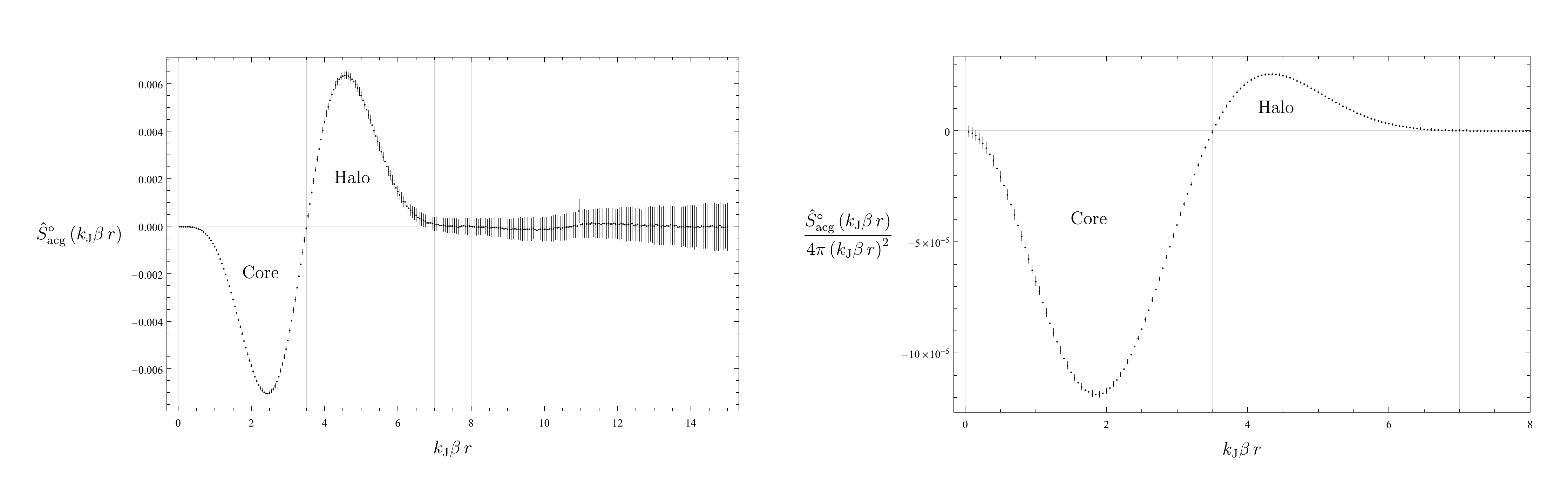}
	\caption{\emph{Left:} The entropy-creation pattern function $\Scgr$ as calculated numerically.  The error bars show the error estimates for the numerical integration.  Note that the pattern function represents the entropy creation in a shell of radius $r.$  Assuming $\kJ$ is given, the plot can be read as showing the core-halo status of shells with varying $r$ for a fixed value of $\beta;$ but it could also be read as showing the status of a fixed shell of radius $r,$ as the scale $\beta$ varies. \emph{Right:} The volume density implied by the entropy-creation pattern function $\Scgr.$  The error bars show the error estimates for the numerical integration, also scaled by $[{4\pi(\kJ\beta\,r)^2}]^{-1}.$  Note that this plot's horizontal scale focuses more tightly than the left-hand plot, on $\kJ\beta\,r\le 8$ to avoid a long, all but featureless, tail of the function.}
	\label{fig:figscgr}
\end{figure*}

Calculations in Appendix~\ref{sec:local} show that, at leading order, 
\bml \label{eq:Scgrr-3}
\pd{^2\Scgr}{t\,\partial r}
=
\frac{\kJ^8\sigma\beta^5\vol^2  \epsilon^2\ \ex^{3\kJ\sigma t}}{8\pi^4}
\,\Shcgr(\kJ\beta r)
\\
=
\frac{2\kJ^2\,\sigma\,N_1^2\ \ex^{3\kJ\sigma t}}{9\pi^2\beta\,n^2\,B^2}
\,\Shcgr(\kJ\beta r)
\,,
\eml
where $\Shcgr$ might be termed the \emph{leading order entropy-creation pattern function} and is shown in Figure~\ref{fig:figscgr}~(Left), from calculations in \citet{mycalcs}. We have defined $\Shcgr$ so that it corresponds to the numerical factor of $-0.0116$ in \eref{eq:dScg-dt}.

To compare the entropy creation density by shell of \eref{eq:Scgrr-3} with the total net entropy creation of \eref{eq:dScg-dt}, we should integrate \eref{eq:Scgrr-3} over $r$ to obtain the entropy creation in a region, for example the total entropy creation in either the core or the halo.  As detailed in Appendix~\ref{sec:local}, comparing \eref{eq:dScg-dt} with the $r$-integrated \eref{eq:Scgrr-3}, we see the size of either the core or the halo is around $\beta^{-2}\gg 1$ times larger than the total net entropy \opb{}creation. 

From \clb{}Figure~\ref{fig:figscgr}~(Left), we can see that there is a \emph{core}, where entropy is destroyed, which spans from $r=0$ to $r\approx \infrac{\num{3.5}}{\kJ\beta}$, and a \emph{halo}, where entropy is created, which spans from  $r\approx \infrac{\num{3.5}}{\kJ\beta}$ to, somewhere around, roughly, $r\approx \infrac{\num{7}}{\kJ\beta}$ to $\infrac{\num{8}}{\kJ\beta}.$  It is plausible that, beyond that radius, volumes of decreasing and increasing entropy alternate, but these are highly suppressed compared with the core and the halo. 

Note that the notions of core and halo are \emph{scale dependent} with respect to $\beta.$  A shell at radius $r,$ which is outside the core and halo at scale $\beta_1,$ will be inside the halo for some smaller $\beta_2<\beta_1,$ and inside the core for some yet smaller  $\beta_3<\beta_2.$  The core-halo pattern is therefore more subtle than a simple core-halo model, with the boundaries of those two regions depending on the scale factor $\beta.$

Recall that the core and halo only indicate local destruction and creation of entropy~-- they take no account of entropy flowing from one place to another with the motion of particles.  The core (resp. halo) is where particles' associated asymptotic coarse-grained entropy tends to be destroyed (resp. created) by collisions.  As an aside, recall that, given the long-range nature of gravity, these collisions are not necessarily close-range, but may be with distant particles outside the core (resp. halo).  Referring to the paragraphs after \eref{eq:S-t-2}, we also see that the core (resp. halo) is where collisions tend gently to enhance (resp. suppress) the perturbation. corresponding to collisional de-relaxation (resp. relaxation).

As confirmed in Appendix~\ref{sec:local}, the absolute value of the total entropy creation in either the core or halo are of very similar size.  This is to be expected, as we know from Subsection~\ref{sec:the-total-entropy-creation} that they must essentially cancel out.  Appendix~\ref{sec:local} also notes that this absolute value is very close to being $\beta^{-2}$ times the size of the total net entropy creation.  Since  $\beta^{-2}\gg 1,$ the core-halo pattern is therefore much more pronounced that the total net entropy creation.

It might be asked if similar calculations to those in this and the previous subsection can be performed for different definitions of asymptotic coarse-grained entropy creation, in particular with different coarse-graining from that set out in the paragraph following \eref{eq:parted-ft}.  That set three  constraints, which may be summarised as $0<k_1,k_2,k_+<\kJ\beta.$ In Appendices~\ref{sec:all-space} and~\ref{sec:local}, we consider the implications of relaxing any one of the three upper constraints.  We get results for entropy creation totalled over all space, but the integrations needed for the distribution over space are no longer so tractable: their integrands no longer contain just the small wave-numbers needed for our analytical calculations. Moreover, Appendix~\ref{sec:all-space} notes that those alternatives match the time dependence of our perturbations less well, again suggesting a preference for the approach to constraints which we have adopted.

\opb{}We could alternatively use a coarse-graining which matches the asymptotic time behaviour of our model yet more closely~-- albeit one which in a physical context we might be unlikely to select if we did not have an analytical expression for that behaviour. As set out in Appendix~\ref{sec:all-space}, this is to choose $\mathcal{K}_2=\left\lbrace(1,2): k_1^2+k_2^2+k_+^2<2k_J^2\beta^2\right\rbrace,$ which has the property that it captures all wave-numbers for which the exponential increase with time is estimated (to second order in $k_j$) as being faster than a given rate. The resulting entropy creation rate is as in \eref{eq:dScg-dt}, but with $\num{-0.0116}$ replaced by $\num{-0.0125}.$

As described in Appendix~\ref{sec:local}, for $\mathcal{K}_2$  the leading order space distribution of the entropy creation rate satisfies the formula of \eref{eq:Scgrr-3}, but with the entropy pattern function shown in Figure~\ref{fig:figscgr} replaced by that of Figure~\ref{fig:figsquares}. We again see a core of entropy destruction surrounded by a halo of entropy creation, here with also small outer shells of entropy destruction and creation (compare with Figure~\ref{fig:figscgr} where there are perhaps such shells but of smaller amplitude than the estimated integration errors). The $\mathcal{K}_2$ coarse-graining, and  another ``taxicab'' coarse-graining outlined at the end of Appendix~\ref{sec:local}, suggest that a core-halo pattern of entropy destruction and creation may be generic for a broad class of coarse-grainings which are symmetrical with respect to $k_1,k_2$ and $k_+.$\clb{}

\subsection{Further avenues for exploration, including of alternative systems}
\label{sec:further-avenues-for-exploration-including-of-alternative-systems}

In Appendix~\ref{sec:modifications-to-the-main-model}, we look at entropy creation when the Maxwellian parameter associated with the initial perturbation differs from that of the underlying perturbation.  Excepting very large parameters, which our approximation methods cannot address, the leading order entropy creation rate is as in \eref{eq:Scgrr-3} and Figure~\ref{fig:figscgr}, with no dependence on the initial perturbation's Maxwellian parameter.  This applies in particular for a perturbation with all its particles initially stationary.

It is also possible to vary the assumption $g_1(1,2,t=0)=0$ that the initial perturbation is uncorrelated. As discussed in Appendix~\ref{sec:modifications-to-the-main-model}, choosing the same correlation as for the underlying distribution, $g_1(1,2,t=0)=g_0(1,2,t=0),$ makes no difference to our results.\\ 

We note some further possible avenues for investigation, looking at alternative systems to the one investigated here.  It is an open question as to how analytically-tractable they might be, or whether approaches which are more numerical than used here might be needed.

This paper examined a single, almost point-like, initial perturbation.  This might be replaced by a ``dipole'' pair of such perturbations, or a ''multipole'' arrangement of many point-like perturbations.

More varied alternatives could also be considered. They might avoid having to impose a \opb{}Jeans swindle\clb{} restriction of the system to a finite volume\opb{}, and the associated total net destruction of entropy\clb{}.

One  system that could be investigated would be the evolution, possibly with a small central perturbation, of a spherically-symmetric, but non-homogeneous, distribution of self-gravitating particles.  This would roughly model a collapsing halo of dark matter, gas or stars. An introduction to such distributions can be found in \citet[section~4.3]{BT}. 

Another alternative is to seek to model the evolution under gravity of a system comprising a small central perturbation in an underlying razor-thin disc of particles, the underlying disc being in equilibrium and also rotating. This might approximate a galactic disc, for example.  A discussion of equilibrium distributions for razor-thin discs can be found in \citet[section~4.5]{BT} or in \citet{1976ApJ...205..751K}.

The assumption of equilibrium  for those two distributions might avoid the requirement for a fixed exterior to keep the initial underlying distribution artificially in equilibrium. This would then perhaps show the dominance of entropy creation over its destruction within the whole system. 

A different avenue of investigation would be to model the expanding universe, with a small perturbation being introduced.  The universe's expansion might then avoid the need for defining an artificial limit to the volume considered, with, perhaps, the expansion's Hubble horizon providing a more natural finite limit.  A start might be made with the formulation of kinetic theory equations set out for a Newtonian expanding universe in \cite{1983Ap&SS..89..143K}.  However, this would not capture general relativity's limitation of gravitational effects to a sphere of influence travelling out from the source at the speed of light, an effect which might define a sharper finite volume limit than the Hubble horizon.  A general relativistic approach would therefore be preferable, see, for example, \citet{2017rkt..book.....V} or \citet{2011LRR....14....4A}.  The post-Newtonian approximation might be useful~-- see, for example, \citet{2014grav.book.....P} for a general introduction, or \citet{2011PhRvD..83l3007A} and \citet{2012PhRvD..86d3008R} for consideration of kinetic theory and the post-Newtonian approximation.

Near the initial central perturbation even a general relativistic system could, presumably, look very like the Newtonian system examined here if the central perturbation were not too dense.  We might see asymptotic coarse-grained entropy destruction locally dominating over time, and, perhaps, being offset by distant entropy creation associated with the initial central perturbation's expanding sphere of influence.

There is also the possibility of the central mass being a black hole. The accreting particles' kinetic theory entropy would be distinct from the Bekenstein-Hawking entropy \citep{1973PhRvD...7.2333B,1976PhRvD..13..191H} of the black hole itself, and there might or might not be any noteworthy relationship between these  entropies. A starting point for such a study might be provided by the very recent consideration of the kinetic theory of collisionless gas accreting on to a Schwarzschild black hole in \citet{2017CQGra..34i5007R}.

\section{Physical discussion}
\label{sec:discussion-of-physical-implications}

\opb{}It is well known that the Universe has a multi-scale hierarchical structure, in which core-halo patterns are ubiquitous.  The identification of observed or simulated astrophysical structure typically involves considering features of especially high or low densities, in physical space, or phase space.  There is no unambiguous definition of structure in this context, which can result in different methods giving different results~-- for example, see \citet{2012MNRAS.423.1200O} on identifying sub-haloes near the centre of dark matter haloes, \citet{2015MNRAS.454.3020B} on  major halo mergers, and  \citet{2018MNRAS.473.1195L} on classifying elements of the cosmic web. This suggests that complementary methods for identifying structure, or structure formation, may be helpful.\\

We\clb{} outline a possible scheme for doing this using kinetic-theory entropy creation. In dealing with observations or simulations, we will need to first construct a phase space distribution function (DF) from the data.  Practically speaking, this will need involve some \emph{smoothing} in phase space.  We then also need to choose a scale of interest in position space for \emph{coarse-graining} in order to help identify the structures we are aiming to explore.  The coarse-graining scale must be at least as large, and might be much larger, than the scale for the practically-necessitated smoothing.

Given now our DF $f(1)$, smoothed for practical reasons, and then coarse-grained to our scale of interest, there are then two alternative approaches:
\begin{enumerate}
	\setlength{\itemsep}{6pt}
	\setlength{\parskip}{6pt}
	\item \label{itm:DF-approach}
	One approach is to rely solely on the DF, starting with the total entropy change at a point and subtracting the entropy flow of \eref{eq:entropy--physical-space-flow}, giving
\bml\label{eq:entropy-creation-ii}
\left(\pd{S_{\xo}}{t}\right)_\text{creation} 
=
- 
N\pd{}{t}\int\!\!f(1) \ln\left[ f(1)\right]\,\dd^3\vo
\\
+
N\pd{}{\xo}\dotp
\int\!\vo 
f(1) \ln\left[f(1)\right]
\,\dd^3\vo
.
\eml
	\item \label{itm:correlation-approach}
The approach closer to that used for our model in Section~\ref{sec:the-entropy-increase-for-the-whole-system} is to construct from the observations or simulation a two-point correlation function $g(1,2),$ coarse-grain to a scale of interest, and then apply \eref{eq:entropy-creation},
	\be \label{eq:entropy-creation-i}
	\left(\pd{S_{\xo}}{t}\right)_\text{creation} 
	=
	\int\! \ln\left[ f(1)\right]\pd{}{\vo}\dotp\int\!\acln(1,2)g(1,2)\,\dd(2)\,\dd^3\vo
	.
	\ee
	 A comparison of methods  for estimating correlations is given in \citet{2000ApJ...535L..13K}, albeit for a correlation function that depends only on the distance between two points, instead of depending, as here, on the position and velocity within phase space.   The estimation will, of course, involve an element of practically-necessitated \opb{}smoothing.\clb{}

\end{enumerate}

\opb{}By\clb{} analogy with our model~-- see the remark after \eref{eq:S-t-2}~-- it is possible that this scheme would characterise under-densities, for example cosmic voids or their centres, as undergoing structure formation.  It is possible that, in some circumstances, transitional regions between over-dense structure and these under-densities might be the regions of maximum entropy creation.

For the model we constructed in Sections~\ref{sec:the-bbgky-hierarchy}-\ref{sec:the-entropy-increase-for-the-whole-system}, there was no question of phase-mixing, as explained at the end of Section~\ref{sec:a-coarse-grained-entropy}. However, once we apply our scheme to observations or a simulation, the practical requirement to smooth the particle distribution implies phase-mixing is a possibility, and so even truly collisionless microscopic processes might give rise to macroscopic entropy \opb{}creation.\\

It\clb{} is beyond the scope of this paper to identify if the above scheme would work in practice.  This might depend upon the details of the particular simulation or set of observations, the approach to smoothing, and the choice of approach and scale for coarse-graining.

An obstacle to robust results~-- how challenging an obstacle is to be determined~-- could well arise from the need to identify differences between possibly relatively similar quantities.  For approach \ref{itm:DF-approach}, this requirement is explicit in the form of \eref{eq:entropy-creation-ii}.  For approach \ref{itm:correlation-approach}, it arises because of the requirement to calculate correlation functions.  Feasibility of our scheme is therefore, perhaps most crucially, dependent on the size of simulation or observation errors relative to the underlying entropy creation and destruction.

The scheme is feasible and useful if, at whatever scale is focused upon, we can robustly detect entropy creation and/or destruction.  It is also feasible, but presumably of less use, if we can robustly exclude entropy creation and/or\opb{} destruction.

Our\opb{} scheme's coarse-graining might help draw out properties of different scales.  For example, as is well known, dark matter is essentially collisionless, whereas dark matter haloes are mutually collisional.  At scales relevant to dark matter haloes, would we see larger entropy creation?  Varying the coarse-graining scale might also help elucidate interaction between various levels of hierarchical structure.  \opb{}If we wanted to include gas in our system, as the dominant baryonic matter component, we might also add terms to encompass hydrodynamic entropy creation in the gas. \clb{}

\section{Conclusion} \label{sec:conclusions}

\opb{}As mentioned in the introduction, a core-halo model was described in \citet[p572]{BT}, making an artificial distinction between the core and the halo during gravitational collapse.  A similar argument \citep[pp377-378]{BT} draws out more clearly that the creation of entropy takes place predominantly within the halo, not the core. If we further assume that the core's structure scales with its radius, then its phase space volume varies like $r^\infrac{3}{2}$ and it is easily seen that entropy is in fact destroyed within the (shrinking) core.

In the current paper, we\clb{} have constructed an analytical kinetic theory perturbation model for the beginning of gravitational collapse.  We introduced an asymptotic coarse-grained entropy, which in our model is associated with the system's fastest-growing modes, and indicates the rate of their collisional relaxation.

Overall for our model, which is not an isolated system, we see net entropy destruction. However, this is a higher order, more suppressed, effect compared with a pattern of entropy destruction and creation.  Entropy destruction occurs in a ``core'' around the central perturbation, with equal and opposite entropy creation in a ``halo'' extending for a finite radius beyond that core, as shown in Figure~\ref{fig:figscgr}. The physical scale for the core-halo pattern depends on the coarse-graining parameter chosen: the coarser the graining, the bigger the physical scale.

In the core, collisions enhance the perturbation in a process of collisional ``de-relaxation.'' Conversely, in the halo, collisional relaxation suppresses the perturbation.  In our linear perturbation model, the effect of such collisional evolution on the perturbation (and hence the entropy creation) is well defined, but small in size compared with its collisionless evolution (and the entropy flow).

A core-halo pattern  of gravitational collapse, well known from simulations and observations, is generally set ``by hand'' in analytical models. As far as the author has been able to determine, this is the first time an analytical kinetic theory model has produced a core-halo pattern.

This motivates a scheme for measuring structure formation in observations or simulations, via patterns of entropy creation and destruction, as set out in \opb{}Section~\ref{sec:discussion-of-physical-implications}. \clb{}The feasibility of this scheme in the contexts of various observations or simulations is the key unanswered question arising from this paper. Because the main difficulty is likely to arise from the size of observation or simulation errors relative to entropy creation and destruction, feasibility is likely to improve along with the precision of observations and simulations.

\section*{Acknowledgement}
 
The author is very grateful to the anonymous referee for his or her extremely helpful suggestions, including on the paper's structure and style, and for prompting inclusion of more discussion of physical implications \opb{}and of alternative coarse-grainings\clb{}.




\bibliographystyle{mnras}
\bibliography{egc-biblio} 

\begin{thebibliography}{}
\makeatletter
\relax
\def\mn@urlcharsother{\let\do\@makeother \do\$\do\&\do\#\do\^\do\_\do\%\do\~}
\def\mn@doi{\begingroup\mn@urlcharsother \@ifnextchar [ {\mn@doi@}
  {\mn@doi@[]}}
\def\mn@doi@[#1]#2{\def\@tempa{#1}\ifx\@tempa\@empty \href
  {http://dx.doi.org/#2} {doi:#2}\else \href {http://dx.doi.org/#2} {#1}\fi
  \endgroup}
\def\mn@eprint#1#2{\mn@eprint@#1:#2::\@nil}
\def\mn@eprint@arXiv#1{\href {http://arxiv.org/abs/#1} {{\tt arXiv:#1}}}
\def\mn@eprint@dblp#1{\href {http://dblp.uni-trier.de/rec/bibtex/#1.xml}
  {dblp:#1}}
\def\mn@eprint@#1:#2:#3:#4\@nil{\def\@tempa {#1}\def\@tempb {#2}\def\@tempc
  {#3}\ifx \@tempc \@empty \let \@tempc \@tempb \let \@tempb \@tempa \fi \ifx
  \@tempb \@empty \def\@tempb {arXiv}\fi \@ifundefined
  {mn@eprint@\@tempb}{\@tempb:\@tempc}{\expandafter \expandafter \csname
  mn@eprint@\@tempb\endcsname \expandafter{\@tempc}}}

\bibitem[\protect\citeauthoryear{{Ag{\'o}n}, {Pedraza}  \&
  {Ramos-Caro}}{{Ag{\'o}n} et~al.}{2011}]{2011PhRvD..83l3007A}
{Ag{\'o}n} C.~A.,  {Pedraza} J.~F.,   {Ramos-Caro} J.,  2011, \mn@doi [\prd]
  {10.1103/PhysRevD.83.123007}, \href
  {http://adsabs.harvard.edu/abs/2011PhRvD..83l3007A} {83, 123007}

\bibitem[\protect\citeauthoryear{{Andr{\'e}asson}}{{Andr{\'e}asson}}{2011}]{2011LRR....14....4A}
{Andr{\'e}asson} H.,  2011, \mn@doi [Living Reviews in Relativity]
  {10.12942/lrr-2011-4}, \href
  {http://adsabs.harvard.edu/abs/2011LRR....14....4A} {14, 4}

\bibitem[\protect\citeauthoryear{{Antonov}}{{Antonov}}{1962}]{antonov1962vest}
{Antonov} V.,  1962, Vest. leningr. gos. Univ., 7, 135

\bibitem[\protect\citeauthoryear{Balescu}{Balescu}{1997}]{balescu1997statistical}
Balescu R.,  1997, Statistical dynamics: matter out of equilibrium.
Imperial College Press, distributed through World Scientific

\bibitem[\protect\citeauthoryear{{Behroozi} et~al.,}{{Behroozi}
  et~al.}{2015}]{2015MNRAS.454.3020B}
{Behroozi} P.,  et~al., 2015, \mn@doi [\mnras] {10.1093/mnras/stv2046}, \href
  {http://adsabs.harvard.edu/abs/2015MNRAS.454.3020B} {454, 3020}

\bibitem[\protect\citeauthoryear{{Bekenstein}}{{Bekenstein}}{1973}]{1973PhRvD...7.2333B}
{Bekenstein} J.~D.,  1973, \mn@doi [\prd] {10.1103/PhysRevD.7.2333}, \href
  {http://adsabs.harvard.edu/abs/1973PhRvD...7.2333B} {7, 2333}

\bibitem[\protect\citeauthoryear{{Bender} \& {Orszag}}{{Bender} \&
  {Orszag}}{1999}]{1978amms.book.....B}
{Bender} C.~M.,  {Orszag} S.~A.,  1999, {Advanced Mathematical Methods for
  Scientists and Engineers: Asymptotic Methods and Perturbation Theory}.
Springer

\bibitem[\protect\citeauthoryear{{Binney} \& {Tremaine}}{{Binney} \&
  {Tremaine}}{2008}]{BT}
{Binney} J.,  {Tremaine} S.,  2008, {Galactic Dynamics: Second Edition}.
Princeton University Press

\bibitem[\protect\citeauthoryear{{Bogolioubov}}{{Bogolioubov}}{1946}]{1946bogbovart}
{Bogolioubov} N.~N.,  1946, J. Phys. USSR, 10, 257

\bibitem[\protect\citeauthoryear{Born \& Green}{Born \&
  Green}{1946}]{born1946general}
Born M.,  Green H.,  1946, \mn@doi [Proc. R. Soc.A] {10.1098/rspa.1946.0093},
  \href {http://adsabs.harvard.edu/abs/1946RSPSA.188...10B} {188, 10}

\bibitem[\protect\citeauthoryear{{Campa}, {Dauxois}  \& {Ruffo}}{{Campa}
  et~al.}{2009}]{2009PhR...480...57C}
{Campa} A.,  {Dauxois} T.,   {Ruffo} S.,  2009, \mn@doi [\physrep]
  {10.1016/j.physrep.2009.07.001}, \href
  {http://adsabs.harvard.edu/abs/2009PhR...480...57C} {480, 57}

\bibitem[\protect\citeauthoryear{Campa, Dauxois, Fanelli  \& Ruffo}{Campa
  et~al.}{2014}]{campa2014physics}
Campa A.,  Dauxois T.,  Fanelli D.,   Ruffo S.,  2014, {Physics of long-range
  interacting systems}.
Oxford University Press

\bibitem[\protect\citeauthoryear{{Chavanis}}{{Chavanis}}{2006}]{2006IJMPB..20.3113C}
{Chavanis} P.~H.,  2006, \mn@doi [International Journal of Modern Physics B]
  {10.1142/S0217979206035400}, \href
  {http://adsabs.harvard.edu/abs/2006IJMPB..20.3113C} {20, 3113}

\bibitem[\protect\citeauthoryear{{Chavanis}}{{Chavanis}}{2012}]{2012PhyA..391.3680C}
{Chavanis} P.-H.,  2012, \mn@doi [Physica A Statistical Mechanics and its
  Applications] {10.1016/j.physa.2012.02.019}, \href
  {http://adsabs.harvard.edu/abs/2012PhyA..391.3680C} {391, 3680}

\bibitem[\protect\citeauthoryear{{Chavanis}}{{Chavanis}}{2013}]{2013A&A...556A..93C}
{Chavanis} P.-H.,  2013, \mn@doi [\aap] {10.1051/0004-6361/201220607}, \href
  {http://adsabs.harvard.edu/abs/2013A&A...556A..93C} {556, A93}

\bibitem[\protect\citeauthoryear{{Chavanis}, {Sommeria}  \&
  {Robert}}{{Chavanis} et~al.}{1996}]{1996ApJ...471..385C}
{Chavanis} P.~H.,  {Sommeria} J.,   {Robert} R.,  1996, \mn@doi [\apj]
  {10.1086/177977}, \href {http://adsabs.harvard.edu/abs/1996ApJ...471..385C}
  {471, 385}

\bibitem[\protect\citeauthoryear{{Chavanis}, {Rosier}  \& {Sire}}{{Chavanis}
  et~al.}{2002}]{2002PhRvE..66c6105C}
{Chavanis} P.-H.,  {Rosier} C.,   {Sire} C.,  2002, \mn@doi [\pre]
  {10.1103/PhysRevE.66.036105}, \href
  {http://adsabs.harvard.edu/abs/2002PhRvE..66c6105C} {66, 036105}

\bibitem[\protect\citeauthoryear{{Cohn}}{{Cohn}}{1980}]{1980ApJ...242..765C}
{Cohn} H.,  1980, \mn@doi [\apj] {10.1086/158511}, \href
  {http://adsabs.harvard.edu/abs/1980ApJ...242..765C} {242, 765}

\bibitem[\protect\citeauthoryear{{Fried} \& {Conte}}{{\relax
  DLMF}}{2014}]{NIST}
{\relax DLMF} 2014, {NIST Digital Library of Mathematical Functions}, Release
  1.0.9 of 2014-08-29, \url {http://dlmf.nist.gov}

\bibitem[\protect\citeauthoryear{{Dehnen}}{{Dehnen}}{2005}]{2005MNRAS.360..892D}
{Dehnen} W.,  2005, \mn@doi [\mnras] {10.1111/j.1365-2966.2005.09099.x}, \href
  {http://adsabs.harvard.edu/abs/2005MNRAS.360..892D} {360, 892}

\bibitem[\protect\citeauthoryear{{Fried} \& {Conte}}{{Fried} \&
  {Conte}}{1961}]{1961pdf..book.....F}
{Fried} B.~D.,  {Conte} S.~D.,  1961, {The Plasma Dispersion Function}.
Academic Press

\bibitem[\protect\citeauthoryear{{Gilbert}}{{Gilbert}}{1968}]{1968ApJ...152.1043G}
{Gilbert} I.~H.,  1968, \mn@doi [\apj] {10.1086/149616}, \href
  {http://adsabs.harvard.edu/abs/1968ApJ...152.1043G} {152, 1043}

\bibitem[\protect\citeauthoryear{{Goodman} \& {Hut}}{{Goodman} \&
  {Hut}}{1985}]{1985IAUS..113.....G}
{Goodman} J.,  {Hut} P.,  eds, 1985, {Dynamics of star clusters; Proceedings of
  the Symposium, Princeton, NJ, May 29-June 1, 1984}  IAU Symposium Vol. 113

\bibitem[\protect\citeauthoryear{{Hawking}}{{Hawking}}{1976}]{1976PhRvD..13..191H}
{Hawking} S.~W.,  1976, \mn@doi [\prd] {10.1103/PhysRevD.13.191}, \href
  {http://adsabs.harvard.edu/abs/1976PhRvD..13..191H} {13, 191}

\bibitem[\protect\citeauthoryear{{Heggie} \& {Hut}}{{Heggie} \&
  {Hut}}{2003}]{2003gmbp.book.....H}
{Heggie} D.,  {Hut} P.,  2003, {The Gravitational Million-Body Problem: A
  Multidisciplinary Approach to Star Cluster Dynamics}.
Cambridge University Press

\bibitem[\protect\citeauthoryear{{Heyvaerts}}{{Heyvaerts}}{2010}]{2010MNRAS.407..355H}
{Heyvaerts} J.,  2010, \mn@doi [\mnras] {10.1111/j.1365-2966.2010.16899.x},
  \href {http://adsabs.harvard.edu/abs/2010MNRAS.407..355H} {407, 355}

\bibitem[\protect\citeauthoryear{{Huang}}{{Huang}}{1987}]{1987stme.book.....H}
{Huang} K.,  1987, {Statistical Mechanics, 2nd Edition}.
Wiley-VCH

\bibitem[\protect\citeauthoryear{Ichimaru}{Ichimaru}{1973}]{ichimaru1973basic}
Ichimaru S.,  1973, Basic principles of plasma physics.
Benjamin

\bibitem[\protect\citeauthoryear{{Irving} \& {Kirkwood}}{{Irving} \&
  {Kirkwood}}{1950}]{1950JChPh..18..817I}
{Irving} J.~H.,  {Kirkwood} J.~G.,  1950, \mn@doi [\jcp] {10.1063/1.1747782},
  \href {http://adsabs.harvard.edu/abs/1950JChPh..18..817I} {18, 817}

\bibitem[\protect\citeauthoryear{{Jeans}}{{Jeans}}{1902}]{1902RSPTA.199....1J}
{Jeans} J.~H.,  1902, \mn@doi [Philosophical Transactions of the Royal Society
  of London Series A] {10.1098/rsta.1902.0012}, \href
  {http://adsabs.harvard.edu/abs/1902RSPTA.199....1J} {199, 1}

\bibitem[\protect\citeauthoryear{{Kalnajs}}{{Kalnajs}}{1976}]{1976ApJ...205..751K}
{Kalnajs} A.~J.,  1976, \mn@doi [\apj] {10.1086/154331}, \href
  {http://adsabs.harvard.edu/abs/1976ApJ...205..751K} {205, 751}

\bibitem[\protect\citeauthoryear{{Kandrup}}{{Kandrup}}{1983}]{1983Ap&SS..89..143K}
{Kandrup} H.~E.,  1983, \mn@doi [\apss] {10.1007/BF01008391}, \href
  {http://adsabs.harvard.edu/abs/1983Ap%26SS..89..143K} {89, 143}

\bibitem[\protect\citeauthoryear{{Katz}}{{Katz}}{1980}]{1980MNRAS.190..497K}
{Katz} J.,  1980, \mn@doi [\mnras] {10.1093/mnras/190.3.497}, \href
  {http://adsabs.harvard.edu/abs/1980MNRAS.190..497K} {190, 497}

\bibitem[\protect\citeauthoryear{{Kerscher}, {Szapudi}  \& {Szalay}}{{Kerscher}
  et~al.}{2000}]{2000ApJ...535L..13K}
{Kerscher} M.,  {Szapudi} I.,   {Szalay} A.~S.,  2000, \mn@doi [\apjl]
  {10.1086/312702}, \href {http://adsabs.harvard.edu/abs/2000ApJ...535L..13K}
  {535, L13}

\bibitem[\protect\citeauthoryear{Kirkwood}{Kirkwood}{1946}]{kirkwood1946statistical}
Kirkwood J.~G.,  1946, \mn@doi [\jcp] {10.1063/1.1724117}, \href
  {http://adsabs.harvard.edu/abs/1946JChPh..14..180K} {14, 180}

\bibitem[\protect\citeauthoryear{Landau}{Landau}{1946}]{landau1946vibrations}
Landau L.~D.,  1946, Zh. Eksp. Teor. Fiz., 10, 25

\bibitem[\protect\citeauthoryear{{Levin}, {Pakter}, {Rizzato}, {Teles}  \&
  {Benetti}}{{Levin} et~al.}{2014}]{2014PhR...535....1L}
{Levin} Y.,  {Pakter} R.,  {Rizzato} F.~B.,  {Teles} T.~N.,   {Benetti}
  F.~P.~C.,  2014, \mn@doi [\physrep] {10.1016/j.physrep.2013.10.001}, \href
  {http://adsabs.harvard.edu/abs/2014PhR...535....1L} {535, 1}

\bibitem[\protect\citeauthoryear{{Libeskind} et~al.,}{{Libeskind}
  et~al.}{2018}]{2018MNRAS.473.1195L}
{Libeskind} N.~I.,  et~al., 2018, \mn@doi [\mnras] {10.1093/mnras/stx1976},
  \href {http://adsabs.harvard.edu/abs/2018MNRAS.473.1195L} {473, 1195}

\bibitem[\protect\citeauthoryear{{Lynden-Bell}}{{Lynden-Bell}}{1967}]{1967MNRAS.136..101L}
{Lynden-Bell} D.,  1967, \mn@doi [\mnras] {10.1093/mnras/136.1.101}, \href
  {http://adsabs.harvard.edu/abs/1967MNRAS.136..101L} {136, 101}

\bibitem[\protect\citeauthoryear{{Lynden-Bell} \& {Wood}}{{Lynden-Bell} \&
  {Wood}}{1968}]{1968MNRAS.138..495L}
{Lynden-Bell} D.,  {Wood} R.,  1968, \mn@doi [\mnras]
  {10.1093/mnras/138.4.495}, \href
  {http://adsabs.harvard.edu/abs/1968MNRAS.138..495L} {138, 495}

\bibitem[\protect\citeauthoryear{{Martys}}{{Martys}}{1999}]{1999IJMPC..10.1367M}
{Martys} N.~S.,  1999, \mn@doi [International Journal of Modern Physics C]
  {10.1142/S0129183199001121}, \href
  {http://adsabs.harvard.edu/abs/1999IJMPC..10.1367M} {10, 1367}

\bibitem[\protect\citeauthoryear{{Merritt}}{{Merritt}}{2013}]{2013degn.book.....M}
{Merritt} D.,  2013, {Dynamics and Evolution of Galactic Nuclei}.
Princeton University Press

\bibitem[\protect\citeauthoryear{{Mo}, {van den Bosch}  \& {White}}{{Mo}
  et~al.}{2010}]{2010gfe..book.....M}
{Mo} H.,  {van den Bosch} F.~C.,   {White} S.,  2010, {Galaxy Formation and
  Evolution}.
Cambridge University Press

\bibitem[\protect\citeauthoryear{Olver, Lozier, Boisvert  \& Clark}{Olver
  et~al.}{2010}]{Olver:2010:NHMF}
Olver F.~W.~J.,  Lozier D.~W.,  Boisvert R.~F.,   Clark C.~W.,  eds, 2010,
  {NIST Handbook of Mathematical Functions}.
Cambridge University Press

\bibitem[\protect\citeauthoryear{{Onions} et~al.,}{{Onions}
  et~al.}{2012}]{2012MNRAS.423.1200O}
{Onions} J.,  et~al., 2012, \mn@doi [\mnras]
  {10.1111/j.1365-2966.2012.20947.x}, \href
  {http://adsabs.harvard.edu/abs/2012MNRAS.423.1200O} {423, 1200}

\bibitem[\protect\citeauthoryear{{Padmanabhan}}{{Padmanabhan}}{1990}]{1990PhR...188..285P}
{Padmanabhan} T.,  1990, \mn@doi [\physrep] {10.1016/0370-1573(90)90051-3},
  \href {http://adsabs.harvard.edu/abs/1990PhR...188..285P} {188, 285}

\bibitem[\protect\citeauthoryear{{Poisson} \& {Will}}{{Poisson} \&
  {Will}}{2014}]{2014grav.book.....P}
{Poisson} E.,  {Will} C.~M.,  2014, {Gravity: Newtonian, Post-Newtonian,
  Relativistic}.
Cambridge University Press

\bibitem[\protect\citeauthoryear{{Ramos-Caro}, {Ag{\'o}n}  \&
  {Pedraza}}{{Ramos-Caro} et~al.}{2012}]{2012PhRvD..86d3008R}
{Ramos-Caro} J.,  {Ag{\'o}n} C.~A.,   {Pedraza} J.~F.,  2012, \mn@doi [\prd]
  {10.1103/PhysRevD.86.043008}, \href
  {http://adsabs.harvard.edu/abs/2012PhRvD..86d3008R} {86, 043008}

\bibitem[\protect\citeauthoryear{{Rioseco} \& {Sarbach}}{{Rioseco} \&
  {Sarbach}}{2017}]{2017CQGra..34i5007R}
{Rioseco} P.,  {Sarbach} O.,  2017, \mn@doi [Classical and Quantum Gravity]
  {10.1088/1361-6382/aa65fa}, \href
  {http://adsabs.harvard.edu/abs/2017CQGra..34i5007R} {34, 095007}

\bibitem[\protect\citeauthoryear{{Staniscia}, {Chavanis}, {de Ninno}  \&
  {Fanelli}}{{Staniscia} et~al.}{2009}]{2009PhRvE..80b1138S}
{Staniscia} F.,  {Chavanis} P.~H.,  {de Ninno} G.,   {Fanelli} D.,  2009,
  \mn@doi [\pre] {10.1103/PhysRevE.80.021138}, \href
  {http://adsabs.harvard.edu/abs/2009PhRvE..80b1138S} {80, 021138}

\bibitem[\protect\citeauthoryear{{Tremaine}, {Henon}  \&
  {Lynden-Bell}}{{Tremaine} et~al.}{1986}]{1986MNRAS.219..285T}
{Tremaine} S.,  {Henon} M.,   {Lynden-Bell} D.,  1986, \mn@doi [\mnras]
  {10.1093/mnras/219.2.285}, \href
  {http://adsabs.harvard.edu/abs/1986MNRAS.219..285T} {219, 285}

\bibitem[\protect\citeauthoryear{{Vereshchagin} \& {Aksenov}}{{Vereshchagin} \&
  {Aksenov}}{2017}]{2017rkt..book.....V}
{Vereshchagin} G.~V.,  {Aksenov} A.~G.,  2017, {Relativistic Kinetic Theory}.
Cambridge University Press

\bibitem[\protect\citeauthoryear{{Villani}}{{Villani}}{2002}]{villani2002review}
{Villani} C.,  2002, in {Friedlander} S.,  {Serre} D.,  eds, Vol.~1, {Handbook
  of Mathematical Fluid Dynamics}.
Elsevier, Chapt.~2, pp 71--305

\bibitem[\protect\citeauthoryear{{Weinberg}}{{Weinberg}}{1993}]{1993ApJ...410..543W}
{Weinberg} M.~D.,  1993, \mn@doi [\apj] {10.1086/172773}, \href
  {http://adsabs.harvard.edu/abs/1993ApJ...410..543W} {410, 543}

\bibitem[\protect\citeauthoryear{Wren}{Wren}{2018}]{mycalcs}
Wren A.~J.,  2018, Calculations for this paper using \emph{Mathematica}, \url
  {https://github.com/AndrewWren/Entropy-and-gravitational-collapse-2018}

\bibitem[\protect\citeauthoryear{Yvon}{Yvon}{1935}]{yvon1935theorie}
Yvon J.,  1935, La th{\'e}orie statistique des fluides et l'{\'e}quation
  d'{\'e}tat.
Actualit{\'e}s scientifiques et industrielles, Hermann {\&} cie

\makeatother
\end{thebibliography}





\onecolumn

\appendix

\section{Approximating the initial delta function perturbation by a Gaussian}
\label{sec:approximating-the-initial-delta-function-perturbation-by-a-gaussian}

As noted after \eref{eq:f1-init}, the formulation of $f_{1,\text{init}}$ via a delta function is strictly speaking not compatible with perturbation theory. This is dealt with by regarding the Dirac delta function as an approximation of a Gaussian in $\xo,$
\be \label{eq:Dirac-Max-approx}
\ddth(\xo)
\approx
\frac{\exp\left[-\infrac{x_1^2}{2 w^2}\right]}{(2\pi w^2)^\infrac{3}{2}}
\ee
for some relatively small width $w>0$ and, having fixed $w,$ then taking $\epsilon$ to be small enough to ensure perturbation theory works. Note that, with our conventions, the Fourier transform of the Dirac delta function of \eref{eq:f1-init} is $1,$ while the Fourier transform of the Maxwellian of \eref{eq:Dirac-Max-approx} which it approximates is $\ex^{-\infrac{k_1^2\,w^2}{2}}
\approx
1
-
\infrac{k_1^2\,w^2}{2}
.$ So, for small $k_1 w,$ the Fourier transform is effectively $1.$   We shall be most interested in wave-numbers $k_1<\kJ\beta,$ for some $\beta\ll 1.$  If we are given $\beta,$ then we need only insist that $w \ll \kJ^{-1}\beta^{-1},$ to get $k_1 w\ll 1.$

From the growth with time of $\bfof$ (and hence $\fof$) set out in \eref{eq:f1-fast-early}, and from Eqs.~\eqref{eq:f0-defn} and~\eqref{eq:Dirac-Max-approx}, we can see that for first order perturbation theory to be valid we need $\epsilon$ and/or $t$ to be of small enough order such that
\be \label{eq:perturb-valid}
\frac{\epsilon\,\ex^{\eta(k_1)\,t}}{w^3}
\ll
\frac{1}{\vol}
\,, 
\quad\text{so, with the quantities defined after \eref{eq:dScg-dt},}\quad
\ex^{\eta(k_1)\,t}
\ll
\frac{n w^3}{N_1}
\ll
\frac{n B}{N_1}
.
\ee

When considering distributions of entropy creation in space, as in Subsection~\ref{sec:the-varying-entropy-change-by-radius}, in order to maintain our Gaussian approximation we must require that radius $r$ considered satisfies $w\ll r\ll R,$ recalling that $R$ is the radius of the large volume $\vol.$ To also satisfy \eref{eq:perturb-valid} again requires small enough $\epsilon$ and/or $t.$

\section{The plasma dispersion function}
\label{sec:behaviour-at-small-wave-numbers}

This appendix explores the properties of the plasma dispersion function defined by 
\be \label{eq:Zz-def}
Z(z)\equiv \frac{1}{\sqrt{\pi}}\int_{-\infty}^\infty\!\frac{\ex^{-s^2}\,\dd s}{s-z}
\ee
for $\Im z>0,$ and by analytic continuation for $\Im z\le 0.$  It is easy to see that this relates to the definition of $Z(k,\omega)$ in \eref{eq:def-Z} via $Z(k,\omega)\equiv Z(\infrac{\omega}{\sqrt{2}k\sigma})\,.$ The key source for this appendix is \citet{1961pdf..book.....F}, and see also \citet[app.~C.3]{BT}. The plasma dispersion function can alternatively be defined by 
\be \label{eq:def-orig-Z}
Z(z)
\equiv
\ii\sqrt{\pi}\,\ex^{-z^2}\left[1+\operatorname{erf}\left(\ii z\right)\right]
=
\ii\sqrt{\pi}\,\ex^{-z^2}\operatorname{erfc}\left(-\ii z\right)
,
\ee
where $\operatorname{erf}$ is the usual error function, and $\operatorname{erfc}$ the usual complementary error function.

We saw definitions of $Y(k,\omega)$ and $P(k,\omega)$ in Eqs.~\eqref{eq:def-Y} and~\eqref{eq:def-P}, which immediately give related definitions of $Y(z)$ and $P(z)$ along the same lines as for $Z(k,\omega)$ and $Z(z).$ The notation $Z$ and $P$ is fairly standard and is used in \citet{1961pdf..book.....F}, but the function $Y$ in \citet{1961pdf..book.....F} is different from ours.  A related function, the \emph{Fadeeva function} $w(z)=\infrac{Z(z)}{\ii\sqrt{\pi}}$ is often considered, and its properties are discussed in, for example, \citet[Ch.~7]{NIST}, which is the online companion to \citet{Olver:2010:NHMF}.

\subsection{Asymptotic series for small wave-numbers}
\label{sec:asymptotic-series-for-small-wave-numbers}

We will be particularly interested in the properties of $Z(k,\omega)$ and the associated functions for small $k,$ that is the properties of $Z(z)$ for large $z.$ As $z\to\infty,$ we have an \emph{asymptotic series} for $Z,$
\bml\label{eq:z-lim-formula}
Z(z)
=
\ii\sqrt{\pi}\,\tau\,\ex^{-z^2}
-
\sum_{j=0}^\infty \frac{(j-\frac{1}{2})!}{\sqrt{\pi}\, z^{2j+1}}
=
\ii\sqrt{\pi}\,\tau\,\ex^{-z^2}
-\frac{1}{z}-\frac{1}{2 z^3}-\frac{3}{4 z^5}-\frac{15}{8z^7}+\order(z^{-9})
\, ,
\qquad
\text{where }
\tau
=
\begin{dcases*}
	0 & for $\Im z > 0$ \\
	1 & for $\Im z = 0$ \\
	2 & for $\Im z < 0$ 
\end{dcases*}
,
\eml
which can be derived from \eref{eq:Zz-def}, and analytic continuation for $\Im z \le 0,$ using standard results for the moments of the Gaussian. The approximation excludes the ``tails'' of the integral as $|s|\to\infty,$ to ensure that $|\infrac{s}{z}|<1$ for the series expansion.  For large $|z|,$ the error is made small by the integral's exponential function.  

As mentioned, the series \eref{eq:z-lim-formula} is not convergent, but \emph{asymptotic}.  As a power series, it does not converge for any finite $z,$ because the ratio of a term over its predecessor is ${\left(j-\infrac{1}{2}\right)}/{z^2},$ which goes to infinity for any finite $z.$  The utility of the series arises because taking the first few terms of the series can give a very good approximation for large but finite $z.$  Heuristically, the first term which is \emph{not} used in the approximation provides an estimate of the error.  Asymptotic series are discussed in depth in, for example, \citet{1978amms.book.....B}. From \eref{eq:z-lim-formula}, we can also find similar asymptotic series for $Z(k,\omega),Y(k,\omega),$ and $P(k,\omega),$ for small $k>0.$

\subsection{The plasma dispersion relation and its zeros}
\label{sec:the-plasma-dispersion-relation-and-its-zeros}

The \emph{plasma dispersion relation}, or here simply \emph{dispersion relation}, is the equation
\be \label{eq:dispersion-relation}
k^2
=
\kJ^2\, P(k,\omega)
\equiv
\kJ^2+\kJ^2\, \frac{\omega}{\sqrt{2 \sigma k}}Z(k,\omega)
,
\ee
and, given $k\ge 0,$ we call a solution $\omega$ a \emph{dispersion zero}.  From the series in \eref{eq:z-lim-formula}, for small $k>0$ and $\Im\omega>0,$ it can be found, order by order, that the  \citep[unique,][]{BT} dispersion zero with positive imaginary part is given by
\be \label{eq:disp-soln-small}
\omega
\equiv
\ii\eta
=
\ii\kJ\sigma
-
\frac{3 \, \ii \sigma k^2}{2 \kJ}
+
\frac{15\, \ii \sigma\,k^4}{8\kJ^3}
-
\frac{147\, \ii \sigma\,k^6}{16\kJ^5}
+
\frac{9531\,\ii \sigma\,k^8}{128\kJ^7}
+
\order\left(k^{10}\right)
,
\ee
which is checked\opb{} in \citet{mycalcs}.

The\clb{} numerically-calculated values of $\eta$  used in  Figure~\ref{fig:figdispersionrelation} are obtained in \citet{mycalcs} by calculating $Z(z)$ using \eref{eq:def-orig-Z} for $-1< \Im z < 1,$ and using a continued fraction method from \citet{1961pdf..book.....F} for $\Im z \ge 1,$ continuing the fraction using $20$ terms.  For $\Im z\le -1,$ the value of $Z(z)$ follows from the continued fraction method and use of the result \citep[eq.~C.25]{BT} that, for real $x$ and $y,$
\be \label{eq:Z-conjugation}
Z(x-\ii y)
=
Z^*(x+\ii y)
+
2\ii\sqrt{\pi}\ex^{-\left(x-\ii y\right)^2}
,
\ee
where $Z^*$ denotes the complex conjugate of $Z.$ (There is a typo omitting that conjugation of $Z$ in the corresponding formula in \citealp{1961pdf..book.....F}.)

\citet{BT} discusses the dispersion zeros. As mentioned, given $k,$ there is at most one dispersion zero with positive imaginary part; this zero only occurs for $0<k<\kJ.$ The only real dispersion zeros occur for $k=0$ and $k=\kJ$ (and have value $0$), while, for any $k>0,$ there are an infinite number of \emph{Landau} dispersion zeros with $\Im z<0$ .  Let $\omega=|\omega|\left[\cos(\theta)-\ii\sin(\theta)\right]$ be a Landau zero,\opb{} where $0<\theta<\pi,$ and\clb{} then using \eref{eq:z-lim-formula} and the dispersion relation \eref{eq:dispersion-relation}, we have
\be \label{eq:disp-lim-formula-3}
0
=
-
\frac{\ii\sqrt{2\pi}\kJ^2\omega}{\sigma k}\,\exp\left[\frac{|\omega|^2\left(-\cos(2\theta)+\ii\sin(2\theta)\right)}{2\sigma^2 k^2}\right]
+
k^2
+
\frac{\kJ^2\sigma^2\,k^2}{\omega^2}
+
\frac{3\kJ^2\sigma^4\,k^4}{\omega^4}
+
\frac{15\kJ^2\sigma^6\,k^6}{\omega^6}
+
\order\!\left(\omega^{-8}\right)
,
\ee
where we have assumed $k$ is fixed and so the final order term is written in terms of $\omega.$

For $0\le\theta<\infrac{\pi}{4}$ or $\infrac{3\pi}{4}<\theta\le\pi$ we have $-\cos(2\theta)<0,$ and so, \opb{}when $|\omega|\gg k\sigma$ the value of the exponential in \eref{eq:disp-lim-formula-3} is close to zero; on the other hand, for $\infrac{\pi}{4}<\theta<\infrac{3\pi}{4},$ when $|\omega|\gg k\sigma$  the value of the exponential in \eref{eq:disp-lim-formula-3} is very large.\footnote{A line, such as $\theta=\infrac{\pi}{4}$ or $\theta=\infrac{3\pi}{4},$ with this kind of behaviour is often described as a \emph{Stokes line}.} Therefore, when $|\omega|\gg k\sigma,$ we\clb{} must have $\theta$ close to either $\infrac{\pi}{4}$ or $\infrac{3\pi}{4}.$  We concentrate on  $\theta\approx\infrac{\pi}{4},$ the other choice being very similar.

\subsection{Landau zeros of relatively large size}
\label{sec:dispersion-zeros-of-relatively-large-size}

In this subsection, we look at the case when we have Landau dispersion zeros of large size, that is where $|\omega|\gg\kJ\sigma.$ This part of Appendix~\ref{sec:behaviour-at-small-wave-numbers} motivates  Appendix~\ref{sec:the-residue-approach-for-the-inverse-laplace-transform-of-the-one-particle-distribution-function}'s key step in verifying that we can apply the residue formula, mentioned before \eref{eq:f1-fast-early}, to the one-particle distribution function.

We assume  $|\omega|\gg\sigma k,$ and \opb{}treat $k>0$ (which is not assumed to be smaller than $\kJ$) as\clb{} fixed. Write $\phi=\infrac{\pi}{4}-\theta,$ giving us $\omega=|\omega|\ex^{\ii\left(\phi-\infrac{\pi}{4}\right)},$ and then, from \eref{eq:disp-lim-formula-3}, we have
\be \label{eq:disp-lim-formula-4}
\exp\left[\frac{|\omega|^2\left(-\sin(2\phi)+\ii\cos(2\phi)\right)}{2\sigma^2 k^2}\right]
=
-
\frac{\ii\sigma k^3\ex^{\ii\left(\frac{\pi}{4}-\phi\right)}}{\sqrt{2\pi}\kJ^2|\omega|}
+
\order\!\left(\frac{\sigma^3 k^3}{|\omega|^3}\right)
=
\frac{\sqrt{2}\sigma k^3\ex^{\ii\left(-\frac{\pi}{4}-\phi\right)}}{\sqrt{\pi}\kJ^2|\omega|}
+
\order\!\left(\frac{\sigma^3 k^3}{|\omega|^3}\right)
.
\ee
Taking the phases, \eref{eq:disp-lim-formula-4} implies
\be \label{eq:disp-lim-formula-6}
\phi
=
-
\frac{\pi}{4}
-
\frac{|\omega|^2}{2\sigma^2 k^2}
+
2\pi n
+
\order\!\left(\frac{\sigma^2 k^2}{|\omega|^2},\phi^2\right)
,
\ee
where, in Appendix~\ref{sec:behaviour-at-small-wave-numbers} only, $n$ is an integer, and we approximated $\cos(2\phi)=1+\order(\phi^2).$  Taking the absolute value of \eref{eq:disp-lim-formula-4}, we then have
\be \label{eq:disp-lim-formula-7}
\frac{\sigma k^3}{\sqrt{2\pi}\kJ^2|\omega|}
=
\exp\left[-\frac{|\omega|^2\phi}{\sigma^2 k^2}\right]
+
\order\!\left(\frac{\sigma^3 k^3}{|\omega|^3},\phi^3\right)
=
\exp\left[-\frac{|\omega|^2\left(
	-
	\frac{\pi}{4}
	-
	\frac{|\omega|^2}{2\sigma^2 k^2}
	+
	2\pi n
	\right)}{\sigma^2 k^2}\right]
+
\order\!\left(\frac{\sigma^3 k^3}{|\omega|^3},\phi^3\right)
.
\ee
although the left-hand side of this equation is small, the factor of $\infrac{|\omega|^2}{\sigma^2 k^2}$ in the exponent on the right-hand side requires that for the equality to hold we must have
\be \label{eq:abs-dispersion-zero}
|\omega|
=
2\sigma k\sqrt{\pi\left(n-\frac{1}{8}\right)}
+
\order\left(\frac{\sigma^2 k^2}{|\omega|^2},\phi^2\right)
,
\ee
to avoid the exponential on the right-hand side of \eref{eq:disp-lim-formula-7} being very large or very small. From \eref{eq:disp-lim-formula-7}, we then have that
\be \label{eq:phi}
\phi
=
\frac{1}{4\pi\left(n-\frac{1}{8}\right)}\ln\left[\frac{2\pi\kJ^2\sqrt{2n-\frac{1}{4}}}{k^2}\right]
+
\order\!\left(\frac{\sigma^3 k^3}{|\omega|^3},\phi^2\right)
.
\ee
The results of Eqs.~\eqref{eq:abs-dispersion-zero} and~\eqref{eq:phi} are similar in form to expressions in \citet[eq.~7.13.4]{NIST} for the zeros of the $\operatorname{erfc}$ operator, which occurs in \eref{eq:def-orig-Z}. Numerical calculations in \citet{mycalcs} confirm the accuracy of these approximations.

Together with the similar results for $\theta\sim\infrac{3\pi}{4},$ \eref{eq:phi} demonstrates the well-known fact that the Landau zeros with negative imaginary parts and large absolute values have real parts of very slightly larger size than their imaginary parts, and that the difference in size between the real and imaginary sizes grows increasingly small as the absolute value of the Landau zero grows. 

\subsection{The residue approach for the inverse Laplace transform of the one-particle distribution function}
\label{sec:the-residue-approach-for-the-inverse-laplace-transform-of-the-one-particle-distribution-function}

We shall now confirm that the (Fourier- and) Laplace-transformed one-particle distribution function $\btfo$ of \eref{eq:f1-model-8} can be inverse Laplace-transformed using the well-known formula 
\be \label{eq:poles-app-2}
\left(\mathcal{L}^{-1}h\right)(t)
=
-\ii\sum_p \residue{\omega_p}\!\!\left[h(\omega)\ex^{-\ii\omega_p t}\right]
,
\ee  
where $p$ ranges over all poles $\omega_p$ of $h.$  The applicability of this well-known formula to $\btfo$ is assumed in similar contexts in, for example, \citet{ichimaru1973basic} and \citet{BT}, but is not demonstrated in those texts. \opb{}We treat $k>0$ as fixed, and, as in the previous subsection, do not need to assume that it is less than $\kJ.$\clb{} 

Recall that the inverse Laplace transform of a function $h(\omega)$ is given by
\be \label{eq:inverse-Laplace}
\left(\mathcal{L}^{-1}h\right)(t)
=
\int_{-\infty+\ii c}^{\infty+\ii c}\!
\frac{\dd\omega}{2\pi}\,h(\omega)\,\ex^{-\ii\omega t}
,
\ee
for $c>0$ such that $h(t)\ex^{-ct}\to 0$ as $t\to 0.$ The standard argument used to justify \eref{eq:poles-app-2} applies to an analytic function with poles, $h(\omega),$ that tends to zero as $|\omega|\to\infty$. Jordan's Lemma then implies that the contour integral around a large semi-circle dropped from the straight line from $-X+\ii c$ to $X+\ii c$ also tends to zero, and therefore
\be \label{eq:inverse-Laplace-closed-contour}
\left(\mathcal{L}^{-1}h\right)(t)
=
\lim_{X\to\infty}\int_{\mathcal{C}_X}\!
\frac{\dd\omega}{2\pi}\,h(\omega)\,\ex^{-\ii\omega t}
,
\ee
where $\mathcal{C}_X$ is the closed contour formed by the straight line from $-X+\ii c$ to $X+\ii c$ followed by the semi-circle dropped below.  The condition for $c$ implies that any pole of $h$ lies within any contour $\mathcal{C}_X$ for large enough $X,$ and \eref{eq:poles-app-2} then follows from the residue theorem.

Returning to the inverse Laplace transform of $\btfo,$ recall that Appendix~\ref{sec:dispersion-zeros-of-relatively-large-size} showed the Landau zeros extend out to infinity.  For fixed wave-number and velocity, the first term of \eref{eq:f1-model-8} clearly tends to zero as $|\omega|\to\infty,$ so we focus entirely on the second term, which, because of the Landau zeros, is not even bounded as $\omega\to\infty$.  This lack of boundedness means we cannot directly apply the standard argument recalled above for $h.$  The residue formula \eref{eq:poles-app-2} will none the less hold if we can \opb{}construct a\clb{} sequence of dropped-below semi-circles growing in radius to infinity, with (the second term of) $\btfo(\k,\v,\omega)$ tending to zero on that sequence of semi-circles. This follows by using the standard argument, but in \eref{eq:inverse-Laplace-closed-contour} confining our attention to that sequence of semi-circles. As in the previous subsection, we will concentrate on dealing with Landau zeros having positive real parts~-- very similar steps deal with the conjugate Landau zeros with negative real parts, and we do not need to set those out explicitly. In the notation of Appendix~\ref{sec:the-residue-approach-for-the-inverse-laplace-transform-of-the-one-particle-distribution-function}, we are assuming that $-\infrac{\pi}{4}\le\phi\le \infrac{3\pi}{4}.$

We have the second term of $\btfo(\k,\v,\omega)$ as
\be \label{eq:second-term}
-
\frac{\ii \kJ^2\,\k\dotp\v\maxwell(\v)\,Y(k,\omega)}{\Big(\k\dotp\v-\omega\Big)\left(k^2-\kJ^2 P(k,\omega)\right)}
=
-
\frac{\ii \kJ^2\,\k\dotp\v\maxwell(\v)\,Y(k,\omega)}{\Big(\k\dotp\v-\omega\Big)\left(k^2-\kJ^2-\kJ^2\omega\, Y(k,\omega)\right)}
=
-
\frac{\ii \kJ^2\,\k\dotp\v\maxwell(\v)}{\Big(\k\dotp\v-\omega\Big)\left[\left(k^2-\kJ^2\right)Y(k,\omega)^{-1}-\kJ^2\omega\right]}
.
\ee
For $\Im\omega\ge 0,$ that is $\infrac{\pi}{4}\le\phi\le\infrac{3\pi}{4}\,,$ we see from \eref{eq:z-lim-formula} that $Y(k,\omega)$ tends to zero as $\omega\to\infty,$ and $\omega\,Y(k,\omega)\to -1,$ so, \eref{eq:second-term}'s middle expression shows that the second term of $\btfo(\k,\v,\omega)$ tends to zero.

Consider next the case $-\infrac{\pi}{4}\le\phi<0.$ From \eref{eq:z-lim-formula}, we see that, for large $\omega,$ we then have $Y(k,\omega)$  dominated by the first, $\tau,$ term of \eref{eq:z-lim-formula}, which is like $\ex^{-\omega^2/(2k^2\sigma^2)},$ with $\omega^2$ having negative real part.  So $Y(k,\omega)$ tends \emph{exponentially} to infinity as $\omega\to\infty,$ and therefore \eref{eq:second-term}'s final expression shows that the second term of $\btfo(\k,\v,\omega)$ again tends to zero as $\omega\to\infty.$

The remaining case is for $0\le\phi\le\infrac{\pi}{4}.$  For this case, we now construct a sequence of semi-circles on which, as $\omega\to\infty,$ the dispersion expression $k^2-\kJ^2P(k,\omega)$ is bounded below.  In constructing this sequence, we will choose the semi-circles to pass between the Landau zeros.  We have from \eref{eq:disp-lim-formula-3},
\bml \label{eq:disp-below}
\left|k^2-\kJ^2 P(k,\omega)\right|
=
\left|
-
\frac{\ii\sqrt{2\pi}\kJ^2\omega}{\sigma k}\,\exp\left[\frac{|\omega|^2\left(-\sin(2\phi)+\ii\cos(2\phi)\right)}{2\sigma^2 k^2}\right]
+
k^2
\right|
+
\order\!\left(\omega^{-2}\right)
\\
=
\left|
\frac{\sqrt{2\pi}\kJ^2|\omega|}{\sigma k}\,\exp\left[-\frac{|\omega|^2\sin(2\phi)}{2\sigma^2 k^2}
+
\ii\left(\phi-\frac{3\pi}{4}+\frac{|\omega|^2\cos(2\phi)}{2\sigma^2 k^2}\right)\right]
+
k^2
\right|
+
\order\!\left(\omega^{-2}\right)
.
\eml
We will call that final expression's first summand, $\left(\infrac{\sqrt{2\pi}\kJ^2|\omega|}{\sigma k}\right)\,\exp\left[\cdots\right],$ the \emph{exponential summand}. We can see that if $\sin(2\phi)>({2\sigma^2k^2}/{|\omega|^2})\ln({2\sqrt{2\pi}\kJ^2|\omega|}/{k^3\sigma}),$ then the exponential summand's absolute value is less than $k^2/2\,,$ implying that $|k^2-\kJ^2 P(k,\omega)|>k^2/2.$ We choose the radius of our semi-circles, $|\omega|,$ to be sufficiently large that the condition on $\sin(2\phi)$ only fails for $\phi$ small enough that $\cos(2\phi)\approx 1$ is a very close approximation.  Note that, for such $\phi,$ we have $\phi\le\order(\ln[|\omega|]\,|\omega|^{-2}).$

To deal with the small $\phi$ case, we now consider the phase from \eref{eq:disp-below}.  Set our large $|\omega|=2\sigma k \sqrt{\pi\left(n-\infrac{1}{8}+\infrac{1}{2}\right)}=2\sigma k \sqrt{\pi\left(n+\infrac{3}{8}\right)}\,,$ where $n$ is a large positive integer.  This puts each of our semi-circles roughly midway between two successive Landau zeros.  Using $\cos(2\phi)=1+\order(\phi^2),$ we now have that the exponential summand's phase is
\be \label{eq:disp-below-2}
\left(\phi-\frac{3\pi}{4}+\frac{|\omega|^2\cos(2\phi)}{2\sigma^2 k^2}\right)
=
\left(\phi-\frac{3\pi}{4}+2\pi\left(n+\frac{3}{8}\right)+\order\left(n\,\phi^2\right)\right)
=
\left(\phi+2\pi n+\order\left(\ln[n]^2\,n^{-1}\right)\right)
,
\ee
which, since $\phi$ is small, and $\ln[n]^2\,n^{-1}\to 0$ as $n\to\infty\,,$ implies the exponential summand's real part is positive, and hence $|k^2-\kJ^2 P(k,\omega)|>k^2>k^2/2.$ We have therefore shown that, on semi-circles of radius $|\omega|=2\sigma k \sqrt{\pi\left(n+\infrac{3}{8}\right)},$ for sufficiently large $n,$ we have $|k^2-\kJ^2 P(k,\omega)|>k^2/2.$ The other factors of the second term of $\btfo(\k,\v,\omega)$ taken together tend to zero as $|\omega|\to\infty\,,$ so we have shown that the second term tends to zero on our sequence of semi-circles for $0\le\phi<\infrac{\pi}{4}.$  We have now shown this in turn for all relevant cases~-- which were $\infrac{\pi}{4}\le\phi\le\infrac{3\pi}{4}\,,$ $-\infrac{\pi}{4}\le\phi<0,$ and $0\le\phi<\infrac{\pi}{4}$~-- giving us the condition we noted in the paragraph before \eref{eq:second-term}, and allowing us to apply the residue formula of \eref{eq:poles-app-2} to $\btfo(\k,\v,\omega).$

We can calculate the residue associated with the dispersion relation in \eref{eq:f1-model-8} using the well-known relation that, if $h(z)$ is any holomorphic function with a simple zero at $z=z_0,$ meaning that $h'(z_0)\ne 0,$ then
\be \label{eq:res-inv-diff}
\res_{z_0}\left[\frac{1}{h(z)}\right]=\frac{1}{h'(z_0)}
.
\ee
Assuming, $\omega\ne 0$ is a dispersion zero, we find
\be \label{eq:omega-deriv}
\pd{\left[k^2-\kJ^2 P(k,\omega)\right]}{\omega}
=
\frac{\sigma^2\left(\kJ^2-k^2\right)+\omega^2}{\omega\,\sigma^2} 
,
\ee 
where we used $Z'(z)=-2-2zZ(z)$ from \citet[eq.~C.26]{BT}, and the dispersion relation itself.

We now check that the dispersion poles are all simple for $k>0.$  The right-hand side of \eref{eq:omega-deriv}  can only be zero if $\omega^2$ is real. From the properties of the dispersion relation discussed in the paragraph after \eref{eq:Z-conjugation}, this is only possible for  $0\le k \le \kJ,$ and when $\omega$ is purely imaginary.  As shown by the blue dotted line in Figure~\ref{fig:figdispersionrelation}, the right-hand side of \eref{eq:omega-deriv} only vanishes for $k=0.$  For $k\ne 0,$ we therefore have only simple poles, which means we can use \eref{eq:res-inv-diff} and also enables us to simplify \eref{eq:poles-app-2} to 
\be \label{eq:poles-app-2-simple}
\bfo(\k,\v,t)
=
-\ii\sum_p \residue{\omega_p}\!\!\left[\btfo(\k,\v,\omega)\right]\ex^{-\ii\omega_p t}
,
\ee  
where the sum ranges over all the poles of $\btfo.$ Asymptotically over time the fastest growing part of $\bfo$ is therefore that associated with the pole $\omega_p=\ii\eta.$\footnote{It is possible for other terms to give initially faster-growing parts.  For example, for $a,b>0,$ suppose $\omega=a-\ii b$ is a Landau zero, and hence so is $-a-\ii b.$  We can also approximate $a\approx b$ if $|\omega|$ is large.  Pairing terms, we get time-dependence like 
	$
		\ii\,\ex^{-a t}\sin\left(a t\right)
	.
	$ 
	It is easy to see that, for $a t \ll 1,$ this quantity is fast growing, while for $t\gg \infrac{1}{a},$ it will be highly suppressed.
} This, with \eref{eq:omega-deriv}, gives us the result for $\bfo$ quoted in \eref{eq:f1-fast-early}.

\section{Asymptotic coarse-grained number density and entropy density of the first order perturbation function}
\label{sec:asymptotic-coarse-grained-number-density-and-entropy-density-of-the-first-order-perturbation-function}

\begin{figure*} 
	\centering
	\includegraphics[width=1.0\textwidth]{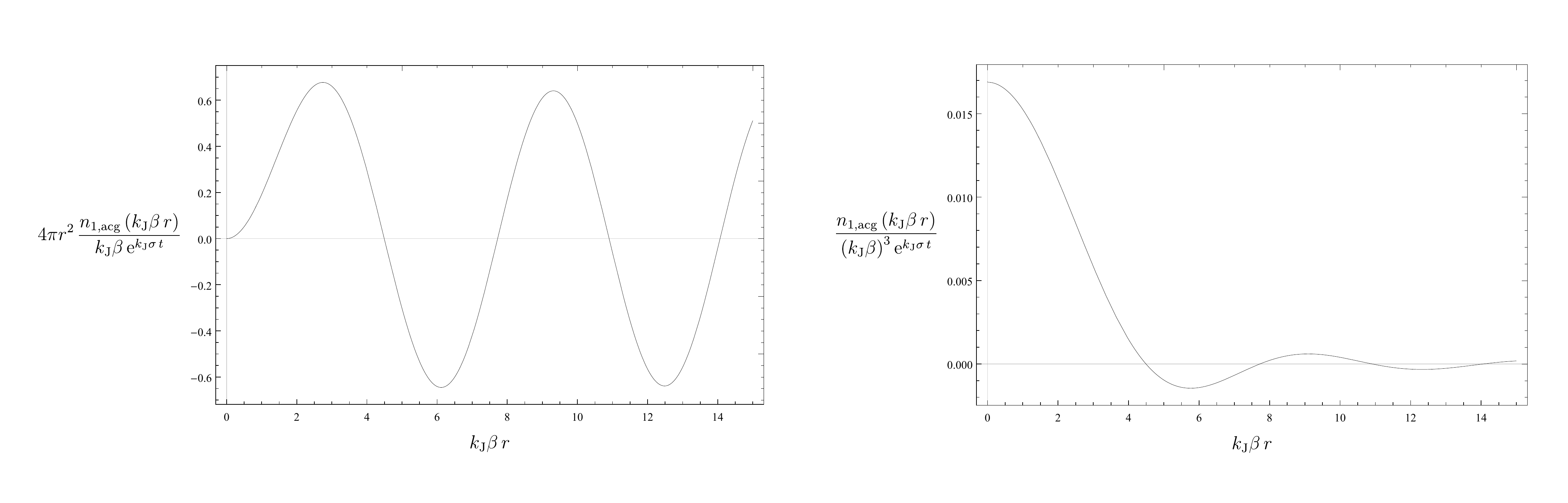}
	\caption{The leading order part of the asymptotic coarse-grained number density $n_{1,\text{acg}}.$ \emph{Left:} The ``number density by shell'' indicating the number of particles in a thin shell at a given  radius.  As an aside, it can easily be shown that the amplitude of the shell density's oscillations tends to a non-zero constant, $4\pi r^2 {n_{1,\text{acg}}}/{(\kJ\beta\,\ex^{\kJ\sigma\,t})}\to {2}/{\pi}$ as $\kJ\beta r\to\infty.$ \emph{Right:} The number density by volume, tending rapidly to zero as $\kJ\beta r\to\infty.$}
	\label{fig:figdensity}
\end{figure*}

Integrating \eref{eq:f1-fast-early} with respect to $\vo,$ we get the (Fourier-transformed) asymptotic number density function,
\be \label{eq:rho1}
\bar{n}_{1,\text{a}}(\ko,\vo,t)
=
\frac{ \sigma^2\,P(\ko,\ii\,\eta)\,\left(\kJ^2-k_1^2\right)\ \ex^{\eta(k_1)\,t}}{\sigma^2\left(\kJ^2-k_1^2\right)-\eta(k_1)^2}
=
\frac{ \sigma^2\,\frac{k_1^2}{\kJ^2}\,\kJ^2\ \ex^{\kJ\sigma\,t}}{\sigma^2\left(\kJ^2-k_1^2\right)-\kJ^2\sigma^2+3\sigma^2k_1^2}
+
\order\left(k_1^2\right)
=
\frac{\ex^{\kJ\sigma\,t}}{2}+\order\left(k_1^2\right)
,
\ee
where we used the dispersion relation, and the series approximation from  \eref{eq:disp-soln-small}. We now make our coarse-graining explicit and treat $\bar{n}_{1,\text{a}}$ as vanishing unless $0<k<\kJ\beta,$ where, as usual, $\beta\ll 1.$  If we then inverse Fourier transform, we get
\be \label{eq:rho1-2}
n_{1,\text{acg}}(\xo,\vo,t)
=
\frac{\left(\kJ\beta\right)^3\left[\sin(\kJ\beta\,x_1)-\kJ\beta\,x_1\cos(\kJ\beta\,x_1)\right]\ex^{\kJ\sigma\,t}}{2\pi^2 \, \left(\kJ\beta\,x_1\right)^3}+\order\left(\beta^4\right)
.
\ee
This is shown in Figure~\ref{fig:figdensity}~(Right).  The pattern of the asymptotic coarse-grained number density (by volume) is a strong positive peak, with a much weaker tail. As described in the caption, Figure~\ref{fig:figdensity}~(Left) shows the number density by shell. 

We now consider the entropy density in space,
\be
s(\xo)
\equiv
-\int\! f(1)\log\left[f(1)\right]\,\dd^3\vo
=
-
\int\!\left(1-\epsilon\right)f_0(1)\log\left[\left(1-\epsilon\right)f_0(1)\right]\,\dd^3\vo
-
\epsilon\int\! 
f_1(1)
\left\lbrace
1
-
\frac{v_1^2}{2\sigma^2}
\right\rbrace
\,\dd^3\vo
+
\order\left(\epsilon^2\right)
,
\ee
and find that the Fourier-transformed asymptotically-dominant entropy density in space, to leading order in $\epsilon,$ is
\bml \label{eq:entropy-density}
\bar{s}_\text{a}(\ko)
=
-
\frac{\epsilon \sigma^2\,\left(\kJ^2-k_1^2\right) \ex^{\eta(k_1)\,t}}{\left(\sigma^2\left(\kJ^2-k_1^2\right)-\eta(k_1)^2\right)}
\int\! 
\frac{\ko\dotp\vo\maxwell(\vo)}{\Big(\ko\dotp\vo-\ii\,\eta(k_1)\Big)}
\left\lbrace
1
-
\frac{v_1^2}{2\sigma^2}
\right\rbrace
\,\dd^3\vo
\\
=
-
\frac{\epsilon \sigma^2\,\left(\kJ^2-k_1^2\right)
	\ex^{\eta(k_1)\,t}}{\left(\sigma^2\left(\kJ^2-k_1^2\right)-\eta(k_1)^2\right)}
\left[
\frac{\eta(k_1)^2}{2\sigma^2 k_1^2} P\left(k_1,\ii\,\eta\right)
-
\frac{1}{2}
\right]
=
\frac{\epsilon \sigma^2\,\kJ^2
\left[\ex^{\kJ\sigma\,t}-1\right]}{2\sigma^2 k_1^2}
\frac{3k_1^2}{2\kJ^2}
+
\order\left(k_1^2\right)
=
\frac{3\epsilon \ex^{\kJ\sigma\,t}}{4}
+
\order\left(k_1^2\right)
,
\eml
where the integration can be done by hand, and is also calculated using Mathematica in \citet{mycalcs}.

Because the the Fourier-transformed asymptotic entropy density in \eref{eq:entropy-density} and number density in \eref{eq:rho1} both have constant leading order in $k_1$, they are proportional, implying that the corresponding asymptotic coarse-grained quantities in position space are also proportional to leading order in $\kJ\beta.$  At leading order therefore, the asymptotic coarse-grained entropy density pattern does not add any new information.

\section{Calculations for the zeroth order correlation function}
\label{sec:calculations-for-the-zeroth-order-correlation-function}

In this appendix, we solve \eref{eq:g0-eq-r}, which was derived from the zeroth order correlation equation. Fourier transforming \eref{eq:g0-eq-r} gives
\bml \label{eq:Gilbert-0-5-app}
\left(\vt-\vo\right)\dotp\k_-\, \bGz(\k_-,\vo,\vt)
+\frac{\kJ^2}{k_-^2}\k_-\dotp\vo \maxwell(\vo)\int  \bGz(\k_-,\u,\vt)\,\dd^3\u
-\frac{\kJ^2}{k_-^2}\k_-\dotp\vt \maxwell(\vt)\int  \bGz(\k_-,\vo,\u)\,\dd^3\u
\\
-\frac{\kJ^2}{k_-^2\,\vol}\left(\vt-\vo\right)\dotp  \k_-\maxwell(\vo)\maxwell(\vt)
=0.
\eml
This can be solved by the ansatz $\bGz(\k_-,\vo,\vt)=\qb(\k_-)\maxwell(\vo)\maxwell(\vt)\,,$ which gives us
\be \label{eq:q-ansatz}
\left(\vt-\vo\right)\dotp\k_-\, \qb(\k_-)
+\frac{\kJ^2}{k_-^2}\k_-\dotp\vo\, \qb(\k_-)
-\frac{\kJ^2}{k_-^2}\k_-\dotp\vt\, \qb(\k_-)
-\frac{\kJ^2}{k_-^2\,\vol}\left(\vt-\vo\right)\dotp  \k_-
=0
,
\ee
implying that our solution is of the form
\be \label{eq:q-ansatz-3}
\qb(\k_-)
=
-\frac{\kJ^2}{\vol\left(\kJ^2-k_-^2\right)}+\lambda(\k_-)\,\delta^{(1)}\left(k_--\kJ\right)+C\,\ddth(\k_-)
,
\ee
where $\lambda(\k_-)$ is an arbitrary function, and $C$ is an arbitrary constant which allows for the possibility that $\k_-=\zvec$ in \eref{eq:q-ansatz}.  Viewing \eref{eq:q-ansatz} as a differential equation in $\r=\xt-\xo,$ the constant $C$ arises because \eref{eq:q-ansatz} represents a total derivative, while the first two summands of \eref{eq:q-ansatz-3}'s right-hand side represent, respectively, a particular integral and a complementary function.

Note that, since $f_0(1,2)=f_0(1)f_0(2)+(\infrac{1}{N})g_0(1,2)$ to first order in $\infrac{1}{N},$ integrating over all $(2)$ gives us
\be
f_0(1)
=f_0(1)+\frac{1}{N}\int\!g_0(1,2)d(2)
=f_0(1)+\frac{1}{N}\int\!G_0(\r,\vo,\vt)\,\dd^3\r\,\dd^3\vt
\ee
and we therefore must have $\int\!q(\r)\,\dd^3\r=0.$  The first and second terms of $\qb(\k_-)$ have well-defined values at $\k_-=\zvec.$ The integral over all space of their inverse Fourier transforms can therefore be evaluated by setting $\k_-=\zvec.$ The result is $-\infrac{1}{\vol}$.  We cannot evaluate the third, $C,$ term at $\k_-=\zvec,$ but we can see that its inverse Fourier transform is $\infrac{C}{(2\pi)^3}.$ Therefore to have $q$ integrating to zero over all space, or more precisely over the volume $\vol,$ we must have that $C=\infrac{(2\pi)^3}{\vol^2}$ and so
\be \label{eq:q-ansatz-result-pre}
\qb(\k_-)
=
\frac{(2\pi)^3}{\vol^2}\ddth(\k_-)-\frac{\kJ^2}{\vol\left(\kJ^2-k_-^2\right)}+\lambda(\k_-)\,\delta^{(1)}\left(k_--\kJ\right)
.
\ee
As the focus in this paper will be on small $k\ll\kJ,$ the function $\lambda(\k_-)$ will not affect our conclusions, so we disregard it for the remainder of this paper, and have the result set out in \eref{eq:q-ansatz-result-main}.

As an aside, we can find the inverse Fourier transform, $q,$ of $\qb,$ directly from inverse Fourier transforming \eref{eq:q-ansatz} to get an equation which can be straightforwardly solved, along lines set out in \citet{1983Ap&SS..89..143K}, to get 
\be \label{eq:g0}
g_0(1,2)
=
G_0(\r,\vo,\vt)
=
q(\r)\maxwell(\vo)\maxwell(\vt)
=
\left[
\frac{1}{\vol^2}
+
\frac{\kJ^2}{4\pi\vol\, r} \,\cos (\kJ \,r)
\right]
\maxwell(\vo)\maxwell(\vt)
.
\ee

\section{The first order correlation equation}\label{sec:calculations-for-the-first-order-correlation-function}

In this appendix, we Fourier transform \eref{eq:g1-eq}, the first order correlation equation, with respect to $\xo$ and $\xt$ which gives us
\bml \label{eq:g1-eq-ft}
\pd{\bbgo(1,2)}{t} 
+\Bigg\lbrace
i\ko\dotp\vo \bbgo(1,2)
+
\frac{i\kJ^2\sigma^2\,\vol \ko}{(2\pi)^3\,k_1^2}\dotp\int\!\bfo(\ko,\u)\dd^3\u*_{\ko}\pd{\bbgz(1,2)}{\vo}
\\
+
\frac{i\kJ^2\sigma^2\,\vol \ko}{k_1^2}\pd{f_0(1)}{\vo}\dotp  \int\! \bbgo(\ko,\u,2)\,\dd^3\u
+
\pd{\bfo(1)}{\vo}\dotp*_{\ko}\frac{i\kJ^2\sigma^2\,\vol \ko}{(2\pi)^3\, k_1^2}  \int\! \bbgz(\ko,\u,2)\,\dd^3\u
\\
+
\frac{i\kJ^2\sigma^2\,\vol \ko}{k_1^2}\dotp\pd{f_0(1)}{\vo}\bfo(\kt,\vt)
-
\frac{i\kJ^2\sigma^2\,\vol \kt}{k_2^2}\dotp\pd{\bfo(\kp,\vo)}{\vo}f_0(2)
-
\frac{i\kJ^2\sigma^2\,\vol \ko}{k_1^2}\dotp\int\!\bfo(\ko,\u)\,\dd^3\u\,\pd{f_0(1)}{\vo}f_0(2)(2\pi)^3\ddth(\kt)
\\
+
\ (1)\leftrightarrow(2)
\Bigg\rbrace
=0
,
\eml
where, in the context of Fourier transforms, we read $(j)$ as $(\k_j,\v_j).$ Using \eref{eq:g0-separation-1} and the $\qb$ ansatz noted before \eref{eq:q-ansatz-result-main}, we have
\bml \label{eq:g1-eq-ft-3-2a}
\pd{\bbgo(1,2)}{t} 
+\Bigg\lbrace
i\ko\dotp\vo \bbgo(1,2)
-
\frac{i\kJ^2\,\vol \kp\dotp\vo}{k_+^2}\int\!\bfo(\kp,\u)\dd^3\u\,\qb(k_2)\maxwell(\vo)\maxwell(\vt)
\\
-
\frac{i\kJ^2\, \ko\dotp\vo}{k_1^2}\maxwell(\vo) \int\! \bbgo(\ko,\u,2)\,\dd^3\u
-
\pd{\bfo(\kp,\vo)}{\vo}\dotp\frac{i\kJ^2\sigma^2\kt}{k_2^2}\, \left(\vol\,  \qb(k_2)+1\right)\maxwell(\vt)
\\
-
\frac{i\kJ^2\,\ko\dotp\vo}{k_1^2}\maxwell(\vo)\bfo(\kp,\vt)
+
\frac{i\kJ^2\,\ko\dotp\vo}{\vol\,k_1^2}\int\!\bfo(\ko,\u)\,\dd^3\u\,\maxwell(\vo)\maxwell(\vt)(2\pi)^3\ddth(\kt)
+
\ (1)\leftrightarrow(2)
\Bigg\rbrace
=0
.
\eml
Recalling, as set out after \eref{eq:f1-init}, that our initial perturbation is uncorrelated, $g_1(1,2,t=0)=0,$ Laplace transforming \eref{eq:g1-eq-ft-3-2a} then gives \eref{eq:g1-eq-flt-main}.

\section{The first order correlation equation using a propagator}
\label{sec:exploring-the-correlation-equation-using-a-propagator}

This appendix solves the BBGKY equation for the first order correlation function, or more precisely the related function $\gof$ from \eref{eq:gamma}, using a propagator approach, which, in essence, derives a Green's function for the equation.  This well-known approach is used in, for example, \citet{ichimaru1973basic} for plasmas and, in an ``angle-action'' approach different to ours, in  \citet{2010MNRAS.407..355H} for self-gravitating particles.

\subsection{Propagators}

Note that the (Fourier-transformed) equation for the correlation function, given in \eref{eq:g1-eq-ft-3-2a}, is of the form 
\be \label{eq:Gilbert-g-eq-concise}
\pd{\bbgo(1,2,t)}{t}
+
H_1 \left[\bbgo(1,2,t)\right]
+
H_2 \left[\bbgo(1,2,t)\right]
=
D_1(1,2,t)
,
\ee
for $t>0.$ For clarity below, we have made the time argument of $\bbgo$ explicit, and we have written
\be \label{eq:H1}
H_1 \left[\bbgo(1,2,t)\right]
\equiv
\ii\ko\dotp\vo \bbgo(1,2)
-
\frac{\ii\kJ^2\sigma^2\,\vol \ko}{k_1^2}\pd{f_0(1)}{\vo}\dotp  \int\! \bbgo(\ko,\u,2)\,\dd^3\u
\,,
\ee
and similarly for $H_2,$ and we have also collected all the other terms, none of which involve $\bbgo,$  as a ``driving'' term $D_1$ on the right-hand side of \eref{eq:Gilbert-g-eq-concise}, 
\bml \label{eq:D1}
D_1(1,2,t)
=
\frac{i\kJ^2\vol\, \kp\dotp\vo}{k_+^2}\int\!\bfo(\kp,\u)\,\dd^3\u\ \qb(k_2)\maxwell(\vo)\maxwell(\vt)
+
\frac{i\kJ^2\sigma^2\, \kt}{k_2^2}\dotp\pd{\bfo(\kp,\vo)}{\vo}\left(\vol\qb(k_2)+1\right)\maxwell(\vt)
\\
+
\frac{i\kJ^2\,\ko\dotp\vo}{k_1^2}\maxwell(\vo)\bfo(\kp,\vt)
-
\frac{i\kJ^2\,\ko\dotp\vo}{\vol\,k_1^2}\int\!\bfo(\ko,\u)\,\dd^3\u\,\maxwell(\vo)\maxwell(\vt)(2\pi)^3\ddth(\kt)
+
\ (1)\leftrightarrow(2)
,
\eml
for $t>0,$ and zero otherwise. Note that $\ko$ and $\kt$ simply parametrise Eqs.~\eqref{eq:Gilbert-g-eq-concise} and~\eqref{eq:H1}, whilst, in contrast, there is differentiation and integration with respect to the velocity variables, which play the active role in the equations.  For brevity, where needed we will write variables $1'\equiv(\ko,\vo')$ and $2'\equiv(\kt,\vt'). $

To solve \eref{eq:Gilbert-g-eq-concise} with a propagator, we adopt the standard approach of writing
\be \label{eq:sep-prop}
\bbgo(1,2,t)
\equiv
\int_0^\infty\!d\tau \int\!\dd^3\vo'\,\dd^3\vt' \ \propa(1,\vo',\tau)\,\propa(2,\vt',\tau)\,D_1(1',2',t-\tau)
,
\ee
with, for $j=1,2,$
\be \label{eq:propagator}
\left[\pd{}{t}+H_j\right]\!\left[\propa(j,\v_j',t)\right]
=
0
,
\ee
and the initial condition $\propa(j,\v_j',0)=\ddth(\v_j-\v_j').$ This equation has exactly the same form as the Fourier-transformed Vlasov equation for $\bfo(\k_j,\v_j,t)\,,$ the only difference being the initial condition. We now solve \eref{eq:propagator}, using the same approach that took us to \eref{eq:f1-model-4} and then \eref{eq:f1-model-8} to find
\be \label{eq:propa0}
\tilde{\propa}(\k_j,\v_j,\v_j',\omega)
=
- \frac{\ii\,\ddth\big(\v_j-\v_j'\big)}{(\k_j\dotp\v_j-\omega) }
-
\frac{\ii\, \kJ^2\,\k_j\dotp\v_j\maxwell(\v_j)}{(\k_j\dotp\v_j-\omega)\,(\k_j\dotp\v_j'-\omega)\left[k_1^2-\kJ^2 P(k_1,\omega)\right]}
.
\ee
The technical discussion of Appendix~\ref{sec:the-residue-approach-for-the-inverse-laplace-transform-of-the-one-particle-distribution-function}, shows that, as for $\btfo,$ we can apply \eref{eq:poles-app-2-simple}, to get,\opb{} for $0<k_j<\kJ,$ that the \clb{}asymptotically-dominant term of $\propa$ is
\be \label{eq:reduced-prop-fast}
\propa_\text{a}(\k_j,\v_j,\v_j',t)
\equiv
-\frac{ \ii\,\kJ^2\,\eta(k_j)\,\sigma^2\,\k_j\dotp\v_j\maxwell(\v_j)\ \ex^{\eta(k_j)\,t}}{\Big(\k_j\dotp\v_j-\ii\,\eta(k_j)\Big)\,\left(\k_j\dotp\v_j'-\ii\,\eta(k_j)\right)\left[\sigma^2\left(\kJ^2-k_j^2\right)-\eta(k_j)^2\right]}
\,,
\ee
where, just as for $\bfof,$ we used Eqs.~\eqref{eq:res-inv-diff} and~\eqref{eq:omega-deriv} to find the square bracketed factor.

\subsection{Time dependence of distribution and correlation functions}
\label{sec:time-dependence-of-distribution-and-correlation-functions}

We now want to look at the dominant time dependence of distribution functions and correlation functions.   Throughout this appendix, we assume all functions of a time argument vanish when that time is less than $0,$  and consider only their behaviour for times greater than or equal to $0.$ We write $y\preccurlyeq z$ to mean that the proportionate growth of $y$ with time is no faster than the proportionate growth of $z$ with time (for times greater than or equal to $0$).  For example, $t\preccurlyeq \ex^t$ or  $\ex^t\preccurlyeq \ex^{2t},$ but also   $\ex^t\preccurlyeq \ex^t,$ and, because we ignore constants of proportionality,  $\ex^t\preccurlyeq 10\,\ex^{2t},$ and even  $10\,\ex^t\preccurlyeq \ex^t.$  Note also that $\num{e100}t\preccurlyeq \ex^t,$ because we are comparing rates of growth, not absolute size.  Factors without a time dependency can be omitted.  

We show that $\bbgo(1,2,t)\preccurlyeq\ex^{2\kJ\sigma\,t}.$  From the discussion of the dispersion relation after \eref{eq:def-Z}, and the detailed discussion in Appendix~\ref{sec:the-plasma-dispersion-relation-and-its-zeros}, we see that  $\propa(\k_j,\v_j,\v_j',t)\preccurlyeq\ex^{\eta(k_j)\,t},$ where, as before, $\eta(k_j)$ is the unique positive imaginary part of a dispersion zero for wave-number $0<k_j<\kJ.$ For a fixed $k_j,$ from \eref{eq:f1-fast-early} we have $\bfo(\k_j,\v_j,t)\preccurlyeq\ex^{\eta(k_j)\,t}.$ From the Fourier-transformed first order correlation equation \eref{eq:g1-eq-ft-3-2a}, we therefore have
\be
D_1(1',2',t)
\preccurlyeq
\ex^{\operatorname{max}\left[\eta(k_1),\eta(k_2),\eta(k_+)\right] t}
\preccurlyeq
\ex^{\kJ\sigma\,t}
,
\ee
the last relation following because $\eta(k_j)\le\kJ\sigma.$

Implicitly assuming still that all functions of time vanish for negative times, we have 
\be \label{eq:time-conv}
\int_0^\infty\!d\tau\,\ex^{b_1\tau}\ex^{b_2\,(t-\tau)}
=
\begin{dcases}
	\frac{\ex^{b_1\,t} - \ex^{b_2\,t}}{b_1-b_2}\preccurlyeq\ex^{\operatorname{max}\left(b_1,b_2\right)\,t}
	& \text{if } b_1\ne b_2\\[9pt]
	t\,\ex^{b_1\,t} &  \text{if } b_1= b_2
\end{dcases}
.
\ee
So, from Eqs.~\eqref{eq:sep-prop} and~\eqref{eq:time-conv}, we have, for wave-numbers $k_j$ in the range $0<k_j<\kJ,$
\bml \label{eq:g1-time-dep}
\bbgo(1,2,t)
=
\int_0^\infty\!d\tau \int\!\dd^3\vo\,\dd^3\vt \  \propa(\ko,\vo,\vo',\tau)\,\propa(\kt,\vt,\vt',\tau)\,D_1(\ko,\vo',\kt,\vt',t-\tau)
\\
\preccurlyeq
\int_0^\infty\!d\tau \, \ex^{\left[\eta(k_1)+\eta(k_2)\right]\,\tau}\,\ex^{\kJ\sigma\,(t-\tau)}
\preccurlyeq
\int_0^\infty\!d\tau  \, \ex^{2\kJ\sigma\,\tau}\,\ex^{\kJ\sigma\,(t-\tau)}
\preccurlyeq
\ex^{2\kJ\sigma\,t}
,
\eml
where in the second line we omitted all the factors without a time dependency.

We therefore have that the relative growth with time $t>0$ of the first order correlation function, $\bbgo(1,2,t)$, is no faster than $\ex^{2\kJ\sigma\,t}.$ If $\eta(k_1)>\infrac{\kJ\sigma}{2}$ and $\eta(k_2)>\infrac{\kJ\sigma}{2},$ which will be the case for small wave-numbers, then we can also see from \eref{eq:g1-time-dep} that $\bbgof(1,2,t),$ for a given $k_1$ and $k_2$ will grow like $\ex^{\left[\eta(k_1)+\eta(k_2)\right]\,t}.$

\subsection{Addressing the correlation equation using propagators}
\label{sec:solving-the-correlation-equation-using-propagators}

From \eref{eq:sep-prop}, and the discussion immediately following \eref{eq:g1-time-dep}, we have that, for $\eta(k_1)>\infrac{\kJ\sigma}{2}$ and $\eta(k_2)>\infrac{\kJ\sigma}{2},$ the asymptotically-dominant part of $\bbgo$ is given by
\bml \label{eq:g1-fast-2}
\bbgof(1,2,t)
=
-
\int\!\dd^3\vo'\,\dd^3\vt'\,\int_0^\infty\, \frac{ \kJ^2\,\eta(k_1)\,\sigma^2\,\ko\dotp\vo\maxwell(\vo)}{\Big(\ko\dotp\vo-\ii\,\eta(k_1)\Big)\,\left(\ko\dotp\vo'-\ii\,\eta(k_1)\right)\left[\sigma^2\left(\kJ^2-k_1^2\right)-\eta(k_1)^2\right]}
\\
\times
\frac{ \kJ^2\,\eta(k_2)\,\sigma^2\,\kt\dotp\vt\maxwell(\vt)}{\Big(\kt\dotp\vt-\ii\,\eta(k_2)\Big)\,\left(\kt\dotp\vt'-\ii\,\eta(k_2)\right)\left[\sigma^2\left(\kJ^2-k_2^2\right)-\eta(k_2)^2\right]}
\,\ex^{\,\left[\eta(k_1)+\eta(k_2)\right]\,\tau}\,D_1(\ko,\vo',\kt,\vt',t-\tau)
\ d\tau 
\\
=
-
\frac{\kJ^4\,\eta(k_1)\,\eta(k_2)\,\sigma^4}{\left[\sigma^2\left(\kJ^2-k_1^2\right)-\eta(k_1)^2\right]\left[\sigma^2\left(\kJ^2-k_2^2\right)-\eta(k_2)^2\right]}
\frac{ \ko\dotp\vo\maxwell(\vo)\,\kt\dotp\vt\maxwell(\vt)}{\Big(\ko\dotp\vo-\ii\,\eta(k_1)\Big)\,\Big(\kt\dotp\vt-\ii\,\eta(k_2)\Big)}
\\
\times
\int\!\frac{\dd^3\vo'\,\dd^3\vt'}{\left(\ko\dotp\vo'-\ii\,\eta(k_1)\right)\,\left(\kt\dotp\vt'-\ii\,\eta(k_2)\right)}\,
\left\lbrace\ex^{\,\left[\eta(k_1)+\eta(k_2)\right]\,t}*D_1(\ko,\vo',\kt,\vt',t)\right\rbrace(t)
,
\eml
where the asterisk denotes a time convolution.  
For brevity, we now write $E=\eta(k_1)+\eta(k_2).$ Note that, for $\omega$ with $\Im\omega>E,$ the Laplace transform of $\ex^{E t}$ is 
\be
\int_0^\infty\!dt\,\ex^{\,\left[E+\ii\omega\right]\,t}
=
\frac{\ii}{\omega-\ii E}
.
\ee
This means that the convolution at the end of \eref{eq:g1-fast-2} is an inverse Laplace transform 
\be \label{eq:convolution-time}
\left\lbrace\ex^{\,\left[\eta(k_1)+\eta(k_2)\right]\,t}*D_1(\ko,\vo',\kt,\vt',t)\right\rbrace(t)
=
\mathcal{L}^{-1}\left[\frac{\ii}{\omega-\ii E}\,\tilde{D}_1(\ko,\vo',\kt,\vt',\omega)\right](t)
.
\ee
Clearly $\infrac{\ii}{\left(\omega-\ii E\right)}$ is bounded as $\omega\to\infty.$ From the discussion of Appendix~\ref{sec:the-residue-approach-for-the-inverse-laplace-transform-of-the-one-particle-distribution-function}, given $\ko$ and $\kt,$ then for $\k_j=\ko,\kt$ or $\kp,$ $\btfo(\k_j,\v,\omega)$ is bounded on a sequence of semi-circles.  Since $\tilde{D}_1$ is the sum of a finite number of terms each depending on $\btfo,$  we can apply the residue formula for the inverse Laplace transform, \eref{eq:poles-app-2-simple}, to $\tilde{D}_1$ and therefore to $\infrac{\ii\,\tilde{D}_1(\ko,\vo',\kt,\vt',\ii E)}{\left(\omega-\ii E\right)}$ overall.  Our assumption that $\eta(k_1)>\infrac{\kJ\sigma}{2}$ and $\eta(k_2)>\infrac{\kJ\sigma}{2},$ implies that $E>\kJ\sigma,$ and so the asymptotically-dominant term comes from the simple pole at $\omega=\ii\,E,$ where the residue is $\ii.$ So, from the residue formula, we have that the asymptotically-dominant term of the convolution is
\be \label{eq:time-conv-2}
\left\lbrace\ex^{\,\left[\eta(k_1)+\eta(k_2)\right]\,t}*D_1(\ko,\vo',\kt,\vt',t)\right\rbrace_\text{a}(t)
=
-\ii\left[\ii\tilde{D}_1(\ko,\vo',\kt,\vt',\ii E)\,\ex^{E t}\right]
=
\tilde{D}_1(\ko,\vo',\kt,\vt',\ii E)\,\ex^{E t}.
\ee
Note that the time-dependency here comes entirely from the $\infrac{\ii}{\left(\omega-\ii E\right)}$ factor on the right-hand side of \eref{eq:convolution-time} -- this means that the driving term enters into \eref{eq:time-conv-2}'s right-hand side solely through its Laplace transform $\tilde{D}_1(\omega),$ and the question of whether to restrict attention to the asymptotically-dominant part of $D_1(t)$ does not arise.
 
Referring back to \eref{eq:g1-fast-2}, we now have
\bml \label{eq:g1-fast-2a}
\bbgof(1,2,t)
=
-
\frac{\kJ^4\,\eta(k_1)\,\eta(k_2)\,\sigma^4}{\left[\sigma^2\left(\kJ^2-k_1^2\right)-\eta(k_1)^2\right]\left[\sigma^2\left(\kJ^2-k_2^2\right)-\eta(k_2)^2\right]}
\frac{ \ko\dotp\vo\maxwell(\vo)\,\kt\dotp\vt\maxwell(\vt)}{\Big(\ko\dotp\vo-\ii\,\eta(k_1)\Big)\,\Big(\kt\dotp\vt-\ii\,\eta(k_2)\Big)}
\\
\times
\int\!\dd^3\vo'\,\dd^3\vt'\,
\frac{\tilde{D}_1(\ko,\vo',\kt,\vt',\ii E)}{\left(\ko\dotp\vo'-\ii\,\eta(k_1)\right)\,\left(\kt\dotp\vt'-\ii\,\eta(k_2)\right)}\,
\,\ex^{E t}
\,.
\eml
We now need to evaluate the velocity integrals in \eref{eq:g1-fast-2a}. To do this we use the Laplace transform of from \eref{eq:D1} and drop the delta function terms, which are of order $V^{-1},$ while the remaining terms are order $V^0.$ 

Using the expression for $\btfo$ in \eref{eq:f1-model-8}, we now have
\bml \label{eq:ID1-1}
I_{D_1}
\equiv
\int\!\frac{\dd^3\vo'\,\dd^3\vt'\ \tilde{D}_1(\ko,\vo',\kt,\vt',\ii E)}{\left(\ko\dotp\vo'-\ii\,\eta(k_1)\right)\,\left(\kt\dotp\vt'-\ii\,\eta(k_2)\right)}
\\
=
\int\!\frac{\dd^3\vo'\,\dd^3\vt'}{\left(\ko\dotp\vo'-\ii\,\eta(k_1)\right)\,\left(\kt\dotp\vt'-\ii\,\eta(k_2)\right)}
\ 
\Bigg\lbrace
-
\frac{\ii\,\kJ^2\,\ko\dotp\vo'}{k_1^2}\maxwell(\vo')
\left[
\frac{i\maxwell(\vt')}{\Big(\kp\dotp\vt'-\ii E\Big)}
+
\frac{i \kJ^2\,\kp\dotp\vt'\maxwell(\vt')\,Y(k_+,\ii E)}{\Big(\kp\dotp\vt'-\ii E\Big)\left(k_+^2-\kJ^2 P(k_+,\ii E)\right)}
\right]
\\
-
\frac{\ii\,\kJ^2\, \kp\dotp\vo'}{k_+^2}
\left[
i Y(k_+,\ii E)
+
\frac{i \kJ^2\,P(k_+,\ii E)
	\,Y(k_+,\ii E)}{\left(k_+^2-\kJ^2 P(k_+,\ii E)\right)}
\right]
\,\vol\,\qb(k_2)\maxwell(\vo')\maxwell(\vt')
\\
-
\frac{\ii\,\kJ^2\sigma^2\, \ko\dotp\kt}{k_2^2}
\,
\left[
\frac{i\maxwell(\vo')}{\Big(\kp\dotp\vo'-\ii E\Big)\,\left(\ko\dotp\vo'-\ii\,\eta(k_1)\right)}
+
\frac{i \kJ^2\,\kp\dotp\vo'\maxwell(\vo')\,Y(k_+,\ii E)}{\Big(\kp\dotp\vo'-\ii E\Big)\left(k_+^2-\kJ^2 P(k_+,\ii E)\right)\,\left(\ko\dotp\vo'-\ii\,\eta(k_1)\right)}
\right]
\,\Big(\vol\,\qb(k_2)+1\Big)\maxwell(\vt')
\\
+
\ (1)\leftrightarrow(2)
\Bigg\rbrace
+
\order(\vol^{-1})\,
,
\eml
where we handled the terms corresponding to velocity derivatives of $\btfo$ via integration by parts.  To help in doing the velocity integrals, write
\be \label{eq:double-Y}
Y(\k,\omega,\kp,\ii E)
\equiv
\int\!\frac{\maxwell(\v')\,\dd^3\v'}{\left(\k\dotp\v'-\omega\right)\,\left(\kp\dotp\v'-\ii E\right)}
\qquad\text{and}\qquad
P(\k,\omega,\kp,\ii E)
\equiv
\int\!\frac{\kp\dotp\v'\,\maxwell(\v')\,\dd^3\v'}{\left(\k\dotp\v'-\omega\right)\,\left(\kp\dotp\v'-\ii E\right)}
.
\ee
We also have a useful relation
\be
\int\!\frac{\dd^3\v'\,\kp\dotp\v'\,\maxwell(\v')}{\k\dotp\v'-\ii\,\eta(k)}
=
\int\!\frac{\left(\k_{+,\parallel\k}\dotp\v'+\k_{+,\perp\k}\dotp\v'\right)\,\maxwell(\v')}{\k\dotp\v'-\ii\,\eta(k)}\,\dd^3\v'
=
\frac{\k\dotp\kp\,P(k,\ii\eta(k))}{k^2}
.
\ee
Doing the velocity integrations from \eref{eq:ID1-1}, we get 
\bml \label{eq:ID1-2}
I_{D_1}
=
\frac{\kJ^2\,P(k_1,\ii\,\eta(k_1))}{k_1^2}
\left[
Y(\kt,\ii\,\eta(k_2),\kp,\ii E)
+
\frac{\kJ^2\,P(\kt,\ii\,\eta(k_2),\kp,\ii E)\,Y(k_+,\ii E)}{\left(k_+^2-\kJ^2 P(k_+,\ii E)\right)}
\right]
+
\frac{\kJ^2\, \ko\dotp\kp\,P(k_1,\ii\eta(k_1))\,Y(k_2,\ii\eta(k_2))\,Y(k_+,\ii E)}{k_1^2\,\left(k_+^2-\kJ^2 P(k_+,\ii E)\right)}
\,\vol\,\qb(k_2)
\\
+
\frac{\kJ^2\sigma^2\, \ko\dotp\kt\,Y(k_2,\ii\eta(k_2))}{k_2^2}
\,
\Bigg[
\pd{Y(\ko,\omega,\kp,\ii E)}{\omega}\Big|_{\omega=\ii\eta(k_1)}
+
\frac{\kJ^2\,Y(k_+,\ii E)}{\left(k_+^2-\kJ^2 P(k_+,\ii E)\right)}
\pd{P(\ko,\omega,\kp,\ii E)}{\omega}\Big|_{\omega=\ii\eta(k_1)}
\Bigg]
\,\Big(\vol\,\qb(k_2)+1\Big)
+
\ (1)\leftrightarrow(2)
.
\eml
We can also recall an expression for $\qb$ from  \eref{eq:q-ansatz-result-main}.\\

From \eref{eq:g1-fast-2}, we have 
\be \label{eq:g1-fast-3}
\bbg_{1,\text{a}}(1,2,t)
=
-
\frac{ \ko\dotp\vo\maxwell(\vo)\,\kt\dotp\vt\maxwell(\vt)}{\Big(\ko\dotp\vo-\ii\,\eta(k_1)\Big)\,\Big(\kt\dotp\vt-\ii\,\eta(k_2)\Big)}
\frac{\kJ^4\,\eta(k_1)\,\eta(k_2)\,\sigma^4\ I_{D_1}\,\ex^{\,\left[\eta(k_1)+\eta(k_2)\right]\,t}}{\left[\sigma^2\left(\kJ^2-k_1^2\right)-\eta(k_1)^2\right]\left[\sigma^2\left(\kJ^2-k_2^2\right)-\eta(k_2)^2\right]}
+
\order(\vol^{-1})\,
.
\ee
Using the dispersion relation, this implies that
\be \label{eq:g1-fast-4}
\gof(\ko,\vo,\kt,t)
=
-
\frac{ \ko\dotp\vo\maxwell(\vo)}{\Big(\ko\dotp\vo-\ii\,\eta(k_1)\Big)}
\frac{\kJ^2\,\eta(k_1)\,\eta(k_2)\,\sigma^4\,k_2^2\  I_{D_1}\,\ex^{\,\left[\eta(k_1)+\eta(k_2)\right]\,t}}{\left[\sigma^2\left(\kJ^2-k_1^2\right)-\eta(k_1)^2\right]\left[\sigma^2\left(\kJ^2-k_2^2\right)-\eta(k_2)^2\right]}
+
\order(\vol^{-1})
\,
.
\ee
The series expansion of $I_{D_1}$ for small $k_j$ is obtained from \eref{eq:ID1-2} via \emph{Mathematica} computer algebra in \citet{mycalcs}, making use of the dispersion relation, and $Z'(z)=-2-2zZ(z)$ from \citet[eq.~C.26]{BT}. Evaluating \eref{eq:g1-fast-4} in \citet{mycalcs}, we then find that, for small $k_j,$ we have 
\bml \label{eq:bigint-result}
\gof(\ko,\vo,\kt,t)
=
\Bigg\lbrace
\frac{\ii\kJ \ko\dotp\vo}{4k_1^2\sigma}
+
\frac{\left(\ko\dotp\vo\right)^2}{4k_1^2\sigma^2}
+
\frac{\ii\ko\dotp\vo}{24 \kJ\sigma k_1^2 k_+^2 }
\left[
-
30 k_+^2 \ko\dotp\kp 
+
12 k_+^4
+
31 k_1^2 k_+^2
+
8 \left(\ko\dotp\kp\right)^2
\right]
-
\frac{\ii\left(\ko\dotp\vo\right)^3}{4 \kJ\sigma^3 k_1^2 }
\\
\\
+
\frac{\left(\ko\dotp\vo\right)^2}{12 \kJ^2\sigma^2 k_1^2 k_+^2 }
\left[
4 \left(\ko\dotp\kp\right)^2
+
20 k_1^2 k_+^2
-
15 \ko\dotp\kp k_+^2
+
6 k_+^4
\right]
-
\frac{\left(\ko\dotp\vo\right)^4}{4 \kJ^2\sigma^4 k_1^2}
+
\order\left(k_j^3\right)
\Bigg\rbrace
\maxwell(\vo)\,
\ex^{\left[\eta(k_1)+\eta(k_2)\right]\,t}
,
\eml
where on the right-hand side we replaced $\kt$ by $\kp-\ko,$ which puts the result in the form most helpful for use in Appendix~\ref{sec:entropy-calculations}.

\section{The Landau approach to the first order correlation equation}
\label{sec:the-landau-approach-to-deriving-the-first-order-correlation-function}

To provide an alternative to, and to check, the propagator method set out in Appendix~\ref{sec:exploring-the-correlation-equation-using-a-propagator} for finding $\gof,$ \eref{eq:g1-eq-flt-main}, this appendix takes an approach based on that used in \cite{landau1946vibrations}, and Subsection~\ref{sec:the-boltzmann-perturbation-equation}, to derive \eref{eq:f1-model-8} for $\btfo.$ Along those lines, we solve \eref{eq:g1-eq-flt-main} by rearranging the equation and integrating it with respect to velocities.  It is lengthier than the derivation of $\btfo,$ requiring calculation of a number of integrals through solving a set of seven simultaneous equations.

We integrate \eref{eq:g1-eq-flt-main} with respect to both $\vo$ and $\vt$ to get
\be \label{eq:mom-0}
\omega\Ato{0}(\ko,\kt)
=
\Ato{1}(\ko,\kt)
+
\Ato{1}(\kt,\ko)
+
\muo{0}(\ko,\kt)
=
\Ato{1}(\ko,\kt)
+
\Ato{1}(\kt,\ko)
,
\ee
where we have written
\be
\Ato{j}(\ko,\kt)
\equiv
\int\!(\ko\dotp\vo)^j\,\ato(\ko,\vo,\kt)\,\dd^3\vo
,
\qquad\qquad
\text{and}
\qquad\qquad
\muo{0}(\ko,\kt)
\equiv
\int\!\ii\,\tilde{D}_1(1,2)\,\dd^3\vo\,\dd^3\vt
=
0
\,,
\ee
where $\tilde{D}_1(1,2)$ is the Laplace transform of the driving term as set out in \eref{eq:D1}. 

Now divide \eref{eq:g1-eq-flt-main} by $1-\frac{\ko\dotp\vo+\kt\dotp\vt}{\omega}$ and integrate with respect to $\vt,$ to get 
\bml \label{eq:g1-eq-div-i}
\omega\,\ato(1,\kt)
=
-
\frac{\kJ^2\, \ko\dotp\vo}{k_1^2}\maxwell(\vo)
\left[
\left(
1
+
\frac{\ko\dotp\vo}{\omega}
+
\frac{(\ko\dotp\vo)^2}{\omega^2}
+
\frac{(\ko\dotp\vo)^3}{\omega^3}
+
\frac{(\ko\dotp\vo)^4}{\omega^4}
\right)
\Ato{0}(\kt,\ko)
\right.
\\
\left.
+
\left(
\frac{1}{\omega}
+
\frac{2\ko\dotp\vo}{\omega^2}
+
\frac{3(\ko\dotp\vo)^2}{\omega^3}
+
\frac{4(\ko\dotp\vo)^3}{\omega^4}
+
\frac{5(\ko\dotp\vo)^4}{\omega^5}
\right)
\Ato{1}(\kt,\ko)
\right.
\\
\left.
+
\left(
\frac{1}{\omega^2}
+
\frac{3\ko\dotp\vo}{\omega^3}
+
\frac{6(\ko\dotp\vo)^2}{\omega^4}
+
\frac{10(\ko\dotp\vo)^3}{\omega^5}
+
\frac{15(\ko\dotp\vo)^4}{\omega^6}
\right)
\Ato{2}(\kt,\ko)
\right.
\\
\left.
+
\left(
\frac{1}{\omega^3}
+
\frac{4\ko\dotp\vo}{\omega^4}
+
\frac{10(\ko\dotp\vo)^2}{\omega^5}
+
\frac{20(\ko\dotp\vo)^3}{\omega^6}
+
\frac{35(\ko\dotp\vo)^4}{\omega^7}
\right)
\Ato{3}(\kt,\ko)
\right]
\\
-
\frac{\kJ^2\sigma^2}{\omega}
\ato(1,\kt)
\left\lbrace
1
+
\frac{2\ko\dotp\vo}{\omega}
+
\frac{3\left[(\ko\dotp\vo)^2+k_2^2\sigma^2\right]}{\omega^2}
+
\frac{4\left[(\ko\dotp\vo)^3+3k_2^2\sigma^2\,\ko\dotp\vo\right]}{\omega^3}
+
\frac{15k_2^4\sigma^4+30k_2^2\sigma^2\,(\ko\dotp\vo)^2+5\,(\ko\dotp\vo)^4}{\omega^4}
\right\rbrace
\\
+
\int\! \ii\,\frac{\tilde{D}_1(1,2)}{\left(1-\frac{\ko\dotp\vo+\kt\dotp\vt}{\omega}\right)}
\,\dd^3\vt
+
\order\left(k_j^4\right)
.
\eml
The leading order of the integral involving the driving term in $k_j$ is $-1,$ which implies this is also the leading order of $\ato(1,\kt).$  As \eref{eq:g1-eq-div-i} suggests, consider $\ato(1,\kt)$ as a power series in $\ko\dotp\vo,$ times $\maxwell(\vo).$  From this power series, and the property that integration of powers of $\ko\dotp\vo$ times the Maxwellian vanishes for odd powers, it can be seen that $\Ato{0}$ and $\Ato{1}$ are both of order zero in $k_j,$ while $\Ato{2}$ and $\Ato{3}$ are both of order two in $k_j$ and $\Ato{4}$ and $\Ato{5}$ are both of order four.

Write
\be
\nuo{j}(\ko,\kt)
\equiv
\int\!\ii\,\left(\ko\dotp\vo\right)^j\,\tilde{D}_1(1,2)\,\dd^3\vo\,\dd^3\vt
.
\ee
Multiplying \eref{eq:g1-eq-div-i} by $\ko\dotp\vo$ and integrating with respect to $\vo,$ we have
\bml \label{eq:mom-1}
\omega\,\Ato{1}(\ko,\kt)
=
-
\kJ^2\sigma^2
\left[
\left(
1
+
\frac{3k_1^2\sigma^2}{\omega^2}
\right)
\Ato{0}(\kt,\ko)
+
\left(
\frac{1}{\omega}
+
\frac{9k_1^2\sigma^2}{\omega^3}
\right)
\Ato{1}(\kt,\ko)
+
\frac{1}{\omega^2}
\Ato{2}(\kt,\ko)
+
\frac{1}{\omega^3}
\Ato{3}(\kt,\ko)
\right]
\\
-
\frac{\kJ^2\sigma^2}{\omega}
\left[
\Ato{1}(\ko,\kt)
+
\frac{2\Ato{2}(\ko,\kt)}{\omega}
+
\frac{3\left(k_2^2\sigma^2\Ato{1}(\ko,\kt)+\Ato{3}(\ko,\kt)\right)}{\omega^2}
\right]
+
\nuo{1}(\ko,\kt)
+
\order\left(k_j^4\right)
.
\eml
Now multiplying \eref{eq:g1-eq-div-i} by $(\ko\dotp\vo)^2$ and integrating with respect to $\vo,$ we get
\bml \label{eq:mom-2}
\omega\,\Ato{2}(\ko,\kt)
=
-
\kJ^2\sigma^2
\left[
\frac{3k_1^2\sigma^2}{\omega}
\Ato{0}(\kt,\ko)
+
\frac{6k_1^2\sigma^2}{\omega^2}
\Ato{1}(\kt,\ko)
\right]
\\
-
\frac{\kJ^2\sigma^2}{\omega}
\left[
\Ato{2}(\ko,\kt)
+
\frac{2}{\omega}
\Ato{3}(\ko,\kt)
\right]
+
\nuo{2}(\ko,\kt)
+
\order\left(k_j^4\right)
.
\eml
Multiplying \eref{eq:g1-eq-div-i} by $(\ko\dotp\vo)^3$ and integrating with respect to $\vo,$ we get
\be \label{eq:mom-3}
\omega\,\Ato{3}(\ko,\kt)
=
-
\kJ^2\sigma^2
\left[
3k_1^2\sigma^2
\Ato{0}(\kt,\ko)
+
\frac{3k_1^2\sigma^2}{\omega}
\Ato{1}(\kt,\ko)
\right]
-
\frac{\kJ^2\sigma^2}{\omega}
\Ato{3}(\ko,\kt)
+
\nuo{3}(\ko,\kt)
+
\order\left(k_j^4\right)
.
\ee

We now solve simultaneously \eref{eq:mom-0}, Eqs.~\eqref{eq:mom-1}-\eqref{eq:mom-3}, and Eqs.~\eqref{eq:mom-1}-\eqref{eq:mom-3} with $\ko$ and $\kt$ swapped, making seven equations in total. We write
\be
\omega\,\boldsymbol{\Gamma}
=
\mathbfss{M}\,\boldsymbol{\Gamma}
+
\boldsymbol{\nu}
+
\order\left(k_j^4\right)
,
\ee
where
\be
\boldsymbol{\Gamma}
=
\begin{pmatrix}
	\Ato{0}(\ko,\kt) & \Ato{1}(\ko,\kt) & \Ato{1}(\kt,\ko) & \Ato{2}(\ko,\kt) & \Ato{2}(\kt,\ko) & \Ato{3}(\ko,\kt) & \Ato{3}(\kt,\ko) 
\end{pmatrix}^{\intercal}
,
\ee
\be
\boldsymbol{\nu}
=
\begin{pmatrix}
	\muo{0}(\ko,\kt) & \nuo{1}(\ko,\kt) & \nuo{1}(\kt,\ko) & \nuo{2}(\ko,\kt) & \nuo{2}(\kt,\ko) & \nuo{3}(\ko,\kt) & \Ato{3}(\kt,\ko) 
\end{pmatrix}^{\intercal}
,
\ee
noting that the first entry in $\boldsymbol{\nu}$ is $\muo{0}$ rather than $\nuo{0},$ and $\mathbfss{M}$ is then defined via terms on the right-hand side of Eqs.~\eqref{eq:mom-0} and~\eqref{eq:mom-1}-\eqref{eq:mom-3}.

Write $\mathbfss{L}=\omega\,\mathbfss{1}-\mathbfss{M},$ where $\mathbfss{1}$ is the $7\times 7$ identity matrix, and we then have
\be\label{eq:bold-Gamma}
\mathbfss{L}\,\boldsymbol{\Gamma}
=
\boldsymbol{\nu}
+
\order\left(k_j^4\right)
,
\qquad\qquad\text{implying}\qquad\qquad
\boldsymbol{\Gamma}
=
\mathbfss{L}^{-1}\,\boldsymbol{\nu}
+
\order\left(k_j^4\right)
,
\ee
giving us, in particular, $\Ato{s}(\kt,\ko)$ for $s=0,..3.$ In \citet{mycalcs}, \emph{Mathematica} is used to make find $\boldsymbol{\Gamma}$ from \eref{eq:bold-Gamma}, and to substitute its components back into \eref{eq:g1-eq-div-i}.  The residue at $\omega=\ii\,\eta(k_1)+\ii\,\eta(k_2),$ is calculated, noting that the driving function $\tilde{D}_1(1,2)$ has residue zero at that value of $\omega.$ We now have a closed equation for $\gof(1,\kt,t),$ explicitly to second order in $k_j,$ which, after some manipulation in \citet{mycalcs}, gives the same result as obtained in Appendix~\ref{sec:exploring-the-correlation-equation-using-a-propagator}'s \eref{eq:bigint-result}.

\section{Entropy creation rate calculations}
\label{sec:entropy-calculations}

The first two parts of this appendix cover detailed calculations needed for, respectively, Subsection~\ref{sec:the-total-entropy-creation} and Subsection~\ref{sec:the-varying-entropy-change-by-radius} in the main text.  The third part checks consistency between total and local entropy creation calculations. The fourth part considers variants to the paper's main model, as mentioned in Subsection~\ref{sec:further-avenues-for-exploration-including-of-alternative-systems}.

\subsection{Calculating the total entropy creation rate}
\label{sec:all-space}

Recall \eref{eq:Scg-defn} for the asymptotic coarse-grained entropy creation rate. An expression for $\gof(-\kp,\vo,\kt)$ can be obtained from \eref{eq:bigint-result} as 
\bml \label{eq:bigint-result-swap-app}
\gof(-\kp,\vo,\kt,t)
=
\Bigg\lbrace
-\frac{\ii\kJ \kp\dotp\vo}{4k_+^2\sigma}
+
\frac{\left(\kp\dotp\vo\right)^2}{4k_+^2\sigma^2}
+
\frac{\ii\left(\kp\dotp\vo\right)^3}{4 \kJ\sigma^3  k_+^2}
-
\frac{\ii\,\kp\dotp\vo}{24 \kJ\sigma k_1^2 k_+^2}
\left[
-
30 k_1^2 \ko\dotp\kp 
+
12 k_1^4
+
31 k_1^2 k_+^2
+
8 \left(\ko\dotp\kp\right)^2
\right]
\\
-
\frac{\left(\kp\dotp\vo\right)^4}{4\kJ^2\sigma^4 k_+^2 }
+
\frac{\left(\kp\dotp\vo\right)^2}{12\kJ^2\sigma^2 k_1^2 k_+^2}
\left[
4 \left(\ko\dotp\kp\right)^2
+
20 k_1^2 k_+^2
-
15 \ko\dotp\kp k_1^2
+
6 k_1^4
\right]
+
\order\left(k_j^3\right)
\Bigg\rbrace
\maxwell(\vo)\,
\ex^{\left[\eta(k_+)+\eta(k_2)\right]\,t}
,
\eml
by mapping $(\ko,\kt)\mapsto(-\kp,\kt),$ which also induces the mapping $\kp\mapsto-\ko.$ From Eqs.~\eqref{eq:factor-we-need} and~\eqref{eq:disp-soln-small}, the factor involving a velocity derivative is 
\bml \label{eq:vel-div-deriv-app}
\pd{\left({\bfof(\ko,\vo)}/{f_0(\vo)}\right)}{\vo}
\\
=
\left[
\frac{\ii \kJ\,\ko}{2\sigma\,k_1^2}
+
\frac{\ko\dotp\vo\,\ko}{\sigma^2\,k_1^2}
-
\frac{\ii\,\left(6\left(\ko\dotp\vo\right)^2-7\sigma^2 k_1^2\right)\,\ko}{4\kJ\sigma^3 k_1^2}
+
\frac{\ko\dotp\vo\left(5\sigma^2k_1^2-2\left(\ko\dotp\vo\right)^2\right)\,\ko}{\kJ^2\sigma^4\,k_1^2}
+
\order\left(k_1^3\right)
\right] 
\,\vol\, \ex^{\eta(k_1)\,t}
.
\eml

The $\vo$ integral in \eref{eq:Scg-defn} is evaluated in \citet{mycalcs}, effectively using identities for a multivariate normal distribution, giving
\bml\label{eq:scg-calc-1}
\d{\,\Scg}{t}
=
-
4\pi G mN\vol \ii \epsilon^2
\int_{\mathcal{K}({\beta\kJ})}
\!\frac{\dd^3\ko}{(2\pi)^3}\,\frac{\dd^3\kt}{(2\pi)^3}
\ \ 
\Bigg\lbrace
-
\frac{\ii\kJ\ko\dotp\kp\,\ko\dotp\kt}{4\sigma k_1^2 k_2^2 k_+^2}
+
\frac{\ii \kJ\,\ko\dotp\kt}{8\sigma\,k_1^2 k_2^2}
\\
-
\frac{\ii\,\ko\dotp\kt}{48 \kJ\sigma k_1^4 k_2^2 k_+^2}
\bigg[
33 k_1^6 - 8 \left(\ko\dotp\kp\right)^2 k_2^2
+
k_1^4 \left(-84 \ko\dotp\kp - 51 k_2^2 + 40 k_+^2\right)
\\
+
2 k_1^2 \left(22 \left(\ko\dotp\kp\right)^2 + 33 \ko\dotp\kp\, k_2^2 + 9 k_2^4 - 9 \ko\dotp\kp\, k_+^2 - 20 k_2^2 k_+^2\right)
\bigg]
+
\order\left(k_j\right)
\Bigg\rbrace
\,
\ex^{\left[\eta(k_1)+\eta(k_+)+\eta(k_2)\right]\,t}
.
\eml
The terms in the braces on the first line of \eref{eq:scg-calc-1}'s final expression are of order $-2$ in $k_j,$ while the other terms in the braces are of order $0.$  Taking account of the $\dd^3\ko\,\dd^3\kt$ factors, the overall integral is therefore of leading order $4$ in $k_j.$ In principle, our approach of not accounting for entropy exterior to the volume $\vol$ means that we should set a lower limit for $k_j$ in the integrals, as well as the upper limit implied by the region $\mathcal{K}({\beta\kJ}).$  However, because the integral is order four in $k_j$ and $R^{-1}\ll\kJ\beta,$ we can safely omit this lower limit with negligible effect on the final answer.

The order $-2$ integrand gives
\be \label{eq:order-minus2}
\int_{\mathcal{K}({\beta\kJ})}
\!\frac{\dd^3\ko}{(2\pi)^3}\,\frac{\dd^3\kt}{(2\pi)^3}
\left[
-
\frac{\ii\kJ\ko\dotp\kp\,\ko\dotp\kt}{4\sigma k_1^2 k_2^2 k_+^2}
+
\frac{\ii \kJ\,\ko\dotp\kt}{8\sigma\,k_1^2 k_2^2}
\right]
=
\frac{\ii \kJ}{8\sigma}\,\int_{\mathcal{K}({\beta\kJ})}
\!\frac{\dd^3\ko}{(2\pi)^3}\,\frac{\dd^3\kt}{(2\pi)^3}\ 
\frac{\ko\dotp\kt\,\left(k_2^2-k_1^2\right)}{k_1^2 k_2^2 k_+^2}
=
0
,
\ee
where the final result of $0$ follows by swapping the variables $\ko$ and $\kt$ in the integral, noting that this preserves $\mathcal{K}({\beta\kJ}).$

From \eref{eq:scg-calc-1}, we can therefore write 
\bml \label{eq:scg-calc-2}
\d{\,\Scg}{t}
=
\frac{4\pi G mN\vol\kJ^5 \beta^6 \epsilon^2}{\kJ\sigma^2}
\frac{4\pi\,2\pi}{(2\pi)^3\,(2\pi)^3}
\int_{\mathcal{K}({1})}
\!\dd k_1\,k_1^2\,\dd k_2\, k_2^2
\int_{-1}^{1}d\left(\cos\left(\theta\right)\right)
\chi\big[\left(\ko,\kt\right)\in \mathcal{K}({1})\big]
\\
\Bigg\lbrace
-
\frac{\ii\,\ko\dotp\kt}{48  k_1^4 k_2^2 k_+^2}
\bigg[
33 k_1^6 - 8 \left(\ko\dotp\kp\right)^2 k_2^2
+
k_1^4 \left(-84 \ko\dotp\kp - 51 k_2^2 + 40 k_+^2\right)
\\
+
2 k_1^2 \left(22 \left(\ko\dotp\kp\right)^2 + 33 \ko\dotp\kp\, k_2^2 + 9 k_2^4 - 9 \ko\dotp\kp\, k_+^2 - 20 k_2^2 k_+^2\right)
\bigg]
\Bigg\rbrace
\,
\ex^{\left[\eta(k_1)+\eta(k_+)+\eta(k_2)\right]\,t}
+
\order\left(\beta^7\right)
,
\eml
where we took advantage of the integrand being homogeneously of order $0$ in $k_j,$ and $\k_j$ was scaled by $\kJ\beta$ to become dimensionless.  The variable $\theta$ is the angle $\kt$ makes with $\ko$: the factor associated with $k_2^2$ is then $2\pi$ rather than $4\pi.$ Note that we can write the pre-factors before the integral in \eref{eq:scg-calc-2} as
\be \label{eq:prefactor2}
\frac{4\pi G mN\vol\kJ^5 \beta^6 \epsilon^2}{\sigma}
\frac{4\pi\,2\pi}{(2\pi)^3\,(2\pi)^3}
=
\frac{\kJ^7\sigma\beta^6\,\vol^2\epsilon^2}{8\pi^4\sigma}
=
\frac{2\kJ\sigma \,N_1^2}{9\pi^2 n^2\,B^2}
\,,
\ee
where $N_1$ and $B$ were defined after \eref{eq:dScg-dt}.

Assuming that $t$ is sufficiently small that we are willing to make the approximation $\left[\eta(k_1)+\eta(k_+)+\eta(k_2)\right] t\approx 3\kJ\sigma\, t,$ the exponential can be factored out from the integral.  The resulting integral is evaluated numerically in \citet{mycalcs}, getting a result of $\num{-0.0115541}$, with an estimated error of $\num{9.94634e-6}.$  The overall result for $\ind{\,\Scg}{t}$ is shown in \eref{eq:dScg-dt}.

\begin{table}
	\caption{Errors in the approximation $\eta(k)=\kJ\sigma,$ as calculated in \citet{mycalcs} for a range of values of  $k/\kJ.$ To two significant figures, the error is as predicted by the $-3\sigma k^2/(2\kJ)$ correction from \eref{eq:disp-soln-small}.  The results make clear that, for any $k\ll\kJ,$ the approximation $\eta(k)t=\kJ\sigma t$ is very good except for extremely large time-scales $t\kJ\sigma.$} 
	\begin{tabular}{r|llll}
		\hline\Tstrut\Bstrut
		$k/\kJ$ & \mc{$10^{-1}$} & \mc{$10^{-2}$} & \mc{$10^{-3}$} & \mc{$10^{-4}$} \\\hline\Tstrut
		Approximation error ${\left[\kJ\sigma-\eta(k)\right]}/{\kJ\sigma}$ & $\num{1.5e-2}$ & $\num{1.5e-4}$ & $\num{1.5e-6}$ & $\num{1.5e-8}$ \\\Tstrut
		Time-scale $t\kJ\sigma$ before significant approximation errors & $\num{6.7e1}$ & $\num{6.7e3}$ & $\num{6.7e5}$ & $\num{6.7e7}$ \Bstrut\\
		\hline
	\end{tabular}
	\label{table:dispersionerrors}
\end{table}

Note that the approximation $\left[\eta(k_1)+\eta(k_+)+\eta(k_2)\right] t\approx 3\kJ\sigma\, t,$ no longer holding at very late times is not a fundamental difficulty~-- it would be straightforward to calculate a time-dependent entropy creation formula which explicitly accounts for accurate values of $\eta(k_j)\,t$ in the numerical integration. It is also straightforward to estimate the maximum error from the approximation $\eta(k_j)\,t=\kJ\sigma\,t,$ using the values of $\eta(\kJ\beta)$ as calculated in \citet{mycalcs} for drawing Figure~\ref{fig:figdispersionrelation}.	Table~\ref{table:dispersionerrors} confirms that, for values of $k\ll\kJ,$ the approximation $\eta(k)\,t=\kJ\sigma\,t$ is very accurate until extremely late times $t\gg \infrac{1}{\kJ\sigma},$ by which time, following \eref{eq:perturb-valid}, $\epsilon$ will need to have been very small indeed for the perturbative regime to remain valid.  \\

As mentioned at the end of Subsection~\ref{sec:the-varying-entropy-change-by-radius}, we can try to alter the definition of asymptotic coarse-grained entropy, by relaxing the constraints that define the region $\mathcal{K}({\beta\kJ}).$   That region sets three constraints, which may be summarised as $0<k_1,k_2,k_+<\kJ\beta.$ In \citet{mycalcs} are calculations  analogous to those in this part of Appendix~\ref{sec:entropy-calculations}, but for the cases where only two of the three upper constraints have effect.  Because, for example, $k_+\le k_1+k_2,$ all three wave-numbers will still be small, allowing our analytical approximations.  The results are as for \eref{eq:dScg-dt}, but \opb{}with $\num{-0.0116}$ replaced by $\num{-1.54e-3}$ if we only have the constraints $0<k_1,k_2<\kJ\beta,$ or by $\num{-0.0240}\beta^{-2}$ if we only have the constraints $0<k_1,k_+<\kJ\beta,$ or by $\num{0.0239}\beta^{-2}$ (that is, positive entropy creation)\clb{} if we only have $0<k_2,k_+<\kJ\beta.$ (Estimated errors in numerical integration are around $\num{1e-5},\num{1e-4}\beta^{-2},$ and $\num{1e-5}\beta^{-2},$ respectively.) The factors of $\beta^{-2}$ arise when the order $-2$ integrand of \eref{eq:order-minus2} no longer vanishes on integration, because the integration limits are no longer symmetric in $k_1$ and $k_2,$ corresponding to having differing coarse-graining approaches for those two wave-numbers.

\opb{}The wave-number $k_1$ corresponds to the entropy creation's physical location. The choice above which leads to positive total net entropy creation, of directly constraining only $0<k_2,k_+<\kJ\beta,$ leads to only an indirect constraint on $k_1$ from $\ko=\kp-\kt$ and is therefore physically a particularly contrived form of coarse-graining with respect to $x_1.$\clb{}

We prefer applying all three upper constraints, as better reflecting coarse-graining that might be done for practical observational or simulation reasons, or to study a particular scale.  The symmetry in treatment of $k_1,k_2$ and $k_+$ also better reflects asymptotic behaviour over time, as seen, for example, in the exponential factor \opb{}of \eref{eq:scg-calc-1}.  Other choices of region $\mathcal{K},$ including the approaches above which directly constrain only two of the three wave-numbers, are therefore inappropriate for defining asymptotic entropy creation.\clb{} We also note another significant advantage \opb{}of our choice of $\mathcal{K}$ at\clb{} the end of Appendix~\ref{sec:local}.

\opb{}Another approach chooses a coarse-graining which matches the asymptotic behaviour yet move closely: for small $k_j$ it captures essentially all wave-numbers which asymptote faster than a given rate. The asymptotic rate is given by the $\eta(k_1)+\eta(k_2)+\eta(k_+)$ factor in \eref{eq:scg-calc-1}'s exponential. From \eref{eq:disp-soln-small} this is approximated to second order in $k_j$ by $3\kJ\sigma-{3 \sigma (k_1^2+k_2^2+k_+^2)}/{2 \kJ},$ and, as shown in Table~\ref{table:dispersionerrors}, this approximation is very good for all $k_j\le 0.1\kJ.$  We choose our coarse-graining to be defined by $0<k_1^2+k_2^2+k_+^2<2\kJ^2\beta^2,$ with the choice of $2$ as a factor for the constraint being somewhat arbitrary, and being made as it gives a similar overall scale to our original constraint $0<k_1,k_2,k_+<\kJ\beta\,,$ for example capturing a similar volume of $k_j$ space \citep{mycalcs}. The part of the total net entropy associated with the order $-2$ integrand of \eref{eq:order-minus2} once again vanishes, because the constraint is symmetrical in $\ko$ and $\kt.$ As calculated in \citet{mycalcs}, this coarse-graining $0<k_1^2+k_2^2+k_+^2<2\kJ^2\beta^2$ produces a total net entropy creation rate as in \eref{eq:dScg-dt}, but with $\num{-0.0116}$ replaced by $\num{-0.0125},$ the latter being subject to an estimated integration error of around $\num{1e-5}.$\clb{}

\subsection{Calculating the distribution of the entropy creation rate over space}
\label{sec:local}

The distribution of the entropy creation rate over space is given by \eref{eq:Scgrr}.  We now evaluate that equation.  We calculated the velocity integral in \eref{eq:scg-calc-1}, expanding  explicitly to order $0$ in $k_j.$ Following that calculation, although there was also an order $-2$ term, \eref{eq:order-minus2} showed that\opb{} it\clb{} vanished on integration by $\ko$ and $\kt.$  However, this need not now be the case for \eref{eq:Scgrr}, because of the new role of $\kz$ and the $\sinc$ factor disrupting the symmetry which ensured this term vanished in the previous subsection.  We therefore look at the product of \eref{eq:bigint-result-swap-app} and \eref{eq:vel-div-deriv-app}, in the latter substituting $\ko\mapsto\kzo,$ to get get the integrand corresponding to that of \eref{eq:scg-calc-1}.  Doing the velocity integral only to order $-2$ in $k_j,$ we find
\be \label{eq:Scgrr-2}
\pd{^2\Scgr(r)}{t\,\partial r}
=
\frac{\kJ^8\sigma\vol^2  \epsilon^2\,\ex^{3\kJ\sigma t}}{8\pi^4}\,\beta^5
\int_{\mathcal{K}'(1)}\!\frac{\dd^3\kz\,\dd^3\ko\,\dd^3\kt}{8\pi^2}\ 
\frac{\left(-\ii\right)\kJ\beta r\,\sin(k_0\, \kJ\beta r)}{2\pi^2  k_0}\,
\left\lbrace
-
\frac{\ii\kJ\kzo\dotp\kot\,\kzo\dotp\kt}{4\sigma k_{01}^2 k_2^2 k_{12}^2}
+
\frac{\ii \kJ\,\kzo\dotp\kt}{8\sigma\,k_{01}^2 k_2^2}
\right\rbrace
+
\order\left(\beta^6\right)
.
\ee
We chose to pull $(8\pi^4)^{-1}$ out of the integral in order to get our first factor of the same form as the corresponding term in \eref{eq:dScg-dt}, choosing the sign to make positive (resp. negative) values of our integral correspond to entropy creation (resp. destruction).  We also kept one factor of $\beta$ inside the integral, to ensure the integral is a dimensionless function $\Shcgr$ of the dimensionless quantity $\kJ\beta r,$ and of order zero in $\beta$ (after the integration).  The numerically-calculated function $\Shcgr$ will be called the \emph{(leading order) entropy-creation pattern function}, and is evaluated in \citet{mycalcs} for varying values of $\kJ\beta r.$ The overall result is shown in \eref{eq:Scgrr-3} and Figure~\ref{fig:figscgr}. If we integrate $\inpd{^2\Scgr(r)}{t\,\partial r}$ with respect to $r$ to get entropy creation in a generic region,  we get a result of order $\beta^4,$ compared with order $\beta^6$ for the total net entropy creation over all space.  In other words, the total net entropy creation is suppressed by a factor of $\beta^2\ll 1$ compared with local entropy creation in a generic region.

Calculations in \citet{mycalcs} confirm that the absolute values of $\Shcgr$ integrated over either the core or halo are of very similar size, $\num{0.0110},$ with the core and halo values agreeing to around $0.01\%.$  This similarity is to be expected, as we know from \eref{eq:order-minus2} that the total net entropy creation over both must essentially vanish.  Note that $\num{0.0110}$ is very close in absolute value to $\num{-0.0116},$ which is the corresponding factor for the total net entropy creation over all space, reinforcing the point made in the previous paragraph that the core-halo pattern is much more prominent than the total net entropy creation. \\

We now look again at the alternative definitions of asymptotic coarse-grained entropy creation which were considered at the end of Appendix~\ref{sec:all-space}.  Comparing Eqs.~\eqref{eq:Scg-defn} and~\eqref{eq:Scgrr}, the three constraints which now apply to our basic definition are $0<k_{01},k_2,k_{12}<\kJ\beta.$  However, if we now relax any one of the upper constraints, then at least one of $k_{01},k_2,k_{12}$ will be unconstrained above.  For example, if we relax the constraint that $k_{01}<\kJ\beta,$ then there is nothing to prevent $k_{01}$ taking any large value, in this case because $0<k_2,k_{12}<\kJ\beta$ provides no constraint on $k_0.$  This means that we cannot apply the approach of this paper, which is based on analytically-tractable series for $\gof(-\kot,\vo,\kt)$ and $\bfof(\kzo,\vo),$ derived from  expansion in their $\k_j$ arguments for small wave-number.  This is the further reason, referred to at the end of Appendix~\ref{sec:all-space}, for preferring our definition of asymptotic coarse-grained entropy creation.

\begin{figure*} 
	\centering
	\includegraphics[width=0.9\textwidth]{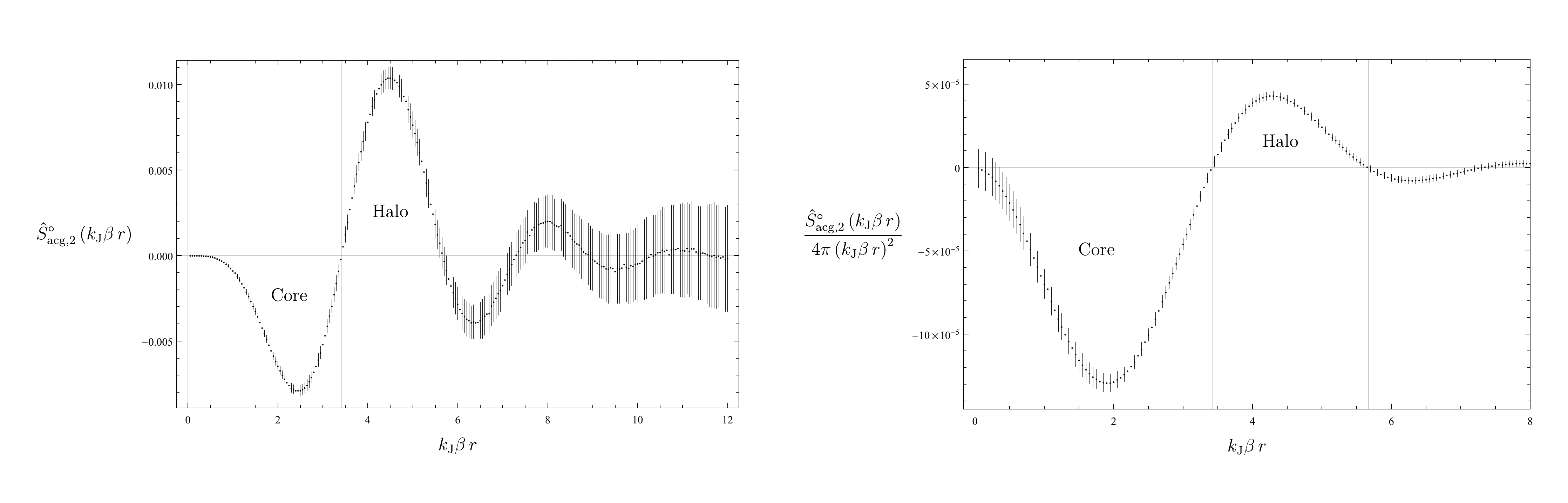}
	\caption{\opb{}\emph{Left:} The entropy-creation pattern function $\Scgrt$ which corresponds to the coarse-graining $0<k_1^2+k_2^2+k_+^2<2\kJ^2\beta^2,$ as calculated numerically.  The error bars show the error estimates for the numerical integration.  \emph{Right:} The volume density implied by the entropy-creation pattern function $\Scgrt.$  The error bars show the error estimates for the numerical integration, also scaled by $[{4\pi(\kJ\beta\,r)^2}]^{-1}.$ \emph{Both:} Note that these plots' horizontal scales are different to those of Figure~\ref{fig:figscgr}, to avoid showing  long tails of increasing error bars.\clb{}}
	\label{fig:figsquares}
\end{figure*}

\opb{}The coarse-graining $0<k_1^2+k_2^2+k_+^2<2\kJ^2\beta^2$ (which corresponds to $0<k_{01}^2+k_2^2+k_{12}^2<2\kJ^2\beta^2$ in \eref{eq:Scgrr-2}'s spatial distribution equation) also gives us a tractable analytical treatment for small wave-numbers.  It yields a spatial distribution as in \eref{eq:Scgrr-3}, but with the resulting entropy pattern function, calculated in \citet{mycalcs}, being $\Scgrt$ of Figure~\ref{fig:figsquares}.  Another ``taxicab'' coarse-graining, which constrains $k_1+k_2+k_+,$ is additionally explored in \citet{mycalcs}. This again gives negative total net entropy creation, at next-to-leading order, and a leading order pattern of an entropy-destroying core, and an entropy-creating halo. Obscured by estimated integration error, there appear to be smaller-amplitude outer shells of entropy
destruction and creation beyond the halo.\clb{}

\subsection{Checking consistency between the total entropy creation and its distribution}
\label{sec:checking-consistency-between-the-total-entropy-creation-and-its-distribution}

\begin{figure*}  
	\centering
		\includegraphics[width=0.9\textwidth]{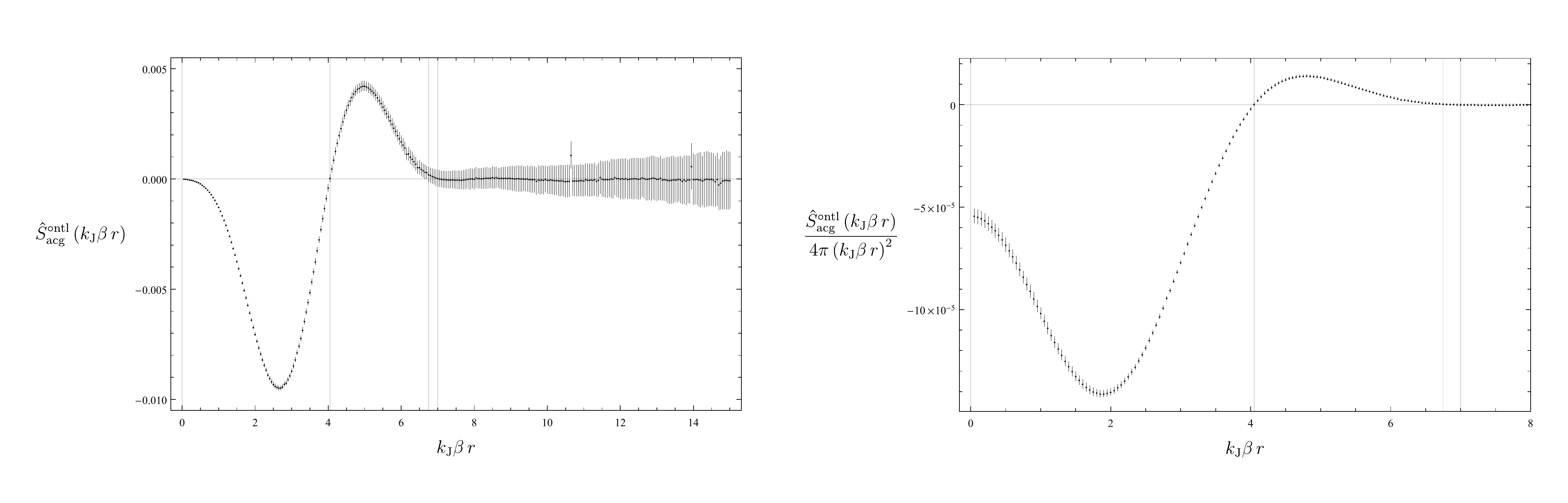}
	\caption{The next-to-leading order equivalent of the leading order  Figure~\ref{fig:figscgr}. The contribution to entropy creation at next-to-leading order is suppressed by a factor of $\beta^2$ relative to the leading order.}
	\label{fig:figscgr20thszerothorder}
\end{figure*}

We can also follow a similar route to that of Appendix~\ref{sec:local} to calculate the next-to-leading order term for the distribution of entropy creation.  This follows from taking the order $0$ terms of \eref{eq:scg-calc-1}, instead of the order $-2$ terms we considered above. It can be found that, at this next-to-leading order, we have
\be \label{eq:Scgrr-3-ntl}
\pd{^2\Scgrntl(r)}{t\,\partial r}
=
\frac{2\kJ^2\,\sigma\,\beta\,N_1^2\ \ex^{3\kJ\sigma t}}{9\pi^2\,n^2\,B^2}
\,\Shcgrntl(\kJ\beta r)
\,,
\ee
where $\Shcgrntl$ is a dimensionless \emph{next-to-leading order entropy-creation pattern function}, which is shown in Figure~\ref{fig:figscgr20thszerothorder} as numerically calculated in \citet{mycalcs}.  Note that, when integrated over a generic region, $\inpd{^2\Scgrntl(r)}{t\,\partial r}$ is of the same order in $\beta$ as the total net entropy creation over all space, and suppressed by a factor of $\beta^2$ compared with its leading order equivalent.

From \eref{eq:dScg-dt}, integrating \eref{eq:Scgrr-3-ntl}'s shell density over all radii should give us
\be \label{eq:constants}
\int_0^\infty\Scgrntl(r')\,\dd r'=-0.0116
\ee
in order to ensure consistency between those equations.  The size of the error bars in Figure~\ref{fig:figscgr20thszerothorder}~(Left) might give us some pause about using our numerical calculations to do the integration in \eref{eq:constants}. None the less, as calculated in \citet{mycalcs}, approximating the integral by summing and appropriately scaling the values shown in Figure~\ref{fig:figscgr20thszerothorder}~(Left) gives a result of $-0.0115,$ surprisingly close to $-0.0116.$  By doing this for a range of radii including only the central sphere of destruction and the surrounding shell of creation we also get a result of $-0.0115,$ suggesting that there is essentially no net entropy destruction outside the sphere and that innermost shell. 

\subsection{Modifications to the main model}
\label{sec:modifications-to-the-main-model}

A simple variant of our model is to allow the initial perturbation $f_{1,\text{init}}$ to have a different Maxwellian parameter $\sigma_1$ from the parameter $\sigma$ of the underlying perturbation $f_0.$  Following the approach of Subsection~\ref{sec:the-boltzmann-perturbation-equation}, we find, compare with \eref{eq:f1-model-8}, 
\be \label{eq:f1-model-8-var}
\btfo(\ko,\vo)
=
-
\frac{\ii\maxwell_1(\vo)}{\Big(\ko\dotp\vo-\omega\Big)}
-
\frac{\ii \kJ^2\,\ko\dotp\vo\maxwell(\vo)\,Y_1(k_1,\omega)}{\Big(\ko\dotp\vo-\omega\Big)\left(k_1^2-\kJ^2 P(k_1,\omega)\right)}
,
\ee
where the $\maxwell_1$ and $Y_1$ (only) are defined using $\sigma_1$ rather than $\sigma.$ Note that the residue with largest positive imaginary part still comes from the dispersion relation $k_1^2-\kJ^2 P(k_1,\omega)=0,$ dependent on $\sigma.$  Hence, we have, compare with \eref{eq:f1-fast-early}, 
\be \label{eq:f1-fast-early-var}
\bfof(\ko,\vo,t)
=
-\frac{\ii\,\kJ \sigma^2\,\ko\dotp\vo\maxwell(\vo)\,Y_1(k_1,\ii\,\eta(k_1))\ \ex^{\eta(k_1)\,t}}{\Big(\ko\dotp\vo-\ii\,\eta(k_1)\Big)\left(\sigma^2\left(\kJ^2-k_1^2\right)-\eta(k_1)^2\right)}
,
\ee
with $\sigma_1$ only entering in through $Y_1(k_1,\ii\,\eta(k_1)).$ By considering the series of \eref{eq:z-lim-formula} for $z=\infrac{\ii\,\eta(k_1)}{\sqrt{2}k_1\sigma_1}\approx\infrac{\ii\,\kJ\sigma}{\sqrt{2}k_1\sigma_1},$ we can see that if $\beta\sigma_1\ll\sigma,$ then that $z$ will be large, and we can safely apply this paper's small $k$ approximation approach.\footnote{If this $\beta\sigma_1\ll\sigma$ condition fails, then $\sigma_1/\sigma$ must be very large, and our perturbation is very different from a point-like perturbation as usually understood~-- the perturbing particles' typical velocity is such that it more resembles an explosion from a point.} Calculations in \citet{mycalcs}, following the approaches of Appendices~\ref{sec:exploring-the-correlation-equation-using-a-propagator} or \ref{sec:the-landau-approach-to-deriving-the-first-order-correlation-function}, and then  Appendices~\ref{sec:all-space} and~\ref{sec:local}, show that, at leading order in $\beta,$ we get a core-halo pattern exactly as in \eref{eq:Scgrr-3}, with the same pattern function shown in Figure~\ref{fig:figscgr}.  In particular, there is no $\sigma_1$ dependency at leading order.

As for the main, $\sigma_1=\sigma,$ model, the total net entropy creation is at a higher order in $\beta$ than the core-halo pattern. Calculations in \citet{mycalcs} give the total net entropy creation as in \eref{eq:dScg-dt}, but with the replacement 
\be \label{eq:dScg-dt-var}
\num{-0.0116}
\mapsto
\num{-8.31e-3}-\num{3.24e-3}\left(\frac{\sigma_1}{\sigma}\right)^2
.
\ee
The estimated integration errors for each of the two numerical coefficients on the right-hand side are around $\num{1e-5}.$ As we would expect, if we set $\sigma_1=\sigma$ in the right-hand side of \eref{eq:dScg-dt-var}, we recover its left-hand side, $\num{-0.0116}.$  Note that we always get negative total net entropy creation.  The requirement that $\beta\sigma_1\ll\sigma$ implies that the size of the total net entropy creation remains much less than the size of the entropy destruction in the core (or the size of its creation in the halo).

\begin{figure*} 
	\centering
	\includegraphics[width=0.9\textwidth]{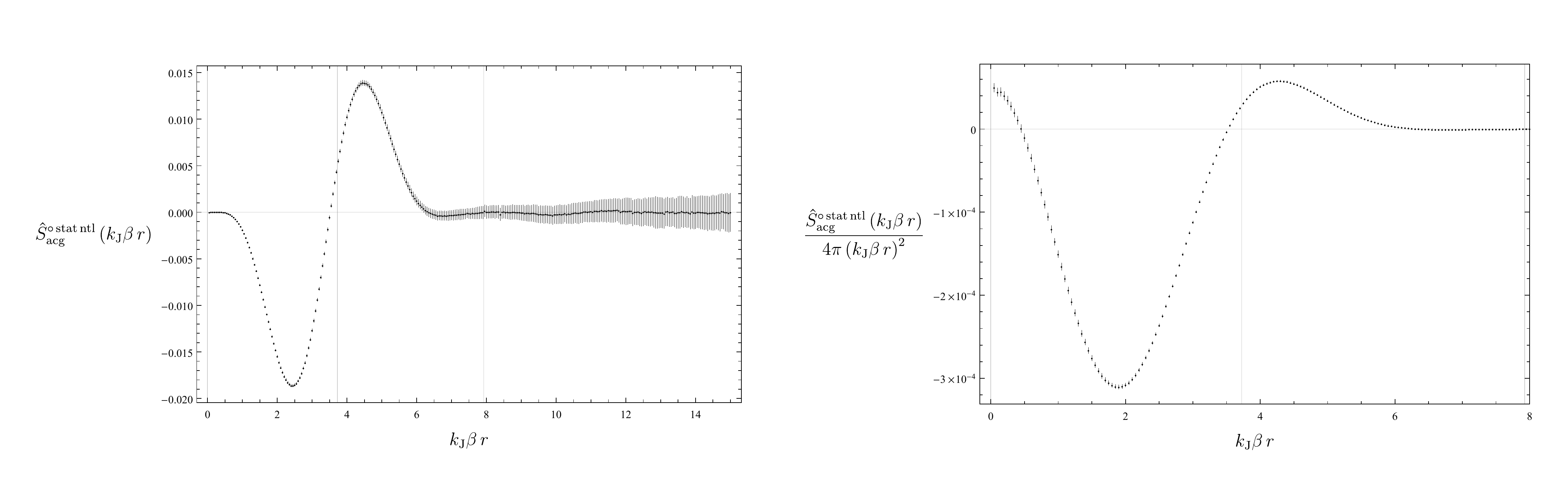}
	\caption{The equivalent of Figure~\ref{fig:figscgr20thszerothorder} for the initially stationary perturbation's next-to-leading order entropy-creation pattern function. This perturbation has all the perturbing particles stationary at time $t=0.$}
	\label{fig:stat}
\end{figure*}

Taking the limit $\sigma_1\to 0$ represents an initial perturbation with all its particles  initially stationary.\footnote{It is also straightforward to derive the equivalent of \eref{eq:f1-fast-early-var} for $\bfino(\ko,\vo)=\ddth(\vo),$ and then see, from the power series expansion of $Y_1(k_1,\omega)$ along the lines of \eref{eq:z-lim-formula}, that this does indeed correspond to the limit $\sigma_1\to 0.$} Since local leading order entropy creation is independent of $\sigma_1,$ we still have exactly the same core-halo equation and pattern as for our main model. The next-to-leading order entropy creation is then given by
\be \label{eq:Scgrr-3-ntl-stat}
\pd{^2S_{\text{acg}}^{\text{stat\,ntl}}}{t\,\partial r}
=
\frac{2\kJ^2\,\sigma\,\beta\,N_1^2\ \ex^{3\kJ\sigma t}}{9\pi^2\,n^2\,B^2}
\,S_{\text{acg}}^{\circ\text{\,stat\,ntl}}(\kJ\beta r)
\,,
\ee
where $S_{\text{acg}}^{\circ\text{\,stat\,ntl}}$ is a dimensionless \emph{next-to-leading order entropy-creation pattern function for the initially stationary perturbation}, which is shown in Figure~\ref{fig:stat} as numerically calculated in \citet{mycalcs}.\\

We can also consider varying the perturbation's initial correlation function  $g_1(1,2,t=0),$ which was assumed to vanish.  Suppose the initial perturbation correlation is altered to be the same as for the underlying DF, so $g_1(1,2,t=0)=g_0(1,2,t=0)$, the latter being set out in, and just before, \eref{eq:q-ansatz-result-main}.  The rule for Laplace transforming time derivatives implies this adds a new term, a constant times $g_0(1,2,t=0),$ to the Laplace transform $\tilde{D}_1,$ as found in \eref{eq:ID1-1}. From \eref{eq:ID1-1}, the dispersion relation, and \eref{eq:q-ansatz-result-main}, it can be seen that the resulting additional term in $I_{D_1}$ is of leading order $4$ in $k_j.$ Via \eref{eq:g1-fast-4}, this gives rise to an order $3$ term in $\gof,$ and does not affect the terms explicitly set out in \eref{eq:bigint-result-swap-app}.  Therefore this new choice of initial perturbation correlation produces the same leading and next-to-leading order entropy creation as our main model with its uncorrelated initial perturbation, and does not affect the results in this paper.

\bsp	
\label{lastpage}
\end{document}